\documentclass[%
reprint,
superscriptaddress,
amsmath,amssymb,
aps,
prx,
]{revtex4-1}

\usepackage{graphicx}
\usepackage{dcolumn}
\usepackage{bm}
\usepackage{color}

\begin{document}

\preprint{APS/123-QED}

\title{Theory of metal-insulator transitions in graphite under high magnetic field}

\author{Zhiming Pan}
\affiliation{International Center for Quantum Materials, School of Physics, Peking University, Beijing 100871, China}
\affiliation{Collaborative Innovation Center of Quantum Matter, Beijing 100871, China}%
\author{Xiao-Tian Zhang}
\affiliation{International Center for Quantum Materials, School of Physics, Peking University, Beijing 100871, China}
\affiliation{Collaborative Innovation Center of Quantum Matter, Beijing 100871, China}%
\author{Ryuichi Shindou}%
\email{rshindou@pku.edu.cn}
\affiliation{International Center for Quantum Materials, School of Physics, Peking University, Beijing 100871, China}%
\affiliation{Collaborative Innovation Center of Quantum Matter, Beijing 100871, China}%

\date{\today}

\begin{abstract}
Graphite under high magnetic field exhibits consecutive metal-insulator (MI) transitions as well as 
re-entrant insulator-metal (IM) transition in the quasi-quantum limit at low temperature.  
In this paper, we identify the low-$T$ insulating phases as excitonic insulators with spin nematic 
orderings. We first point out that graphite under the relevant field regime is in the charge neutrality region, 
where electron and hole densities compensate each other. Based on this observation, 
we introduce interacting electron models with electron pocket(s) and hole pocket(s) and  
enumerate possible umklapp scattering processes allowed under the charge neutrality. 
Employing effective boson theories for the electron models and renormalization group (RG) analyses for  
the boson theories, we show that there exist critical interaction 
strengths above which the umklapp processes become  
relevant and the system enter excitonic insulator phases with long-range order of spin superconducting  
phase fields (``spin nematic excitonic insulator"). We argue that, when a pair of electron and hole 
pockets get smaller in size, a quantum fluctuation of the spin superconducting phase becomes larger 
and destabilizes the excitonic insulator phases, resulting in the re-entrant IM transitions. We also 
show that an odd-parity excitonic pairing between the electron and hole pockets 
reconstruct surface chiral Fermi arc states of electron and hole into a 2-dimensional 
helical surface state with a gapless Dirac cone. We discuss field- and temperature-dependences 
of in-plane resistance by surface transports via these surface states.   
\end{abstract}


\maketitle

\section{introduction}
Graphite under high magnetic field exhibits a metal-insulator transition at low temperature 
($H\ge H_{c,1}\simeq 30 \!\ {\rm T}$)~\cite{tanuma81,iye82}. The transition has been often 
considered as a prototype of one-dimensional 
Peierls density-wave instability associated with the $2k_F$ logarithmic singularity in the 
Lindhard response 
function~\cite{fukuyama78a,yoshioka81,iye84,iye85,takahashi94,takada98,sugihara84,
tesanovic87,macdonald87,yakovenko93}. A transition temperature $T_c$ of the density wave 
ordering is determined by a BCS type gap equation, $\ln T_c \propto -1/\rho(0)$. 
The density of states at the Fermi level $\rho(0)$ is proportional to the magnetic field $H$, 
so that $T_c$ increases monotonically in the magnetic field~\cite{fukuyama78a,yoshioka81,iye84,iye85}.     
Further experiments discovered that graphite shows 
another metal-insulator transition ($H \ge H_{0}\simeq 53$ T)~\cite{ochimizu92,yaguchi98a,yaguchi98b,yaguchi01,fauque13,akiba15,arnold17,zhu17,zhu18} 
as well as an insulator-metal re-entrant transition at higher magnetic 
field ($H =H_{c,2}\simeq 75$ T)~\cite{fauque13,akiba15,arnold17,zhu17,zhu18}. So far, there exist at 
least two distinct low-temperature insulating phases in graphite under high magnetic field: one insulating 
phase ranges in $H_{c,1}<H<H_0$  and the other ranges in 
$H_0<H<H_{c,2}$. The re-entrant transition at $H=H_{c,2}$ indicates 
a presence of a normal metal phase with pristine electron and hole pockets above the transition 
field, bringing about a skepticism against the density wave scenarios. Namely, the 
transition temperature of the density wave phase would increase monotonically in the 
field, until the electron and hole pockets that would form the Peierls 
density wave leave the Fermi level~\cite{fukuyama78a,yoshioka81,iye84,iye85}. 

\begin{figure}
	\centering
	\includegraphics[width=0.9\linewidth]{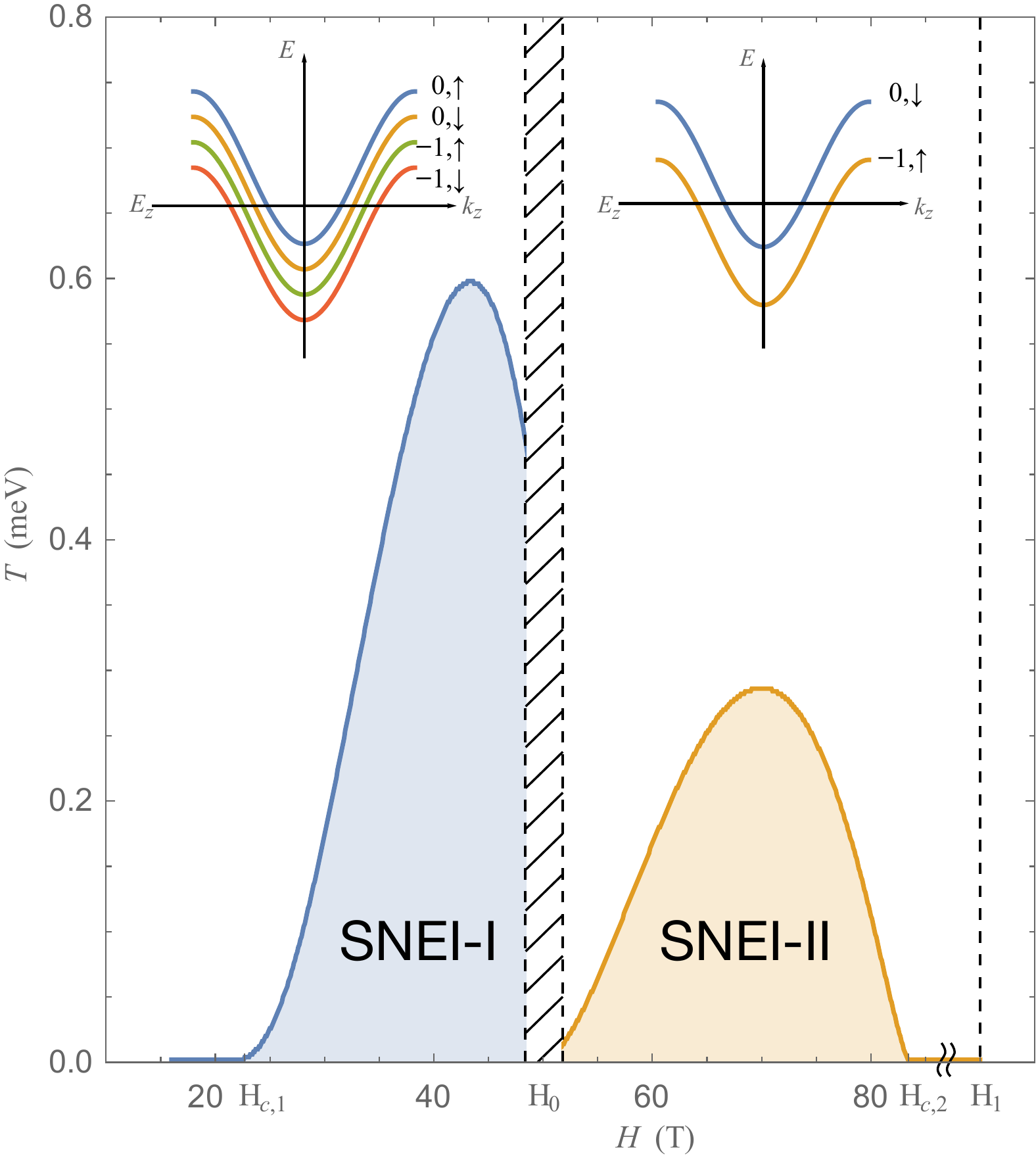}
	\caption{(color online) Theoretical phase diagram for graphite under high magnetic field.  
The phase diagram is obtained from the RG equations, Eqs.~(\ref{rg1a},\ref{rg2a},\ref{rg3a}) for $H<H_0$ and 
Eqs.~(\ref{rg1c},\ref{rg2c},\ref{rg3c}) for $H_0<H<H_1$. ``SNEI-I" 
and ``SNEI-II" stand for two distinct spin nematic excitonic insulator phases 
(strong-coupling phase). For $H<H_0$, the 
electronic state near the Fermi level comprises of two electron pockets ($n=0$ LL with $\uparrow$ 
spin and $\downarrow$ spin) and two hole pockets  ($n=-1$ LL with $\uparrow$ spin and 
$\downarrow$ spin). At $H=H_0$, the outer two pockets ($n=0$ LL 
with $\uparrow$ spin and $n=-1$ LL with $\downarrow$ spin) leave the 
Fermi level. For $H_0<H<H_1$, the electronic state has one electron pocket ($n=0$ LL 
with $\downarrow$) and one hole pocket ($n=-1$ LL with $\uparrow$). 
We choose $H_0=50$ T and  $H_1=120$ T. For a detailed parameter set of the RG 
equations, see Appendix C4. Our theory may not be able to predict much about 
a transition between SNEI-I and SNEI-II phases (a shaded area around 
$H=H_0$); see a discussion in Sec.~XB. $T=0$ metal-insulator 
transition at $H=H_{c,1}$ and insulator-metal transition at $H=H_{c,2}$ are the quantum 
phase transition with the dynamical exponent $z=1$.} 
	\label{fig:1}
\end{figure}

Theoretically, stabilities of the Peierls density wave phases against random single-particle 
backward scatters depend crucially on a commensurability condition 
of an electron filling~\cite{imry76,sham76,fukuyama78b,zhang17}.  
From preceding ab-initio band calculations of graphite under high magnetic field~\cite{takada98,arnold17},  
a sequence of specific values of the field in a range of $30\!\ {\rm T} \lesssim H \lesssim 50 \!\ {\rm T}$ 
satisfy the commensurability condition. Nonetheless, experimental transition temperatures  
of the two insulating phases do not show any dramatic sensitivities on 
certain values of the field in the range. Both of the insulating phases 
range rather broadly in field (over 20 Tesla)~\cite{fauque13,akiba15,arnold17,zhu17,zhu18}.

In this paper, we explain these two low-$T$ insulating phases 
in graphite under the high field 
as manifestation of excitonic insulators with spin nematic orderings. We first argue 
that graphite under high magnetic field ($H \gtrsim 20$T) is in the charge neutrality region, 
where electron density and hole density compensate each other. Based on this observation, 
we begin with interacting electron models with electron pockets and hole pockets, to 
enumerate possible umklapp scattering processes allowed under the charge neutrality condition. 
Using perturbative renormalization group (RG) analyses on their effective boson theories, 
we show that the umklapp terms have critical interaction strength above/below 
which they become relevant/irrelevant on the renormalization. Above the critical interaction 
strength, the umklapp term locks the total displacement field as well as 
spin superconducting phase field. The former locking causes the insulating behavior along the 
field direction, while the latter results in a long-range order of spin quadrupole moment. 
We explain the re-entrant insulator-metal transition in graphite, through a quantum 
fluctuation of the spin superconducting phase field. 
We characterize the spin nematic excitonic insulator phases by out-of-plane (infrared optical)  
conductivity as well as in-plane transport property [out-of-plane current is parallel to 
the field]. The field and temperature dependences of the 
transport properties are consistent with experimental observation in graphite. 

\subsection{issues to be addressed in this paper} 
\subsubsection{direct metal-insulator transition}
Under the magnetic field $H$ ($\parallel z$), kinetic energy part of the three-dimensional semimetal  
takes a form of decoupled one-dimensional quantum chains (or quantum wires). Namely, the kinetic 
energy within the $xy$ plane is quenched by the Landau quantization, while the kinetic energy 
along the field direction remains intact, forming one-dimensional momentum-energy dispersion. 
As a result, the RPA density correlation function is characterized by the Lindhard response function 
in the one-dimension~\cite{fukuyama78a,yoshioka81}. 
The function has the logarithmic singularity at $k_z=2k_F$, where $2k_{F}$ 
is a distance between the left and right Fermi points in the same energy band~\cite{gruner}. Thereby, the system 
has a generic instability toward the charge density wave ordering, that breaks the spatially translational 
symmetry along the field direction~\cite{fukuyama78a,yoshioka81,iye84,iye85,takahashi94,takada98,sugihara84,
tesanovic87,macdonald87,yakovenko93}. 

Meanwhile, graphite under the relevant field regime has four bands that 
run across the Fermi level (two electron pockets and 
two hole pockets; $ H \le H_0 \simeq 53 \!\ {\rm T}$) or two bands 
(one electron pocket and one hole pocket; $H_0 \le H$)~\cite{takada98,arnold17}. 
When each of these four (or two) bands would undergo the Peierls density wave (DW) 
instabilities individually, the respective instabilities would appear at different critical temperatures or 
critical fields. In other words, the graphite transport experiment would observe a step-wise 
increase of the (out-of-plane) resistance $R_{zz}$ on lowering temperature or on increasing 
the magnetic field. 

Nonetheless, the graphite experiment observed a {\it direct} phase transition from high-$T$ normal metal 
phase to the low-$T$ insulating phase~\cite{yaguchi01,fauque13,akiba15,zhu17,zhu18}. 
Around the transition, the resistance along the field 
direction $R_{zz}$ continuously increases~\cite{yaguchi01,fauque13,akiba15,arnold17,zhu17,zhu18} 
and it increases by 100 times within windows of several 
Kelvin or Tesla~\cite{yaguchi01,fauque13,akiba15,zhu18}. 
These experimental observations clearly dictate that all the energy bands 
(pockets) at the Fermi level are gapped out {\it simultaneously} at the transition point. Thereby, 
the key question to be asked here is; {\it what is a ``talking-channel" among these four (or two) 
bands that enables this direct metal-insulator transition ?} 

In this paper, we consider this channel as {\it umklapp scattering terms} and construct a mean-field theory that 
explains this direct metal-insulator transition. To be more specific, all the excitonic insulator phases discussed 
in this paper are stabilized by the umklapp terms that lock a {\it total} displacement field along the field direction, 
a sum of the displacement fields of the four (or two) bands. When the umklapp terms 
become relevant in the standard RG argument sense, the total displacement field 
(electric polarization) is locked, resulting in the electrically insulating behaviour along the 
field direction. By calculating an optical conductivity along the field direction, we 
explicitly demonstrate the presence of finite mobility gaps in the excitonic insulator phases. 

\subsubsection{re-entrant insulator-metal transition}
The second issue is the re-entrant insulator-metal transition observed at 
the higher field region in the graphite experiment~\cite{yaguchi98a,yaguchi98b,yaguchi01,fauque13,akiba15,arnold17,zhu17,zhu18},   
that can hardly be explained by the conventional Peierls DW scenarios. Namely, the 
RPA density correlation function at finite temperature suggests that the transition temperature 
of the Peierls DW phase increases monotonically in the field, until the electron and/or hole 
pockets that would form the DW leave the Fermi level. When they leave the Fermi level, however, 
the electronic state simply ends up in semiconductor phase rather than metallic 
phase. Contrary to this, the graphite experiments clearly observe the 
insulator-{\it metal} re-entrant transition in the higher-field region. The low-$T$ electric 
transport along the field direction above the critical field ($H > H_{c,2} \simeq 75 \!\ {\rm T}$) is 
as metallic as the electric transport in the high-$T$ normal metal phase~\cite{fauque13,akiba15,zhu18}. 
The experiment clearly indicates 
a presence of pristine electron and hole pockets at the Fermi level above the critical field. 

In this paper, we explain this re-entrant insulator-metal transition as a consequence 
of quantum {\it spin} fluctuation enhanced by raising the magnetic field. To be more specific, 
we first point out that the umklapp terms lock not only the total displacement field but also 
a {\it spin superconducting phase field}, a difference between a superconducting phase field of an 
electron/hole pocket with $\uparrow$ spin and hole/electron pocket with $\downarrow$ spin 
respectively. The higher magnetic field makes the electron and hole pockets to be 
smaller in size in the $k_z$ space. In the presence of the repulsive electron-electron interaction, 
the smaller pockets make their Luttinger parameters to be smaller than the unit. Smaller 
Luttinger parameters mean larger quantum fluctuation of superconducting phase field 
as well as the spin superconducting phase field. Thus, we can naturally argue that, in the 
presence of such smaller electron and hole pockets, the umklapp terms suffer from the 
enhanced quantum {\it spin} fluctuation, and become irrelevant in the RG argument sense. 
When the umklapp terms become irrelevant, the spin superconducting phase field as well 
as the total displacement field are unlocked, resulting in the re-entrant insulator-metal 
transition. Importantly, the electronic state still possesses electron and hole pockets  
above the critical field, though their sizes in the $k_z$ space might be small. 

\subsubsection{field-dependence of in-plane resistance}
The third issue to be addressed in this paper is an unusual field-dependence of the 
electric transport in the directions transverse to the magnetic field~\cite{yaguchi98a,yaguchi98b,yaguchi01,fauque13,akiba15,arnold17,zhu17,zhu18}. 
Generally, the bulk electric transport perpendicular to the field is quenched in the clean limit at low 
temperature ($T\ll h\omega_0$; $h\omega_0$ is the cyclotron frequency). Nonetheless, 
the system still has low-$T$ electric transport perpendicular to the field through the so-called 
surface chiral Fermi arc (SCFA) states~\cite{halperin87,balents96}. The associated surface resistance is inversely proportional 
to a length of the arc in the $k_z$ space. The length is approximately equal to the size of the 
respective electron (or hole) pocket in the bulk. The size of the pocket generally 
decreases in the field. 
Thereby, the surface resistance perpendicular to the field is expected to increase in the 
field. Contrary to this theory expectation, the in-plane resistance $R_{xx}$ in the graphite 
under the field $(H\parallel z)$ shows an unusual field-dependence. The low-$T$ 
resistance $R_{xx}$ shows a broad peak around $15\!\ {\rm T} < H < 30\!\ {\rm T}$~\cite{tanuma81,iye82,iye84,iye85,ochimizu92,yaguchi98a,yaguchi98b,yaguchi01,fauque13,akiba15,arnold17,zhu17,zhu18}. 
From $H=30 \!\ {\rm T}$ to $H=H_0 \simeq 53 \!\ {\rm T}$, 
$R_{xx}$ typically reduces by half~\cite{yaguchi98a,yaguchi98b,yaguchi01,fauque13,akiba15,arnold17,zhu17,zhu18}. 
Inside the high-field-side insulating phase ($H_0 < H< H_{c,2} \simeq 75\!\ {\rm T}$), 
the low-$T$ in-plane resistance $R_{xx}$ stays nearly constant in the field~\cite{fauque13,akiba15,arnold17,zhu17,zhu18}. 
For $H_{c,2} < H$, $R_{xx}$ starts increasing in the field again~\cite{zhu18}. 

Field-(nearly) independent and metallic $R_{xx}$ in the high-field-side  
insulating phase can be naturally explained by a novel surface reconstruction of the 
surface chiral Fermi arc (SCFA) states due to the excitonic pairing 
in the bulk. To be more specific, we will show that an {\it odd-parity} 
excitonic pairing between 
electron and hole pockets in the bulk reconstructs the SCFA state of 
electron and that of hole into a $(2+1)$-d helical surface state with a gapless Dirac cone. $R_{xx}$ 
through such a Dirac-cone surface state is determined by carrier density doped in the 
surface region, that is typically independent from the magnetic field. 
Namely, unlike `decoupled' SCFA states of electron and hole, 
the reconstructed Dirac-cone surface state barely changes 
its shape as a function of the magnetic field. At the zeroth order approximation, 
the field only changes a `depth' of a band inversion between electron and hole pockets, 
while the shape of the Dirac-cone 
surface state is mainly determined by the excitonic pairing strength 
inside the inverted band gap. Thereby, one can naturally expect that the surface resistance 
due to the reconstructed Dirac-cone surface state is much less field-dependent than 
that of the decoupled SCFA states of electron and hole.  

\subsection{structure of the paper} 
The structure of the paper is as follows. In the next section with a help of appendix A, 
we argue that the graphite under the relevant field regime ($ 20\!\ {\rm T}< H$) is in 
the charge neutrality region, where electron and hole densities compensate each other. 
Based on this observation, we enumerate in Sec.~III possible umklapp terms 
that are allowed under the charge neutrality condition in the four pockets
model (a model with two electron pockets and two hole pockets; 
$H \le H_{0} \simeq 53\!\ {\rm T}$). Employing a Hartree-Fock 
approximation, we construct effective field theories for possible insulating phases 
that can be stabilized by these umklapp terms (Sec.~IV). There are three such phases;  
spin-nematic excitonic insulator, magnetic Mott insulator and plain excitonic insulator phases. 
Using renormalization group (RG) analyses, we argue typical 
field-dependences of the respective transition temperatures of these three 
phases and conclude that the spin nematic 
excitonic insulator (SNEI-I) phase could naturally fit in the phenomenology of the 
low-field-side out-of-plane insulating phase ($H_{c,1} < H < H_{0}$) 
in the graphite experiment (Sec.~V). In Sec.~VI, we enumerate possible umklapp terms 
that are allowed under the charge neutrality condition in the two-pockets model 
(one electron and one hole pocket; $H_0 < H$). We construct effective field 
theories for the possible insulating phases that can be stabilized by the umklapp terms. 
We found two such phases; a phase with two superposed charge density waves and a spin nematic 
excitonic insulator (SNEI-II) phase. Using the RG analyses, we conclude that the SNEI-II phase 
can naturally explain the high-field-side out-of-plane insulating phase 
($H_0 < H < H_{c,2}$).  In Secs.~VII and VIII, 
we argue field-dependences of the in-plane 
resistance in the graphite experiment by the surface electric transports. Especially, 
we show in Sec.~VIII that the odd-parity excitonic pairing in the two-pockets model reconstructs 
the surface chiral Fermi arc (SCFA) states of electron and hole into a $(2+1)$-d helical surface 
state with a gapless Dirac cone. The surface Dirac-cone state could naturally explain 
field-(nearly) independent and metallic behaviour of the in-plane resistance inside the high-field-side 
insulating phase. After a brief summary in Sec.~IX, we give a 
discussion with complementary viewpoint (Sec. X). 

\section{charge neutrality regime in graphite under high magnetic field}
Low-temperature transport properties of graphite are dominated by four $\pi$-orbital bands around 
zone boundaries of the first Brillouin zone~\cite{wallace47,slonczewski55,mcclure57}. 
Graphite is a three-dimensional $AB$ stacking of graphene 
layers. A unit cell has two graphene layers and it has four inequivalent carbon 
sites. Call them as $A$, $A'$, $B$ and $B'$. $A$ and $B$ share the same layer, and so 
do $A'$ and $B'$.  $A$ comes right above $A'$ in the cell. The electronic band structure near the Fermi 
level of graphite is composed by $\pi$ orbitals of carbon atoms that are odd under the mirror with respect 
to the layer, for example, $2p_z$ orbital~\cite{wallace47,slonczewski55,mcclure57}.  
$\pi$ orbitals of $A$ and $A'$ carbon atoms hybridize rather 
strongly, forming two $\pi$ orbital bands at the zone boundaries that have large momentum-energy dispersions 
along the $c$-axis ($4000$ K). Call these $\pi$ orbitals as $\pi_A$ and $\pi_{A'}$ respectively. 
$\pi$ orbitals of $B$ and $B'$ hybridize much weakly, as 
$B$ and $B'$ locate right above the centers of the hexagon in their neighboring layers.  
These two, which we call $\pi_{B}$ and $\pi_{B'}$ henceforth, form 
two degenerate bands at the zone boundaries that 
have a weaker energy-momentum dispersion along the $c$-axis ($400$ K). 

Under the field along the $c$-axis, the four bands in the zone boundaries are 
split into Landau 
levels (LLs)~\cite{mcclure57,inoue62,g-dresselhaus65,nakao76,takada98,arnold17}. For 
$H \gtrsim 30$ T, the $n=0$ LLs with $\uparrow$ spin and $\downarrow$ spin 
 form two electron pockets around $k_z=0$, and the $n=-1$ LLs with $\uparrow$ spin 
and $\downarrow$ spin form two hole 
pockets around $k_z=\pi/c_0$. Here $c_0$ is a lattice constant along the $c$-axis. 
According to the band calculation, the outer electron pocket ($n=0$ LL with $\uparrow$ spin) 
and the outer hole pocket ($n=-1$ LL with $\downarrow$ spin) leave the 
Fermi level at $H=H_0 \simeq 53$ T.

The Hall conductivity measurements in a regime of 
$20\!\ {\rm T} \lesssim H \lesssim 60 \!\ {\rm T}$~\cite{uji98,kopelevich09,kumar10,akiba15} 
suggest that the number of the electron states and that of the hole states compensate each 
other almost completely. An estimation gives $N_e-N_h:L_z/c_0=10^{-4}:1$, where $L_z$ is 
a linear dimension along the $c$-axis, $N_e$ and $N_h$ are numbers of the $k_z$ points 
within the electron pockets and hole pockets respectively 
[$k_z$ is a crystal momentum along the $c$-axis] (see Appendix A for 
a validity of the estimation).

\begin{figure}[b]
	\centering
	\includegraphics[width=0.9\linewidth]{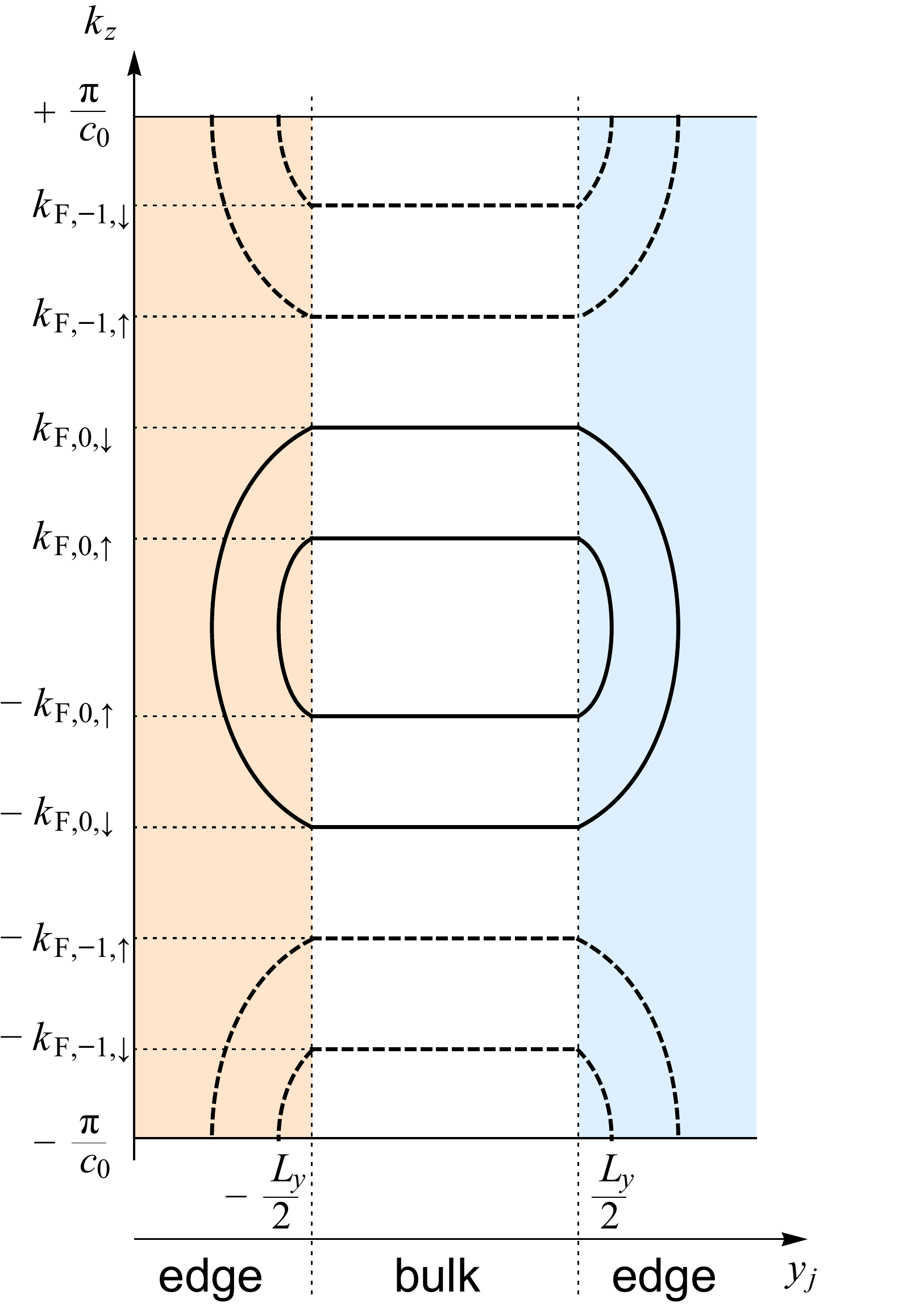}
	\caption{(color online) Schematic picture of electronic states of graphite under 
high field ($H<H_0$). Solid/dotted lines describe Fermi surfaces of two electron/hole pockets 
in both bulk and edge regions. Two electron/hole pockets in the bulk region are terminated by   
electron/hole-type surface chiral Fermi arc states in edge regions respectively. Namely, 
$E_{0,\sigma}(k_z,y_j)/E_{1,\sigma}(k_z,y_j)$ goes higher/lower in energy, when $y_j$ goes from 
the bulk region to the edge regions (see Appendix A).} 
	\label{fig:2}
\end{figure}

\section{four pockets model ($H<H_0$)}
Based on this observation, we consider an electron model with two electron pockets 
($n=0$ LL with $\uparrow$ spin and that with $\downarrow$ spins) 
and two hole pockets ($n=-1$ LL with $\uparrow$ spin and that with $\downarrow$ spins) with 
the charge neutrality condition ($N_e=N_h$); 
\begin{align}
H_{\rm kin}&=\sum_{k_z,j}  
\sum_{n=0,-1}\sum_{\sigma=\uparrow,\downarrow} 
E_{n,\sigma}(k_z) c^{\dagger}_{(n,\sigma),j}(k_z) c_{(n,\sigma),j}(k_z). \label{Hkin}
\end{align}
The two electron pockets encompass $k_z=0$ and two hole pockets are around the 
zone boundary $k_z=\pm \pi/c_0$,  
\begin{align}
E_{n,\sigma}(k_z) &= - 2\gamma_2 \big[\cos(k_z c_0) - \cos(k_{F,n,\sigma} c_0)\big], \label{Enk}
\end{align}
with $0<k_{F,0,\uparrow}<k_{F,0,\downarrow}<\pi/(2c_0)<k_{F,-1,\uparrow}<k_{F,-1,\downarrow}<\pi/c_0$ 
\cite{yoshioka81}. 
An index $j$ $(=1,2,\cdots,(L_xL_y)/(2\pi l^2))$ counts degenerate electron states within each LL. 
$l$ denotes a magnetic length, $l \equiv \sqrt{\hbar c/(eH)}$. Since the kinetic energy part 
takes the exactly same one-dimensional momentum-energy dispersion along $k_z$ direction 
for different $j$, we regard the system as coupled chains and call  $j$ as a `chain index'~\cite{biagini01,tsai02a,tsai02b}. 
The charge neutrality condition takes a form of 
\begin{eqnarray}
k_{F,0,\uparrow}+k_{F,0,\downarrow}+
k_{F,-1,\uparrow}+k_{F,-1,\downarrow}=\frac{2\pi}{c_0}. \label{CNP}
\end{eqnarray}

For low-temperature electric transports at those temperature much below the band width ($2\gamma_2 \simeq 400$ K), 
the kinetic energy part can be linearized around the Fermi points of each 
pockets ($k_{z} \simeq \pm k_{F,n,\sigma}$);
\begin{align}
H_{\rm kin} = \sum_{j} \sum_{a,\tau} \tau v_{F,a} \int dz \psi^{\dagger}_{a,\tau,j}(z) 
i\partial_z \psi_{a,\tau,j}(z) + \cdots. 
\end{align}
A chirality index $\tau$ specifies left mover ($\tau=-1$) or right mover ($\tau=+1$). 
$v_{F,a}$ is a bare Fermi velocity of each pocket with $a\equiv (n,\sigma)$. 
For simplicity, we label $(n,\sigma)=(0,\uparrow),(0,\downarrow), (-1,\uparrow)$ and 
$(-1,\downarrow)$ as $a=1,2,3$ and $4$ respectively throughout this paper, e.g. 
$k_{F,a} \equiv k_{F,n,\sigma}$, $c_{a,j}(k_z) \equiv c_{(n,\sigma),j}(k_z)$, 
and $\psi_{a,\pm,j}(z) \equiv \psi_{(n,\sigma),\pm,j}(z)$. 
$\psi_{a,\pm,j}(z)$ is a slowly-varying 
Fourier transform of those $c_{a,j}(k_z)$ around $k_z \simeq \pm k_{F,a}$;
 \begin{eqnarray}
\psi_{a,\tau,j}(z) \equiv \frac{1}{\sqrt{L_z}} \sum_{|k_z -\tau k_{F,a}|<\Lambda} c_{a,j}(k_z) e^{i(k_z - \tau k_{F,a})z}.  
\label{slow}
\end{eqnarray}

A short-ranged repulsive interaction is considered;
\begin{align}
H_{\rm int} &=\sum_{\sigma,\sigma'} \sum_{c,d=A,A',B,B'} 
\int d{\bm r} \int d{\bm r}' V({\bm r}-{\bm r}') \nonumber \\ 
& \hspace{0.8cm}
\times \psi^{\dagger}_{\sigma}({\bm r},c)\psi^{\dagger}_{\sigma'}({\bm r}',d) 
\psi_{\sigma'}({\bm r}',d)\psi_{\sigma}({\bm r},c), \label{Hint1} 
\end{align}
where 
\begin{align}
V({\bm r}) &\equiv \frac{g}{(\sqrt{2\pi})^3 l_{0,x}l_{0,y} l_{0,z}} 
e^{-\frac{z^2}{2l^2_{0,z}}--\frac{x^2}{2l^2_{0,x}}-\frac{y^2}{2l^2_{0,y}}}, \label{Hint2} 
\end{align} 
${\bm r}\equiv (x,y,z)$, $\sigma,\sigma'=\uparrow,\downarrow$, and  
$g>0$. $l_{0,\mu}$ denotes an interaction length along 
the $\mu$-direction. $\psi^{\dagger}_{\sigma}({\bm r},c)$ denotes 
an electron creation at $\pi$-orbital $\pi_{c} \!\ (c=A,A',B,B'$) of carbon atom 
at ${\bm r}$ with spin $\sigma$. The creation 
field can be expanded in term of single-particle bases of the $n=0$ and $n=-1$ LLs in the Landau 
gauge; 
\begin{widetext}
\begin{align}
\left(\begin{array}{c}
\psi_{\sigma}({\bm r},A) \\
\psi_{\sigma}({\bm r},A') \\
\psi_{\sigma}({\bm r},B) \\
\psi_{\sigma}({\bm r},B') \\
\end{array}\right) = \sum_{j} \frac{ e^{ik_j x}}{\sqrt{L_x}}\sum_{\tau=\pm} 
\bigg\{ 
\left(\begin{array}{c} 
\gamma_{A,\sigma} Y_{0,j}(y) \\
\gamma_{A',\sigma} Y_{0,j}(y) \\
\gamma_{B,\sigma} Y_{1,j}(y) \\
\gamma_{B',\sigma} Y_{1,j}(y) \\ 
\end{array}\right) e^{i\tau k_{F,0,\sigma}z} \psi_{(0,\sigma),\tau,j}(z) +  
\left(\begin{array}{c} 
0 \\
0 \\
\eta_{B,\sigma} Y_{0,j}(y) \\
\eta_{B',\sigma} Y_{0,j}(y) \\ 
\end{array}\right) e^{i\tau k_{F,-1,\sigma}z} \psi_{(-1,\sigma),\tau,j}(z) \bigg\}, \label{exp1} 
\end{align}
\end{widetext}
where 
\begin{align}
Y_{0,j}(y) &\equiv \frac{1}{\sqrt{\sqrt{\pi}l}} e^{-\frac{(y-y_j)^2}{2l^2}}, \label{exp2} \\ 
Y_{1,j}(y) &\equiv \sqrt{2}l \frac{d}{dy_j} Y_{0,j}(y) = \frac{\sqrt{2}(y-y_j)}{\sqrt{\sqrt{\pi}l^3}} 
 e^{-\frac{(y-y_j)^2}{2l^2}}, \label{exp3}  
\end{align}
with $y_j \equiv k_j l^2$ and $k_j \equiv 2\pi j/L_x$.  
The slowly varying field $\psi_{(n,\sigma),\tau,j}(z) \equiv \psi_{a,\tau,j}(z)$ was 
defined in Eq.~(\ref{slow}) with 
$a\equiv (n,\sigma)$. $\gamma_{c,\sigma}$ ($c=A,A',B,B'$) comprises 
an eigenvector of a $4$ by $4$ 
SWM (Slonczewski-Weiss-McClure) Hamiltonian at 
$k_z = \pm k_{F,0,\sigma}$~\cite{slonczewski55,mcclure57,inoue62,g-dresselhaus65,nakao76}. 
$\eta_{c,\sigma}$ ($c=B,B'$) comprises the eigenvector at $k_z = \pm k_{F,-1,\sigma}$.  
$L_x$ is a linear dimension of the system size along the $x$-direction.  
A substitution of Eqs.~(\ref{exp1},\ref{exp2},\ref{exp3})
into Eq.~(\ref{Hint1}) and expansion in $\psi_{(n,\sigma),\tau,j}(z) \equiv \psi_{a,j}(z)$  
lead to scatterings between different pockets (inter-pocket scattering) 
and scatterings within the same pocket (intra-pocket scattering).

\begin{figure}[h]
	\centering
	\includegraphics[width=0.75\linewidth]{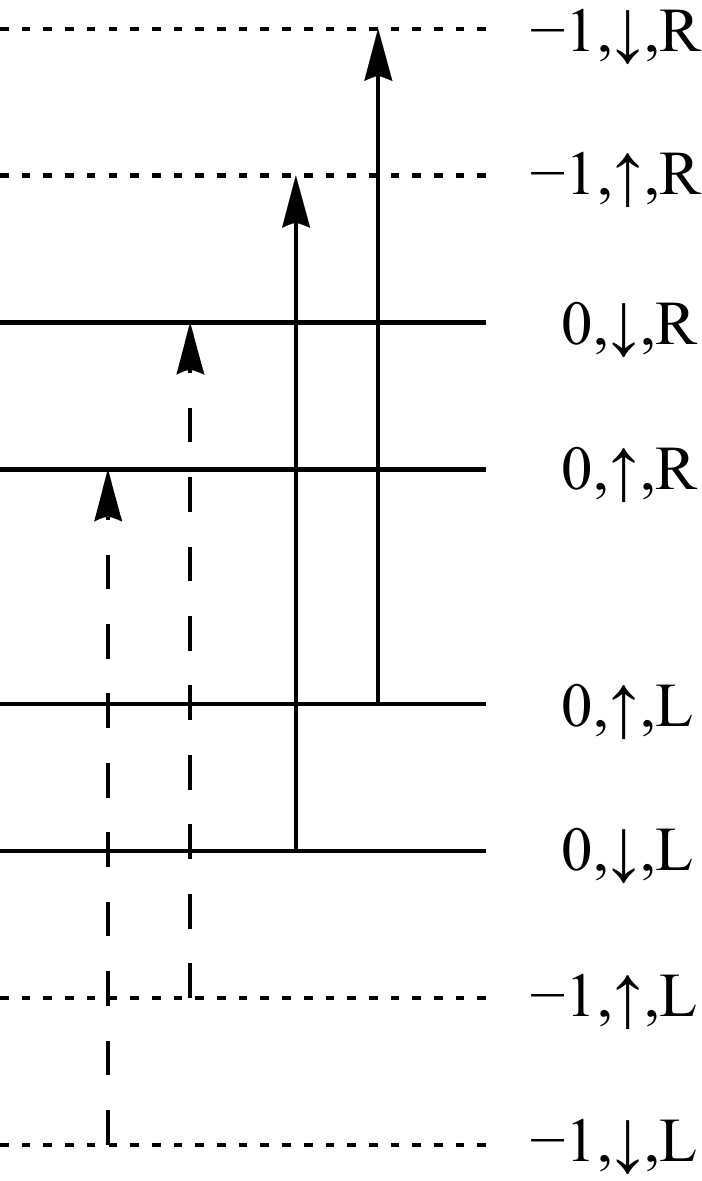}
	\caption{Schematic pictures of one of the umklapp scattering, $H_{{\rm u},2}$. 
As in Fig.~\ref{fig:2}, the vertical axis denotes the momentum along the field direction ($k_z$), 
while the horizontal axis denotes the chain index $y_j=k_j l^2$ with $k_j\equiv 2\pi j/L_x$ 
 $(j=1,2,\cdots,L_xL_y/(2\pi l^2))$. 
The two-particle scatterings with solid/dotted arrows are the exchange processes ($m=n$) 
of the first/fourth terms in Eq.~(\ref{Hu}) with $(0,\uparrow) \equiv 1$, $(0,\downarrow)\equiv 2$, 
$(-1,\uparrow) \equiv 3$ and $(-1,\downarrow) \equiv 4$.} 
	\label{fig:3a}
\end{figure}

\begin{figure}[h]
	\begin{minipage}[t]{0.5\linewidth} 
		\centering 
		\includegraphics[width=1.5in]{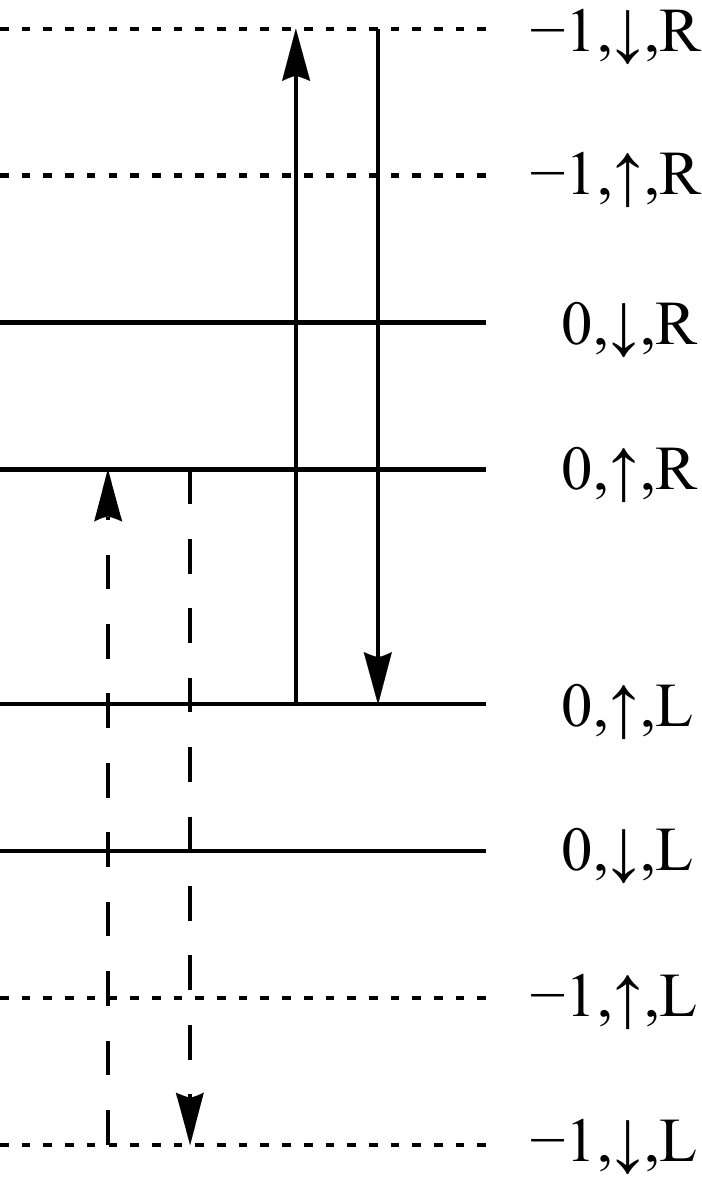} 
	\end{minipage}%
	\begin{minipage}[t]{0.5\linewidth} 
		\centering 
		\includegraphics[width=1.5in]{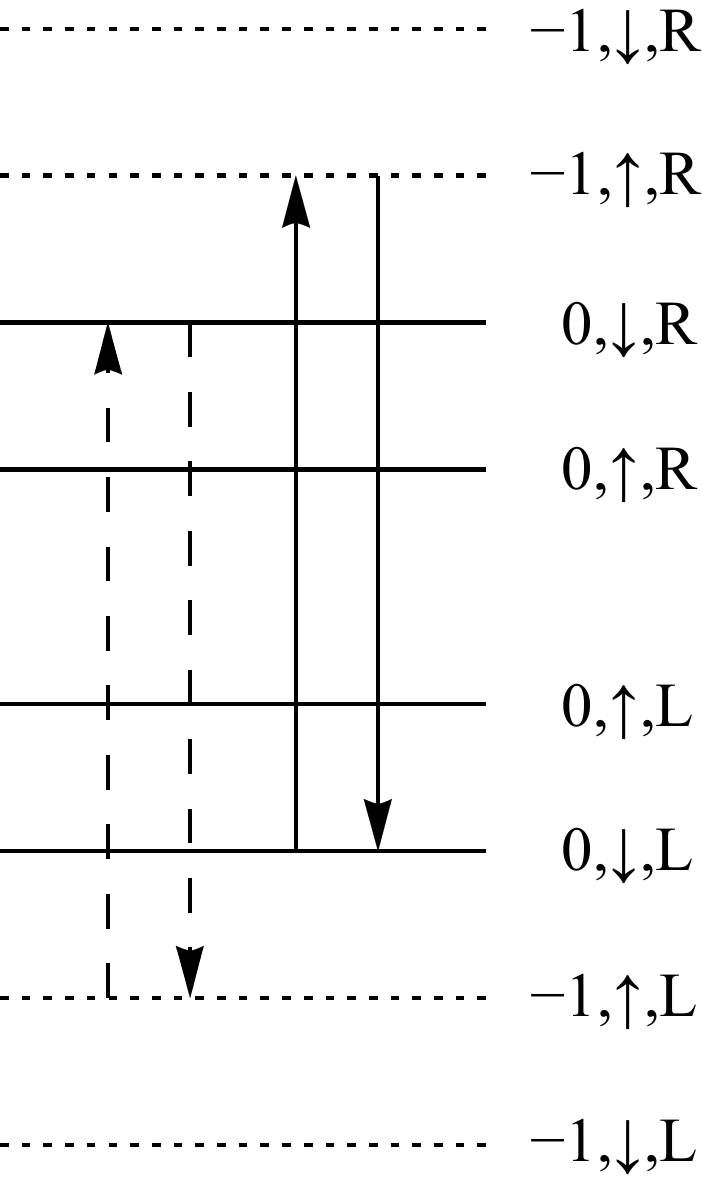} 
	\end{minipage} 
	\caption{Schematic pictures of one of the inter-pocket scatterings, $H_{{\rm b},2}$. 
They are the exchange processes ($m=n$) of the first two terms in Eq.~(\ref{Hb}) with 
$(0,\uparrow) \equiv 1$, $(0,\downarrow)\equiv 2$, 
$(-1,\uparrow) \equiv 3$ and $(-1,\downarrow) \equiv 4$. As in Fig.~\ref{fig:2}, the vertical axis denotes 
the momentum along the field direction ($k_z$), while the horizontal axis denotes the chain 
index $y_j=k_j l^2$ with $k_j\equiv 2\pi j/L_x$ $(j=1,2,\cdots,L_xL_y/(2\pi l^2))$. } 
	\label{fig:3b}
\end{figure}

In this paper, we take into consideration only umklapp scattering terms that 
are allowed under the charge neutrality condition (Fig.~\ref{fig:3a}), 
inter-pocket scattering terms between {\it opposite} chiralities (Fig.~\ref{fig:3b}), 
and intra-pocket scatterings $H_{\rm f}$. This is because, in fermionic functional 
renormalization group analyses~\cite{abrikosov70,brazovskii71,yakovenko93}, these  
scatterings are coupled with one another at the one-loop level; they 
have larger chances to become relevant upon the renormalization than those scattering 
terms omitted. 

Under the charge neutrality condition (Eq.~(\ref{CNP})), 
the interaction allows the following four umklapp terms 
and their hermitian conjugates;
\begin{align}
H_{\rm u} = \sum_{j,m,n}\left\{\begin{array}{l}
\psi^{\dagger}_{4,+,n}\psi^{\dagger}_{3,+,j+m-n} \psi_{1,-,m} \psi_{2,-,j}, \\ 
\psi^{\dagger}_{2,+,n}\psi^{\dagger}_{3,+,j+m-n} \psi_{1,-,m}\psi_{4,-,j}, \\
\psi^{\dagger}_{4,+,n}\psi^{\dagger}_{1,+,j+m-n} \psi_{3,-,m} \psi_{2,-,j}, \\
\psi^{\dagger}_{2,+,n}\psi^{\dagger}_{1,+,j+m-n} \psi_{3,-,m}\psi_{4,-,j}. \  
\end{array}\right. \label{Hu}
\end{align} 
Due to the translational symmetry along $x$ in the Landau gauge, the scattering processes 
conserve a momentum $k_j \equiv 2\pi j/L_x$ that is conjugate to $x$. 
In eq.~(\ref{Hu}), integrals over the spatial coordinate $z$, and the scattering matrix 
elements that depend on $z$ and $j,m,n=1,2,\cdots,(L_xL_y)/(2\pi l^2)$ are omitted 
for clarity. For example, the first and fourth terms in Eq.~(\ref{Hu})  take the following 
explicit form with their Hermitian conjugates,
\begin{align}
&({\rm 1st} \  {\rm and} \ {\rm 4th} \ {\rm terms} \  {\rm in} \ {\rm Eq}.~(\ref{Hu})) \nonumber \\ 
& \ \ = 2 \sum_{j,m,n} V^{(12)}_{m-n,j-n} \int dz \int dz' \!\ 
e^{-\frac{(z-z')^2}{2l^2_{0,z}}} \nonumber \\
& \hspace{-0.cm} \bigg\{ e^{-ik_{F,3}z-ik_{F,4}z'-ik_{F,2}z'-ik_{F,1}z} \nonumber \\
& \hspace{0.3cm} \Big( 
\psi^{\dagger}_{4,+,n} \psi^{\dagger}_{3,+,j+m-n} \psi_{1,-,m} \psi_{2,-,j} \nonumber \\
& \hspace{0.4cm} + \psi^{\dagger}_{2,+,n}\psi^{\dagger}_{1,+,j+m-n} \psi_{3,-,m}\psi_{4,-,j} \Big) 
+ {\rm h.c.}\bigg\}. \label{scat-u-12a}
\end{align} 
The matrix element in Eq.~(\ref{scat-u-12a}) are given by a dimensionless function $f^{(12)}(x,y)$ as 
\begin{eqnarray}
V^{(12)}_{m,n} \equiv \frac{g}{L_x} \frac{1}{2\pi l_{0,z} l} 
f^{(12)}(y_{m}/l,y_n/l). \label{scat-u-12}
\end{eqnarray}
The function $f^{(12)}(x,y)$ can be calculated 
by the direct substitution of Eqs.~(\ref{Hint2},\ref{exp1},\ref{exp2},\ref{exp3}) 
into Eq.~(\ref{Hint1}). $g$ is from Eq.~(\ref{Hint2}). 

In addition to $H_{\rm u}$, we consider the inter-pocket scatterings between the 
{\it opposite} chirality ($H_{\rm b}$)  
as well as the intra-pocket scatterings ($H_{\rm f}$). They are   
\begin{align}
H_{\rm b} = \sum_{j,m,n} \left\{\begin{array}{l}
\psi^{\dagger}_{4,\pm,n}\psi^{\dagger}_{1,\mp,j+m-n} \psi_{1,\mp,m}\psi_{4,\pm,j}, \\  
\psi^{\dagger}_{3,\pm,n}\psi^{\dagger}_{2,\mp,j+m-n} \psi_{2,\mp,m}\psi_{3,\pm,j}, \\  
\psi^{\dagger}_{4,\pm,n}\psi^{\dagger}_{2,\mp,j+m-n} \psi_{2,\mp,m}\psi_{4,\pm,j}, \\   
\psi^{\dagger}_{3,\pm,n}\psi^{\dagger}_{1,\mp,j+m-n} \psi_{1,\mp,m}\psi_{3,\pm,j}, \\ 
\psi^{\dagger}_{4,\pm,n}\psi^{\dagger}_{3,\mp,j+m-n} \psi_{3,\mp,m}\psi_{4,\pm,j}, \\  
\psi^{\dagger}_{2,\pm,n}\psi^{\dagger}_{1,\mp,j+m-n} \psi_{1,\mp,m}\psi_{2,\pm,j}, \
\end{array}\right. \label{Hb}  
\end{align}
and 
\begin{align}
H_{\rm f} = \sum_{a} \sum_{j,m,n} 
\psi^{\dagger}_{a,n} \psi^{\dagger}_{a,j+m-n} \psi_{a,m} \psi_{a,j},  \label{Hf}
\end{align}
with $\psi_{a,n}(z) \equiv 
e^{ik_{F,a}z}\psi_{a,+,n}(z) + e^{-ik_{F,a}z}\psi_{a,-,n}(z)$ ($a=1,2,3,4$; 
see Eq.~(\ref{Hfa}) for an actual form of $H_{\rm f}$). We do not take into account 
the inter-pocket scatterings between the {\it same} chirality, because, at the one-loop level 
of the fermionic renormalization group equations~\cite{abrikosov70,brazovskii71,yakovenko93}, they are decoupled 
from $H_{\rm u}$, $H_{\rm b}$ and $H_{\rm f}$, and do not grow up into larger values upon the 
renormalization.   

\section{effective boson theory}
In this section, we construct effective field theories of possible insulating phases that 
are stabilized by the umklapp scattering 
terms in Eq.~(\ref{Hu}). To this end, we first assume that the low-$T$ insulating 
phases in the graphite experiment do not break the translational symmetries within the graphene 
plane [the graphene plane is perpendicular to the field ($z$)]. The assumption apparently does not 
contradict with any experimental observations in the past literatures
~\cite{tanuma81,iye82,iye84,iye85,ochimizu92,yaguchi98a,yaguchi98b,yaguchi01,fauque13,
akiba15,arnold17,zhu17,zhu18}.  We thus introduce as `mean fields' the pairings among 
electron creation/annihilation operators {\it within the same chain}, and treat the 
inter-chain electron-electron interactions within a Hartree-Fock approximation. To be more specific, we keep only the 
direct process (Hartree term: $j=n$) and exchange process (Fock term: $m=n$) in Eqs.~(\ref{Hu},\ref{Hb},\ref{Hf}). 
Within such effective theory framework, the metal-insulator transitions in the graphite experiment are described by 
a phase transition between a phase with the mean fields being zero and a phase with the mean fields being finite. 
The former phase corresponds to the high-$T$ normal metal phase and the latter 
corresponds to the low-$T$ insulating phases.
 
To do this construction transparently, we first bosonize the slowly-varying 
fermion field in terms of two phase variables defined for each pocket $a=(n,\sigma)$ and each chain 
$j=1,\cdots,(L_xL_y)/(2\pi l^2)$~\cite{giamarchi03,zhang17}; 
\begin{align}
\left\{\begin{array}{l}
\psi_{a,+,j}(z) \equiv \frac{\eta_{a,j}}{\sqrt{2\pi \alpha}} 
e^{-i(\phi_{a,j}(z)-\theta_{a,j}(z))}, \\ 
\psi_{a,-,j}(z) \equiv \frac{\eta_{\overline{a},j}}{\sqrt{2\pi \alpha}} 
e^{-i(-\phi_{a,j}(z)-\theta_{a,j}(z))}. \\ 
\end{array}\right. \label{bosonization1}
\end{align}
Here $(n,\sigma)=(0,\uparrow),(0,\downarrow),(-1,\uparrow)$ and $(-1,\downarrow)$ 
are abbreviated as $a=1,2,3$ and $4$ respectively. 
$\alpha$ is a short-range cutoff for the spatial coordinate $z$. $\phi_{a,j}(z)$, $\theta_{a,j}(z)$ 
and $\partial_z \theta_{a,j}(z)$ are a displacement field along the field direction ($z$), 
superconducting phase field, and current density field along the field respectively.  
They are associated with the pocket $a$ and the $j$-th chain. The displacement field and 
superconducting phase field cannot be simultaneously definite; they are canonical 
conjugate to each other; 
\begin{eqnarray}
[ \phi_{a,j}(z), \partial_{z'} \theta_{b,m}(z') ] = i\delta_{a,b} \delta_{j,m} \delta(z-z'). \label{bosonization2}
\end{eqnarray}
$\eta_{a,j}$ and $\eta_{\overline{a},j}$ in Eq.~(\ref{bosonization1}) are Klein factors ensuring 
the anticommutation relation 
among fermion fields on different chains ($j$), pockets ($a$) and chiralities ($\tau=\pm$);  
$\{ \eta_{a,j}, \eta_{b,m} \} = \{ \eta_{\overline{a},j}, \eta_{\overline{b},m} \} 
= \delta_{a,b} \delta_{j,m}$, and $\{ \eta_{a,j}, \eta_{\overline{b},m} \} = 0$. 
Due to the Klein factor, the interaction parts given in Eqs.~(\ref{Hu},\ref{Hb},\ref{Hf}) {\it cannot} be 
fully bosonized without approximation. 

To obtain the effective boson theories of the insulating phases, 
we thus employ the Hartree-Fock approximation for the inter-chain 
interactions in Eqs.~(\ref{Hu},\ref{Hb},\ref{Hf}), to keep only direct process ($j=n$) and 
exchange process ($m=n$) in Eqs.~(\ref{Hu},\ref{Hb},\ref{Hf}). 
This leads to a fully-bosonized Hamiltonian  
\begin{align} 
H_{\rm kin} + H_{\rm u} &+ H_{\rm b} + H_{\rm f} 
=  H_0 + \sum^4_{i=1} H_{{\rm u},i} + \sum^4_{i=1} H_{{\rm b},i} + \cdots, \label{H00} \\ 
H_0 &= \sum_m \sum^4_{a=1} 
\int dz \Big\{ \frac{u_aK_a \pi }{2} [\Pi_{a,m}(z)]^2 \nonumber \\ 
&\hspace{2cm} +\frac{u_a}{2\pi K_a} [\partial_z \phi_{a,m}(z)]^2 \Big\}, \label{H0} 
\end{align}
and $\pi \Pi_{a,j}(z) \equiv \partial_z \theta_{a,j}(z)$. $K_a$ and $u_a$ are Luttinger parameter and 
Fermi velocity of a pocket with $a=(n,\sigma)$  that are renormalized by the intra-pocket 
forward scatterings $H_{\rm f}$ (see appendix B for its details). 
As in the standard bosonization, the renormalizations are given by two parameters 
$g_{2,a} (>0)$ and $g_{4,a} (>0)$ as,   
\begin{align}
\frac{u_{a}}{v_{F,a}} &= \sqrt{\Big(1 + \frac{g_{4,a}}{2\pi v_{F,a}}\Big)^2 -\Big(\frac{g_{2,a}}{2\pi v_{F,a}}\Big)^2}, \label{luttinger1} \\
K_a &= \sqrt{\frac{2\pi v_{F,a} + g_{4,a} - g_{2,a}}
{2\pi v_{F,a} + g_{4,a} + g_{2,a}}}. \label{luttinger2}
\end{align} 
See also Appendix B for expressions of $g_{2,a}$ and $g_{4,a}$ in terms of $g$ in 
Eq.~(\ref{Hint2}). The Hartree and Fock terms in the umklapp scatterings of Eq.~(\ref{Hu}) are bosonized as
\begin{align}
H_{{\rm u},1}& = \sum_{j,m} M^{(1)}_{j-m} \int dz \!\ \Big\{ \sigma_{3\overline{1},j}  \sigma_{4\overline{2},m}  
\cos \big[Q^{13}_{+,j} + Q^{24}_{+,m}\big] \nonumber \\ 
& \hspace{1.5cm}  + \!\ \sigma_{\overline{3}1,j} \sigma_{\overline{4}2,m}  
\cos \big[Q^{13}_{-,j}+Q^{24}_{-,m}\big] \Big\}, \label{Hu1} \\
H_{{\rm u},2}& = \sum_{j,m} M^{(2)}_{j-m} \int dz \!\ \Big\{ \sigma_{3\overline{2},j} \sigma_{4\overline{1},m}  
\cos \big[Q^{23}_{+,j} + Q^{14}_{+,m}\big] \nonumber \\ 
& \hspace{1.5cm}  + \!\ \sigma_{\overline{3}2,j} \sigma_{\overline{4}1,m} 
\cos \big[Q^{23}_{-,j}+Q^{14}_{-,m}\big] \Big\}, \label{Hu2} \\
H_{{\rm u},3}& = \sum_{j,m} M^{(3)}_{j-m} \int dz \!\ \Big\{ \sigma_{3\overline{1},j} \sigma_{2\overline{4},m}  
\cos \big[Q^{13}_{+,j} + Q^{24}_{-,m}\big] \nonumber \\ 
& \hspace{1.5cm}  + \!\ \sigma_{\overline{3}1,j} \sigma_{\overline{2}4,m} 
\cos \big[Q^{13}_{-,j}+Q^{24}_{+,m}\big] \Big\}, \label{Hu3} \\
H_{{\rm u},4}& = \sum_{j,m} M^{(4)}_{j-m} \int dz \!\ \Big\{ \sigma_{3\overline{4},j} \sigma_{2\overline{1},m}  
\cos \big[Q^{34}_{-,j}+Q^{12}_{+,m}\big] \nonumber \\ 
& \hspace{1.5cm}  + \!\ \sigma_{\overline{3}4,j} \sigma_{\overline{2}1,m}  
\cos \big[Q^{34}_{+,j}+Q^{12}_{-,m}\big] \Big\}, \label{Hu4} 
\end{align}
where 
\begin{align}
Q^{ab}_{\pm,j} \equiv \phi_{a,j}+\phi_{b,j}\pm (\theta_{a,j}-\theta_{b,j}), \label{Qab}
\end{align}
with $a,b=1,2,3,4$. $\sigma_{a\overline{b},j}$ and $\sigma_{\overline{a}b,m}$ are Ising variables 
associated with the Klein factors 
within the same chain, $\sigma_{a\overline{b},j} \equiv i \eta_{a,j} \eta_{\overline{b},j}$, and 
$\sigma_{\overline{a}b,m} \equiv i \eta_{\overline{a},m} \eta_{b,m}$. The Ising variables take $\pm 1$.

The Fock term ($m=n$) of the inter-pocket scatterings, Eq.~(\ref{Hb}), are bosonized as 
\begin{align}
H_{{\rm b},13}& = \sum_{j,m} H^{(13)}_{j-m} \int dz \!\ \Big\{ \sigma_{3\overline{1},j} \sigma_{3\overline{1},m}  
\cos\big[\Delta_{jm}Q^{13}_{+}\big] \nonumber \\
&\hspace{2.1cm} +  \sigma_{\overline{3}1,j}\sigma_{\overline{3}1,m} 
\cos\big[\Delta_{jm}Q^{13}_{-}\big]\Big\} \nonumber \\ 
&   \hspace{0.1cm}  + \sum_{j,m}\overline{H}^{(13)}_{j-m} \int dz \!\ \Big\{  \sigma_{4\overline{2},j}\sigma_{4\overline{2},m}  
\cos\big[\Delta_{jm}Q^{24}_{+}\big] \nonumber \\
&\hspace{2.1cm} +  \sigma_{\overline{4}2,j} \sigma_{\overline{4}2,m} 
\cos\big[\Delta_{jm}Q^{24}_{-}\big] \Big\}, \label{Hb13} \\
H_{{\rm b},2} & = \sum_{j,m} H^{(2)}_{j-m} \int dz \!\ \Big\{ \sigma_{3\overline{2},j} \sigma_{3\overline{2},m}  
\cos\big[\Delta_{jm}Q^{23}_{+}\big] \nonumber \\
&\hspace{2.1cm} +  \sigma_{\overline{3}2,j} \sigma_{\overline{3}2,m} 
\cos\big[\Delta_{jm}Q^{23}_{-}\big]\Big\} \nonumber \\ 
&  \hspace{0.1cm}  + \sum_{j,m}\overline{H}^{(2)}_{j-m} \int dz \!\ \Big\{ 
\sigma_{4\overline{1},j} \sigma_{4\overline{1},m}  
\cos\big[\Delta_{jm}Q^{14}_{+}\big] \nonumber \\
&\hspace{2.1cm} +  \sigma_{\overline{4}1,j} \sigma_{\overline{4}1,m} 
\cos\big[\Delta_{jm}Q^{14}_{-}\big] \Big\}, \label{Hb2} \\ 
H_{{\rm b},4} &= \sum_{j,m} H^{(4)}_{j-m} \int dz \!\ \Big\{ \sigma_{3\overline{4},j}\sigma_{3\overline{4},m} 
\cos\big[\Delta_{jm}Q^{34}_{-}\big]  \nonumber \\
&\hspace{2.1cm} +  \sigma_{\overline{3}4,j} \sigma_{\overline{3}4,m} 
\cos\big[\Delta_{jm}Q^{34}_{+}\big] \Big\} \nonumber \\ 
&  \hspace{0.1cm} + \sum_{j,m} \overline{H}^{(4)}_{j-m} \int dz \!\ \Big\{ 
\sigma_{2\overline{1},j} \sigma_{2\overline{1},m}  
\cos\big[\Delta_{jm}Q^{12}_{+}\big] \nonumber \\ 
&\hspace{2.1cm} +  \sigma_{\overline{2}1,j} \sigma_{\overline{2}1,m}
\cos\big[\Delta_{jm}Q^{12}_{-}\big] \Big\}, \label{Hb4} 
\end{align}
with $\Delta_{jm} f  \equiv f_{j} - f_{m}$. The Hartree term ($j=n$) of the inter-pocket scatterings 
in Eq.~(\ref{Hb}) could also renormalize the Luttinger parameters and Fermi 
velocities. For simplicity, however, we consider the renormalizations of $K_a$ and $u_{a}$ only 
by the intra-pocket scatterings $H_{\rm f}$ as given in appendix B. 
Figs.~\ref{fig:3a} and \ref{fig:3b} schematically 
show the inter-pocket scattering processes that lead to $H_{{\rm u},2}$ and $H_{{\rm b},2}$ 
respectively.  

The inter-chain interactions in $H_{{\rm u},i}$ and $H_{{\rm b},i}$  
range over the magnetic length;
\begin{align}
M^{(n)}_{j-m} & \equiv \frac{g}{L_x \alpha^2 l} {\cal M}^{(n)}((y_j-y_m)/l), \label{Mn}  \\ 
H^{(n)}_{j-m} & \equiv  \frac{g}{L_x \alpha^2 l} {\cal H}^{(n)}((y_j-y_m)/l),  \label{Hn} \\ 
\overline{H}^{(n)}_{j-m} & \equiv \frac{g}{L_x \alpha^2 l} \overline{\cal H}^{(n)}((y_j-y_m)/l), \label{Hnd}
\end{align}  
with $n=1,2,3,4$ and $13$. ${\cal M}(y)$ and ${\cal H}(y)$ as well as 
$\overline{\cal H}(y)$ are dimensionless functions. 
For example, Eqs.~(\ref{Hu1},\ref{Hu2},\ref{Mn}) are obtained 
from the direct ($j=n$) and exchange processes ($m=n$) 
of Eq.~(\ref{scat-u-12a}) respectively with  
\begin{align}
{\cal M}^{(1)}(x) &= - e^{-\frac{1}{8}(k_{F,1}+k_{F,3}-k_{F,2}-k_{F,4})^2 l^2_{0,z}}
\frac{f^{(12)}(x,0)}{\sqrt{2\pi} \pi^2}, \label{m1x} \\
{\cal M}^{(2)}(x) &= e^{-\frac{1}{8}(k_{F,2}+k_{F,3}-k_{F,1}-k_{F,4})^2 l^2_{0,z}} 
\frac{f^{(12)}(0,x)}{\sqrt{2\pi}\pi^2}  . \label{m2x} 
\end{align}
For the repulsive interaction case ($g>0$), integrals of Eqs.~(\ref{Hn},\ref{Hnd}) 
over $y \equiv (y_j-y_m)/l$ give negative values for any $n=13,2,4$. 

In the next section, we will use a perturbative renormalization group (RG) analyses on the effective 
boson model, $H_0+\sum_{i} H_{{\rm u},i} + \sum_{i} H_{{\rm b},i}$, where $H_{{\rm u},i}$ and 
$H_{{\rm b},i}$ are treated perturbatively (appendix C). We show that, at the one-loop level of the perturbative 
RG equations, $H_{{\rm b},13}$, $H_{{\rm u},1}$ and $H_{{\rm u},3}$ are coupled with one another and stabilize 
(what we call) a plain excitonic insulator phase. Meanwhile, $H_{{\rm u},2}$ and $H_{{\rm b},2}$ 
stabilize spin nematic excitonic insulator phase,  $H_{{\rm u},4}$ and $H_{{\rm b},4}$ 
stabilize magnetic Mott insulator phase.  

\section{renormalization group analyses}   
\subsection{spin-nematic excitonic insulator (SNEI-I) phase}
We begin with the spin-nematic excitonic insulator phase stabilized by 
$H_{{\rm b},2}$ and $H_{{\rm u},2}$. The renormalization group (RG) equations 
for the inter-chain interaction functions in $H_{{\rm b},2}$ and $H_{{\rm u},2}$   
take following forms at the one-loop level; 
\begin{align}
\frac{dM^{(2)}_{j-m}}{d{\rm ln}b} &= \frac{A_{23}+A_{14}}{2} M^{(2)}_{j-m}  \nonumber \\ 
&\hspace{-1.1cm} - 2C_{23} \sum_{n} M^{(2)}_{j-n} H^{(2)}_{n-m} - 2C_{14} \sum_{n} M^{(2)}_{j-n} \overline{H}^{(2)}_{n-m}, 
\label{rg1} \\ 
\frac{dH^{(2)}_{j-m}}{d{\rm ln}b} &= A_{23} H^{(2)}_{j-m}  \nonumber \\ 
&\hspace{-0.9cm} - \frac{1}{2} \sum_{n} \big(C_{14} M^{(2)}_{j-n}M^{(2)}_{n-m} + 4 C_{23} H^{(2)}_{j-n} H^{(2)}_{n-m} \big), 
\label{rg2} \\ 
\frac{d\overline{H}^{(2)}_{j-m}}{d{\rm ln}b} &= A_{14} \overline{H}^{(2)}_{j-m}  \nonumber \\ 
&\hspace{-0.9cm} - \frac{1}{2} \sum_{n} \big(C_{23} M^{(2)}_{j-n} M^{(2)}_{n-m} 
+ 4C_{14} \overline{H}^{(2)}_{j-n} \overline{H}^{(2)}_{n-m}\big). \label{rg3} 
\end{align} 
${\rm ln}b >0 $ is a scale change of the RG equations [see appendix C for their derivations]. 
The temperature $T$ increases monotonically on renormalization;  
$dT/d{\rm ln}b = T$. $A_{23}$, $A_{14}$ and their linear combination are the scaling 
dimensions of $H_{j-m}$, $\overline{H}_{j-m}$ and $M_{j-m}$ at the tree-loop level;
\begin{align}
A_{ab} \equiv 2 - \frac{1}{2} 
\sum_{c=a,b} \big(K_c + K^{-1}_c\big) \coth \Big(\frac{u_c \Lambda}{2T}\Big) <0. \label{Aab}
\end{align}  
$a,b=1,2,3,4$ are the pocket indice, where $1 \equiv (0,\uparrow)$, $2 \equiv (0,\downarrow)$, 
$3\equiv (-1,\uparrow)$ and $4\equiv (-1,\downarrow)$.  
$\Lambda$ is a short-range cutoff in the momentum space, $\Lambda= \alpha^{-1}$. 
$C_{ab}$ in Eqs.~(\ref{rg1},\ref{rg2},\ref{rg3}) is always finite positive definite constant 
for any $a,b=1,2,3,4$ (see Appendix C3). 
We assume that $C_{ab}$ has no dependence on temperature and magnetic field. 
Eqs.~(\ref{rg1},\ref{rg2},\ref{rg3}) are functional RG equations under which inter-chain 
interactions change their functional forms. To gain a simpler idea of these functional RG equations, 
we take a sum of the inter-chain interactions over their chain indices. The sum reduces the inter-chain 
coupling {\it functions} into coupling {\it constants} as follows,
\begin{align}
m_{(2)} &\equiv 2\pi l^2 \sum_{j} M^{(2)}_{j} = \frac{g}{\alpha^2} \int {\cal M}^{(2)}(y) \!\ dy,  \label{m2-i} \\ 
h_{(2)} &\equiv 2\pi l^2 \sum_{j} H^{(2)}_{j} = \frac{g}{\alpha^2} \int {\cal H}^{(2)}(y) \!\ dy <0, \label{h2-i} \\
\overline{h}_{(2)} &\equiv 2\pi l^2 \sum_{j} \overline{H}^{(2)}_j = \frac{g}{\alpha^2} \int 
\overline{\cal H}^{(2)}(y) \!\ dy <0. \label{h2d-i}   
\end{align} 
As mentioned above, the inequalities in Eqs.~(\ref{h2-i},\ref{h2d-i}) 
hold true for the repulsive interaction case. 
Considering the repulsive interaction case, we assume the  
negative bare values of $h_{(2)}$ and $\overline{h}_{(2)}$ in the followings.

The RG equations for the coupling constants take forms of 
\begin{align}
\frac{dm_{(2)}}{d{\rm ln}b} &= \frac{A_{23}+A_{14}}{2} m_{(2)} - \frac{1}{\pi l^2} 
m_{(2)} \big(C_{23} h_{(2)} + C_{14} \overline{h}_{(2)}\big), 
\label{rg1a} \\ 
\frac{dh_{(2)}}{d{\rm ln}b} &= A_{23} h_{(2)}  - \frac{1}{4\pi l^2} 
\big(C_{14} m^2_{(2)} + 4 C_{23} h^2_{(2)} \big), \label{rg2a} \\ 
\frac{d\overline{h}_{(2)}}{d{\rm ln}b} &= A_{14} \overline{h}_{(2)}  
- \frac{1}{4\pi l^2}  \big(C_{23} m^2_{(2)} + 4C_{14} \overline{h}^2_{(2)} \big). \label{rg3a} 
\end{align} 
The equations dictate that the umklapp term as well as the inter-chain backward scattering 
are irrelevant at the tree-loop level, as $A_{ab}$ is negative semi-definite (Eq.~(\ref{Aab})). 
Smaller $m_{(2)}$, $h_{(2)}$ and $\overline{h}_{(2)}$ are always renormalized 
into zero (`weak coupling phase'; normal metal phase). 

$C_{ab}$ is positive definite. Thus, the bare repulsive interaction 
$g$ has a critical strength, above which $m_{(2)}$, $h_{(2)}$ and $\overline{h}_{(2)}$ help 
one another to grow up into larger values 
(`strong coupling phase'). The critical strength 
decreases not only on increasing the magnetic field through a dependence of the one-loop terms 
on the magnetic length $l$, but also on decreasing the temperature through a dependence of 
$A_{ab}$ on the temperature. This suggests that the strong coupling phase generally 
appears in low temperature side and a transition temperature of the strong coupling phase 
increases in larger magnetic field (e.g. see a field-dependence of the transition temperature of 
the SNEI-I phase in Fig.~\ref{fig:1} in a region of $H<40\!\ {\rm T}$). 

The transition temperature can also 
{\it decrease} when the Luttinger parameters $K_{a}$ ($a=1,\cdots,4$) deviate largely from the unit.  
$|A_{ab}|$ has a global minimum at $K_a=K_b=1$ and $T=0$. When $K_{a}$ deviates away from $1$, 
$A_{ab}$ becomes negatively larger and thus the critical strength for $g$ increases;  
the transition temperature decreases. Physically speaking, $K_a$ being greater/smaller than the unit 
means stronger quantum fluctuation of the displacement field/superconducting phase 
field of the $a$-th pocket [Eqs.~(\ref{H0},\ref{bosonization2})]. The enhanced quantum 
fluctuations generally destabilize the strong coupling phase.  

This observation readily lets us propose a new microscopic mechanism for the re-entrant transition from 
the strong-coupling to weak-coupling phases; the transition induced by {\it raising} the magnetic field. 
The higher magnetic field generally makes the electron pocket ($a$) and hole pocket ($b$) to be smaller 
in size in the $k_z$ space. This makes their bare Fermi velocities, $v_{F,a}$,$v_{F,b}$, to be 
smaller with respect to the electron interaction energy scale. Thus, in the presence of the repulsive interaction, 
$g_{2,a},g_{2,b}>0$ in Eq.~(\ref{luttinger2}), the smaller Fermi velocities make their Luttinger 
parameters to be smaller than the unit, $K_{a},K_{b} < 1$. Especially, for $H<H_0$, $K_1$ and 
$K_4$ are expected to be much smaller than the unit near $H=H_0$, where the electron pocket 
with $a=1\!\ [(n,\sigma)=(0,\uparrow)]$ and the hole pocket with $b=4\!\ [(n,\sigma)=(-1,\downarrow)]$ 
are about to leave the Fermi level. Thus, the transition temperature 
of the strong coupling phase reduces dramatically near $H=H_0$ through an enhancement of 
$K^{-1}_1$ and $K^{-1}_{4}$ in Eq.~(\ref{Aab}) (e.g. see the field-dependence of $T_c$ of 
the SNEI-I phase in Fig.~\ref{fig:1} in a region of $40 \!\ {\rm T} < H<50\!\ {\rm T}$). 
Physically speaking, this reduction is nothing but a consequence of the enhanced quantum fluctuation 
of spin superconducting phase variable. 
    
When the bare repulsive interaction is greater than the critical value (strong coupling phase), 
the umklapp and inter-pocket backward scattering terms grow up into larger 
values; 
\begin{align}
h_{(2)}, \overline{h}_{(2)} \rightarrow -\infty, \ \ \ m_{(2)} \rightarrow \pm \infty. \nonumber 
\end{align}
The following argument does not depend on the sign of $m_{(2)}$, so that we set $m_{(2)}>0$ 
henceforth. In the strong coupling regime, $H_{{\rm u},2}$ and $H_{{\rm b},2}$ are maximally minimized by 
\begin{align}
&\phi_{3,m} + \phi_{2,m} = \Phi_{-}, \ \ \phi_{4,m} + \phi_{1,m} = \left\{\begin{array}{l} 
2n\pi - \Phi_{-}  \\
(2n+1) \pi - \Phi_{-} \\
\end{array}\right. \label{phi-lock} \\ 
&\theta_{3,m} - \theta_{2,m} = \Theta_{-}, \ \ \theta_{4,m} - \theta_{1,m} = \left\{\begin{array}{l}
(2n+1)\pi - \Theta_{-} \\ 
2n\pi - \Theta_{-} \\
\end{array}\right. , \label{theta-lock}
\end{align}
with 
\begin{align}
\sigma_{3\overline{2},m} = \sigma_{4\overline{1},m} = \sigma_{\overline{3}2,m} = \sigma_{\overline{4}1,m} 
= \sigma.   
\label{Ising-lock}
\end{align}  
The locking of the total displacement field, $\phi_{3,m}+\phi_{2,m}+\phi_{4,m}+\phi_{1,m}=2n \pi$ 
or $(2n+1)\pi$, dictates that the system is electrically insulating along the field direction. Meanwhile, any  
electron densities, $\langle\rho({\bm r},c)\rangle \equiv \sum_{\sigma=\uparrow,\downarrow} 
\langle \psi^{\dagger}_{\sigma}({\bm r},c)\psi_{\sigma}({\bm r},c)\rangle$ with $c=A,A',B,B'$, do not break 
the translational symmetry along the field direction ($z$), because 
\begin{align}
&\langle \psi^{\dagger}_{(n,\sigma),\tau,j}(z) \psi_{(n,\sigma),\overline{\tau},j}(z) \rangle  = 0, \nonumber \\ 
& \langle \psi^{\dagger}_{(0,\sigma),\tau,j}(z) \psi_{(-1,\sigma),\tau^{\prime},j}(z) \rangle = 0,  \nonumber  
\end{align}
with $n=0,-1$, $\tau,\tau^{\prime}=\pm$, $\overline{\tau}=-\tau$, 
$\sigma=\uparrow,\downarrow$. 
Due to the charge neutrality condition, the mean electron density is 2 
per two LLs, $n=0$ and $n=-1$ LLs, and per the unit cell along the $c$-axis. 
Besides, the insulating phase is associated with particle-hole pairings between $n=0$ LL (electron pocket) 
and $n=-1$ LL (hole pocket). Thus, we regard this phase as 
excitonic insulator~\cite{akiba15,zhu17,jerome67,fenton68,abrikosov70,brazovskii71} 
instead of charge density wave phase.

An insulating property is manifested by the optical conductivity along the $c$-axis, 
$\sigma_{zz}(\omega)$. In the strong coupling phase with large $m_{(2)}$, 
we may employ a Gaussian approximation for the cosine terms in $H_{{\rm u},2}$. 
$\sigma_{zz}(\omega)$ is calculated within the linear response 
theory as $\sigma_{zz}(\omega) = (e^2 u K)/(2\pi l^2) \delta(\omega-\omega_g)$, 
where $uK \equiv \sum_{a}u_a K_a$. $\omega^2_g \equiv 
2 \pi uK \sum_{j} M^{(2)}_{j}$ defines a gap for collective particle-hole excitation 
associated with a fluctuation of the total displacement field.  An inclusion of a 
short-ranged dielectric disorder renormalizes the gap into a smaller 
value $\omega_{*}$ with a smaller spectral weight for the delta function 
(see Appendix D). Meanwhile, it adds 
a continuum spectrum in higher energy region. The continuum spectra compensate 
the reduced spectral weight of the delta function. The observation concludes 
that the excitonic insulator phase is robust against any small dielectric disorder, provided 
that the renormalized gap size and the spectral weight of the delta function 
remains finite (see appendix D). 

\begin{figure}[t]
	\centering
	\includegraphics[width=0.9\linewidth]{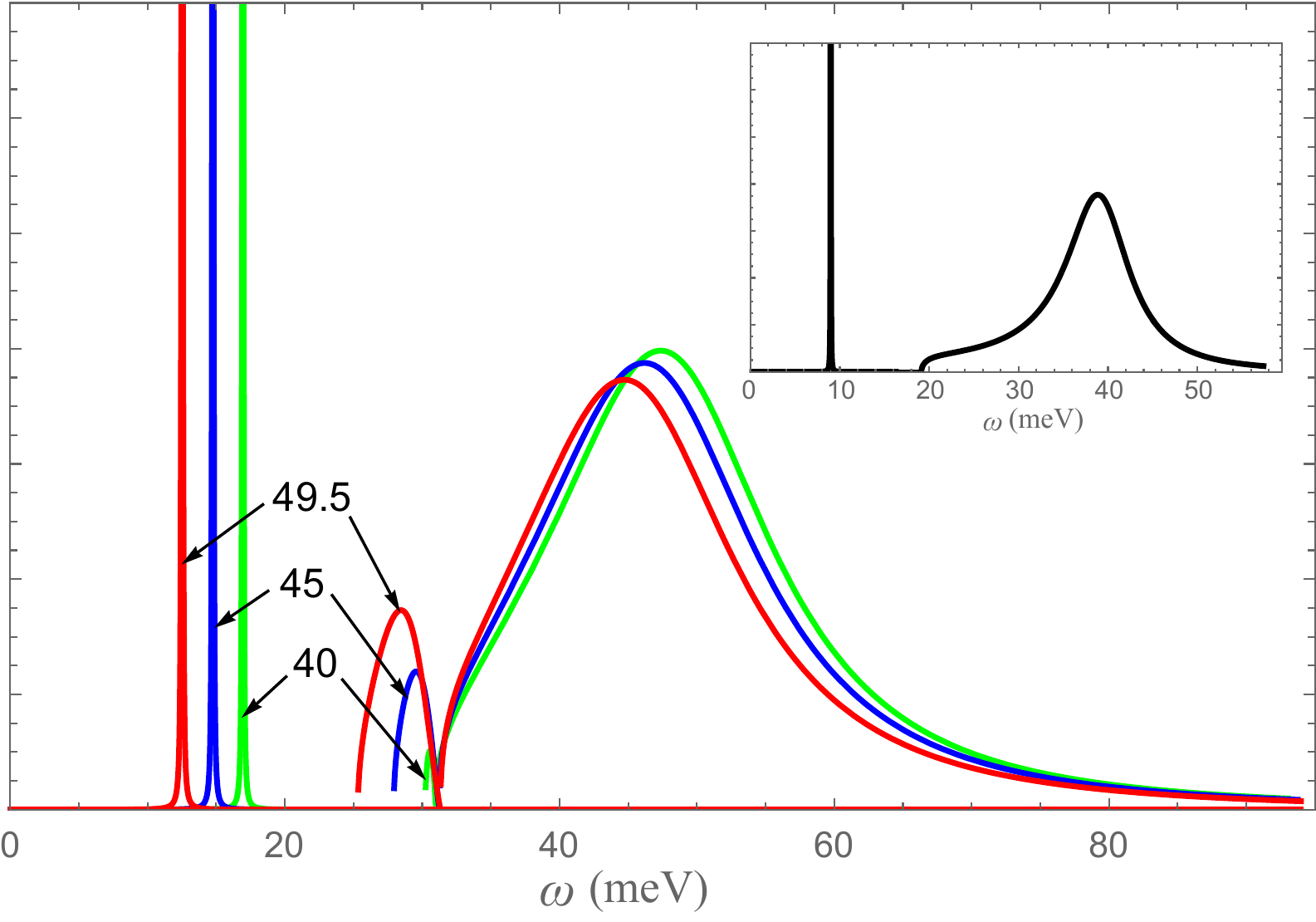}
	\caption{(color online) Theoretical calculation results of the 
optical conductivity $\sigma_{zz}(\omega)$ 
in the SNEI-I phase ($H=40$, $45$, $49.5$ T). (Inset) $\sigma_{zz}(\omega)$ 
in the SNEI-II phase ($H=55$ T). We use the same parameter sets as 
in Fig.~\ref{fig:1}. For its details, see the appendix D. Unlike its appearance 
in the figures, the delta function at $\omega=\omega_{*}$ is the most 
prominent in amplitude, while the continuum spectrum is much less significant. 
The renormalized gap $\omega_{*}$ is on the order of 
$\sqrt{E_{\rm int} E_{\rm bw}}$, where $E_{\rm int}$ is an interaction energy 
scale, $E_{\rm int} \sim e^2/(\epsilon l)$, and $E_{\rm bw}$ is a band width energy 
scale (see appendix C4).} 
	\label{fig:4}
\end{figure}

The long-range order of the spin superconducting phases such as $\theta_{3}-\theta_{2}$, 
$\theta_{4}-\theta_{1}$ in Eq.~(\ref{theta-lock})
breaks the U(1) spin-rotational symmetry around the field direction. The breaking 
of the continuous spin-rotational symmetry is manifested by a long-range   
ordering of spin quadrupole moment (`spin-nematic excitonic insulator'). 
The quadrupole moment that exhibits the order is a symmetric part of a 
2nd-rank spin tensor composed by two spin-$1/2$ moments [see appendix E]. 
One spin-$1/2$ is from the $\pi$ orbital of $A$ or $A'$ carbon atom, 
while the other spin-$1/2$ is from the $\pi$ orbital of $B$ or $B'$ carbon atom. 
The 2nd rank spin tensor is defined as 
\begin{align}
Q^{cd}_{\mu\nu}({\bm r}) &\equiv \langle S_{c,\mu}({\bm r}) S_{d,\nu}({\bm r}) 
+ S_{c,\nu}({\bm r}) S_{d,\mu}({\bm r})\rangle \nonumber \\
& \hspace{1cm} - \delta_{\mu\nu} 
\langle {\bm S}_{c,\perp}({\bm r})\cdot {\bm S}_{d,\perp}({\bm r}) \rangle, \label{quadrupole}
\end{align} 
with $c=A,A'$, $d= B,B'$, $\mu,\nu=x,y$, ${\bm S}_{c,\perp} \equiv (S_{c,x},S_{c,y})$,  
$2S_{c,\mu}({\bm r}) \equiv \psi^{\dagger}_{\sigma'}({\bm r},c) 
[\sigma_{\mu}]_{\sigma^{\prime}\sigma^{\prime\prime}} 
\psi_{\sigma^{\prime\prime}}({\bm r},c)$ and 
$\sigma^{\prime},\sigma^{\prime\prime}=\uparrow,\downarrow$. 
The order of the spin superconducting phase [Eq.~(\ref{theta-lock})]  
leads to a ferro type as well as density-wave type ordering of the 2nd rank spin tensor, e.g.  
\begin{align}
Q^{AB}_{xx} ({\bm r}) + i Q^{AB}_{xy} ({\bm r}) &= e^{2i \Theta_{-}}\Big(u + u 
\cos(\Delta K z-2\Phi_{-})\Big), \nonumber 
\end{align} 
where $\Delta K \equiv k_{F,3} + k_{F,2} - k_{F,4} - k_{F,1}$. $u$ is a complex-valued 
coefficient. Symmetry-wise speaking, the long-range order given in Eq.~(\ref{theta-lock}) can be 
also accompanied by a helical magnetic order whose magnetic moment lies in the $xy$ plane. 
The helical order has two spatial 
pitches along the $c$-axis, $(2\pi)/(k_{F,3}+k_{F,2})$ and $(2\pi)/(k_{F,1}+k_{F,4})$. Microscopically 
speaking, however, an amplitude of the magnetic moment is tiny and, if any, it 
appears only in those spatial regions in the cell where two neighboring $\pi$ orbitals 
in the same graphene layer overlap [Appendix E].       

On increasing the magnetic field, the outer electron pocket with 
$(n,\sigma)=(0,\uparrow)$ and hole pocket with $(-1,\downarrow)$ leave the Fermi level  
at $H=H_0$. Ab-initio electronic band structure calculations evaluate $H_0$ 
around $53$ T~\cite{takada98}. For $H\rightarrow H_0 \!\ (H<H_0)$, the bare Fermi velocities 
of the two pockets $v_{F,1}$ 
and $v_{F,4}$ become smaller. So do the Luttinger parameters of the two pockets 
$K_{1}$ and $K_4$ [Eq.~(\ref{luttinger2})]. The reduction of the Luttinger parameters 
makes the tree-level scaling dimension $A_{14}$ negatively very large [Eq.~(\ref{Aab})]. Thus, 
according to Eqs.~(\ref{rg1a},\ref{rg3a}), $m_{(2)}$ and $\overline{h}_{(2)}$ are renormalized 
into smaller values at an early stage of the RG flow for $H\lesssim H_0$, irrespective of 
bare values of $m_{(2)}$ and $\overline{h}_{(2)}$. Meanwhile, 
$A_{23}$ as well as $v_{F,2}$ and $v_{F,3}$ 
remain rather constant around $H=H_0$. Thus, according to Eq.~(\ref{rg2a}),  
$h_{(2)}$ grows up to a larger value and eventually diverges, provided that its bare (initial) 
value is greater than a critical value (see below for the critical value).
Larger $h_{(2)}$ then helps $m_{(2)}$ and $\overline{h}_{(2)}$ to grow up  
at a late stage of the RG flow, by way of the one-loop terms in 
Eqs.~(\ref{rg1a},\ref{rg3a}). The argument so far concludes that, 
for $H\lesssim H_0$, the transition temperature of the strong coupling phase  
is determined only by Eq.~(\ref{rg2a}) with $m_{(2)}=0$; 
\begin{align}
\frac{dh_{(2)}}{d{\rm ln}b} &= A_{23} h_{(2)}  - \frac{C_{23}}{\pi l^2} h^2_{(2)} . \label{rg2aa}  
\end{align}
At the zero temperature, Eq.~(\ref{rg2aa}) gives the critical value for $h_{(2)}$ as,     
\begin{align}
h_{(2),c} & \equiv \frac{\pi l^2}{C_{23}} (A_{23})_{|T=0} \nonumber \\
&= \frac{\pi l^2}{C_{23}} \Big[2- \sum_{c=2,3} \big(K_c + K^{-1}_c\big)\Big] < 0. \label{rg2aaa}
\end{align}     
When $h_{(2)}<h_{(2),c}<0$, 
the spin nematic excitonic insulator phase always appears below a finite critical temperature 
$T_c$ at $H \le H_0$ (Fig.~1). The situation is consistent with the experimental 
phase diagram of the graphite under high field. Meanwhile, RG phase 
diagrams of the other insulators stabilized by $H_{{\rm u},1}$, $H_{{\rm u},3}$ or $H_{{\rm u},4}$ 
are not consistent with the graphite experiment. 

\subsection{magnetic Mott insulator and plain excitonic insulator phases}
To see this, let us next consider a nature and a RG phase diagram of the  
magnetic Mott insulator phase stabilized by $H_{{\rm u},4}$ and $H_{{\rm b},4}$. By exchanging 
$2$ and $4$ in Eqs.~(\ref{rg1a},\ref{rg2a},\ref{rg3a}), we can readily obtain corresponding 
one-loop RG equations for their coupling constants;
\begin{align}
\frac{dm_{(4)}}{d{\rm ln}b} &= \frac{A_{34}+A_{12}}{2} m_{(4)} - \frac{1}{\pi l^2} 
m_{(4)} \big(C_{34} h_{(4)} + C_{12} \overline{h}_{(4)}\big), 
\label{rg1b} \\ 
\frac{dh_{(4)}}{d{\rm ln}b} &= A_{34} h_{(4)}  - \frac{1}{4\pi l^2} 
\big(C_{12} m^2_{(4)} + 4 C_{34} h^2_{(4)} \big), \label{rg2b} \\ 
\frac{d\overline{h}_{(4)}}{d{\rm ln}b} &= A_{12} \overline{h}_{(4)}  
- \frac{1}{4\pi l^2}  \big(C_{34} m^2_{(4)} + 4 C_{12} \overline{h}^2_{(4)} \big). \label{rg3b} 
\end{align}     
Here, the coupling constants are integrals of the inter-chain coupling functions in $H_{{\rm b},4}$ and 
$H_{{\rm u},4}$;  
\begin{align}
m_{(4)} &\equiv 2\pi l^2 \sum_{j} M^{(4)}_{j} = \frac{g}{\alpha^2} \int {\cal M}^{(4)}(y) \!\ dy,  \nonumber \\ 
h_{(4)} &\equiv 2\pi l^2 \sum_{j} H^{(4)}_{j} = \frac{g}{\alpha^2} \int {\cal H}^{(4)}(y) \!\ dy <0, \nonumber \\
\overline{h}_{(4)} &\equiv 2\pi l^2 \sum_{j} \overline{H}^{(4)}_j = \frac{g}{\alpha^2} \int 
\overline{\cal H}^{(4)}(y) \!\ dy <0. \nonumber   
\end{align} 
The inequalities hold true for bare values of $h_{(4)}$ and $\overline{h}_{(4)}$ in the presence of 
the repulsive interaction $g$ ($>0$). 

The RG equations tell that the bare value of the repulsive interaction $g$ has a critical strength 
above/below which $m_{(4)}$ as well as $h_{(4)}$ and $\overline{h}_{(4)}$ become 
relevant/irrelevant on the renormalization. In the strong coupling phase with $m_{(4)} \rightarrow 
\pm \infty$ and $h_{(4)},\overline{h}_{(4)}\rightarrow -\infty$, $H_{{\rm b},4}$ and $H_{{\rm u},4}$ 
are maximally minimized by 
\begin{align}
&\phi_{3,j} + \phi_{4,j} = \Phi_{-}, \ \ \phi_{2,j} + \phi_{1,j} = \left\{\begin{array}{l} 
2n\pi - \Phi_{-}  \\
(2n+1) \pi - \Phi_{-} \\
\end{array}\right. \label{phi-lock-2} \\ 
&\theta_{3,j} - \theta_{4,j} = \Theta_{-}, \ \ \theta_{2,j} - \theta_{1,j} = \left\{\begin{array}{l}
(2n+1)\pi - \Theta_{-} \\ 
2n\pi - \Theta_{-} \\
\end{array}\right. , \label{theta-lock-2}
\end{align}
with 
\begin{align}
\sigma_{3\overline{4},m} = \sigma_{2\overline{1},m} = \sigma_{\overline{3}4,m} = \sigma_{\overline{2}1,m} 
=\sigma,  
\end{align}  
for $m_{(4)}>0$. The locking of the total displacement field results in an electrically insulating behavior 
along the field direction, while the long-range order 
of the spin-superconducting phases leads to a long-range   
helical magnetic order, e.g. 
\begin{align}
&\langle S_{A,x}({\bm r}) \rangle + i \langle S_{A,y} ({\bm r}) \rangle = v^{\prime} e^{i\Theta_{-}} \cos((k_{F,1}+k_{F,2})z), \nonumber \\   
& \langle S_{B,x}({\bm r}) \rangle + i \langle S_{B,y} ({\bm r}) \rangle = v^{\prime\prime} 
e^{i\Theta_{-}} \cos\big((k_{F,1}+k_{F,2})z\big) \nonumber \\ 
&\hspace{3cm} + w^{\prime\prime} e^{i\Theta_{-}} \cos\big((k_{F,3}+k_{F,4})z\big). \nonumber
\end{align}
As for the charge degree of freedom, the insulating phase does not break the translational symmetry; 
$\langle \rho({\bm r},c)\rangle$ always respects the translational symmetry for $c=A,B,A',B'$. 
The phase is stabilized by the pairings with the same LL but between the different spins, so that  
we call this phase as a magnetic Mott insulator. 

Unlike the spin-nematic excitonic insulator, a transition temperature of 
the magnetic Mott insulator goes to zero at a certain critical field below $H_0$. 
For $H\rightarrow H_0 \!\ (H<H_0)$, where $K_1$ and $K_4$ become very small, {\it both} 
$A_{34}$ {\it and} $A_{12}$ in Eqs.~(\ref{rg1b},\ref{rg2b},\ref{rg3b}) 
become negatively very large. Accordingly, unlike in the spin nematic excitonic insulator case 
in the previous section, {\it all} of the three coupling constants, $m_{(4)}$, $h_{(4)}$ and 
$\overline{h}_{(4)}$, are renormalized to zero for those $H$ sufficiently close to $H_0$ ($H < H_0$). 
In other words, the transition temperature of the magnetic Mott insulator 
always goes to zero at a certain 
critical field below $H_0$. This is also the case with the plain excitonic 
insulator stabilized by $H_{{\rm u},1}$, $H_{{\rm u},3}$ and $H_{{\rm b},13}$. 
These RG phase diagrams are {\it not} consistent 
with the experimental phase diagram of graphite under the high 
field~\cite{ochimizu92,yaguchi98a,yaguchi98b,yaguchi01,fauque13,akiba15,zhu17}. 

Besides, the helical magnetic order in the Mott insulator is expected to be weak 
against {\it magnetic} disorders. Considering an anisotropy of $g$-factor in graphite~\cite{matsubara91},  
it is natural to assume that the high magnetic field allows the system to have 
single-particle backward scatterings between two electron pockets with 
$(n,\sigma)=(0,\uparrow)$ and $(0,\downarrow)$, and also that between two hole pockets 
with $(n,\sigma)=(-1,\uparrow)$ and $(-1,\downarrow)$. The backward scatterings do exist, 
especially when graphite contains those graphene layers whose normal vectors ($c$-axis) have  
non-zero angles with respect to the field direction. Such graphene layers can appear 
anywhere and randomly along the $c$-axis, so that the backward scatterings are 
generally accompanied by random U(1) phases $\lambda_{j,\pm}(z)$;
\begin{align}
H^{\prime}_{\rm imp} &= 
\sum_{j} \int dz A_{j,+}(z) \big\{ e^{i\lambda_{j,+}(z)} \psi^{\dagger}_{1,+,j}(z) \psi_{2,-,j}(z) 
+ {\rm h.c.}\big\}  \nonumber  \\
& \hspace{-0.5cm} + \sum_{j} \int dz A_{j,-}(z) \big\{ e^{i\lambda_{j,-}(z)} \psi^{\dagger}_{1,-,j}(z) 
\psi_{2,+,j}(z) 
+ {\rm h.c.}\big\} + \cdots \nonumber 
\end{align} 
When bosonized, these single-particle backward scatterings add random $U(1)$ phases 
into $\Phi_{-}\pm \Theta_{-}$ in Eqs.~(\ref{phi-lock-2},\ref{theta-lock-2}) respectively;
\begin{align}
H^{\prime}_{\rm imp} &=  
\sum_{j} \int dz A_{j,+}(z) \sigma_{1\overline{2},j}\nonumber \\ 
& \hspace{0.6cm} \times \cos\big[\phi_{2,j} + \phi_{1,j} 
-\theta_{2,j} + \theta_{1,j} + \lambda_{+,j}(z) \big] \nonumber \\ 
& \hspace{0.2cm} +  \sum_{j} \int dz A_{j,-}(z) \sigma_{\overline{1}2,j} \nonumber \\ 
&\hspace{0.4cm} \times \cos\big[\phi_{2,j} + \phi_{1,j} 
+ \theta_{2,j} - \theta_{1,j} + \lambda_{-,j}(z) \big] + \cdots. \nonumber 
\end{align}
Since $\Phi_{-}$ and $\Theta_{-}$ comprise gapless Goldstone modes in 
the magnetic Mott insulator, the added random U(1) phases readily kill the 
long-range orders of $\Phi_{-}$ and $\Theta_{-}$, however small 
the amplitudes $A_{j,\pm}(z)$ are~\cite{imry76,sham76,fukuyama78b,zhang17}. 
Likewise, the plain excitonic 
insulator phase stabilized by $H_{{\rm u},1}$, $H_{{\rm u},3}$ and $H_{{\rm b},13}$ is expected 
to be weak against short-ranged charged disorders. The short-ranged disorder causes single-particle type 
backward scatterings between $(0,\uparrow)$ and $(-1,\uparrow)$ pockets and those between 
$(0,\downarrow)$ and $(-1,\downarrow)$ pockets. From these reasonings as well as 
inconsistency between their RG phase diagrams and the experimental phase diagram of 
graphite, we conclude that the magnetic Mott insulator as well as the plain excitonic insulator 
can hardly explain the graphite experiment coherently.  

One may expect that the spin-nematic excitonic insulator could also suffer from  
random single-particle backward scatterings between $(0,\uparrow)$ and 
$(-1,\downarrow)$ pockets or those between $(0,\downarrow)$ and $(-1,\uparrow)$ pockets.  
Nonetheless, these scatterings unlikely exist in the real system. 
Or, if any, they are much 
smaller than the others, because the relativistic spin-orbit interaction is needed for 
them, and it is extremely small in graphite~\cite{g-dresselhaus65,matsubara91}. Without the 
spin-orbit interaction, these backward scatterings need both the 
magnetic scatter and the short-ranged charged scatter on the same spatial point. 
Microscopically, however, these two types of the scatters are of different origins 
and they have no correlation at all. From these reasonings as well as the generic 
consistency between the RG phase diagram ($H<H_0$ in Fig.~\ref{fig:1}) and the experimental phase diagram, 
we conclude that an insulating phase in graphite at $H<H_0$ is  
the spin-nematic excitonic insulator stabilized by the interplay 
between $H_{{\rm u},2}$ and $H_{{\rm b},2}$.

\section{two pockets model ($H>H_0$)}
For $H > H_0$, both the electron pocket with $(n,\sigma)=(0,\uparrow)$ and hole pocket 
with $(n,\sigma)=(-1,\downarrow)$ leave the Fermi level~\cite{takada98,arnold17}.  
The low-energy electronic system for $H>H_0$ comprises only 
of the electron pocket with $(n,\sigma)=(0,\downarrow)$ and 
the hole pocket with $(n,\sigma)=(-1,\uparrow)$. As before, we call $(n,\sigma)=(0,\downarrow)$ 
as $a=2$ and $(n,\sigma)=(-1,\uparrow)$ as $a=3$. The charge neutrality condition is given by  
$k_{F,0,\downarrow} + k_{F,-1,\uparrow} = \pi/c_0$. Under the condition, the interaction 
allows the following umklapp term; 
\begin{align}
H^{\prime}_{\rm u} = \sum_{j,m,n} \psi^{\dagger}_{3,+,n} \psi^{\dagger}_{2,+,j+m-n} 
\psi_{2,-,m} \psi_{3,-,j} + {\rm h.c.}, \label{huu1}
\end{align}
where the integrals over $z$ and scattering matrix elements are omitted. 
Other two-particle interaction terms that are linked with the umklapp term at the one-loop level  
of the fermionic RG equations are inter-pocket and intra-pocket scatterings between 
different chiralities~\cite{brazovskii71}. They are 
\begin{align}
H^{\prime}_{\rm b} = \sum_{j,m,n} \psi^{\dagger}_{3,\pm,n} \psi^{\dagger}_{2,\mp,j+m-n}
\psi_{2,\mp,m} \psi_{3,\pm,j}, \label{hbd}
\end{align}
and 
\begin{align}
H^{\prime}_{\rm d} = \sum_{j,m,n} \left\{\begin{array}{c} 
\psi^{\dagger}_{2,\pm,n} \psi^{\dagger}_{2,\mp,j+m-n} \psi_{2,\mp,m} \psi_{2,\pm,j}, \\ 
\psi^{\dagger}_{3,\pm,n} \psi^{\dagger}_{3,\mp,j+m-n} \psi_{3,\mp,m} \psi_{3,\pm,j}, \\
\end{array}\right. \label{hdd}
\end{align}
respectively.

\begin{figure}[t]
	\begin{minipage}[t]{0.33\linewidth} 
		\centering 
		\includegraphics[width=1.1in]{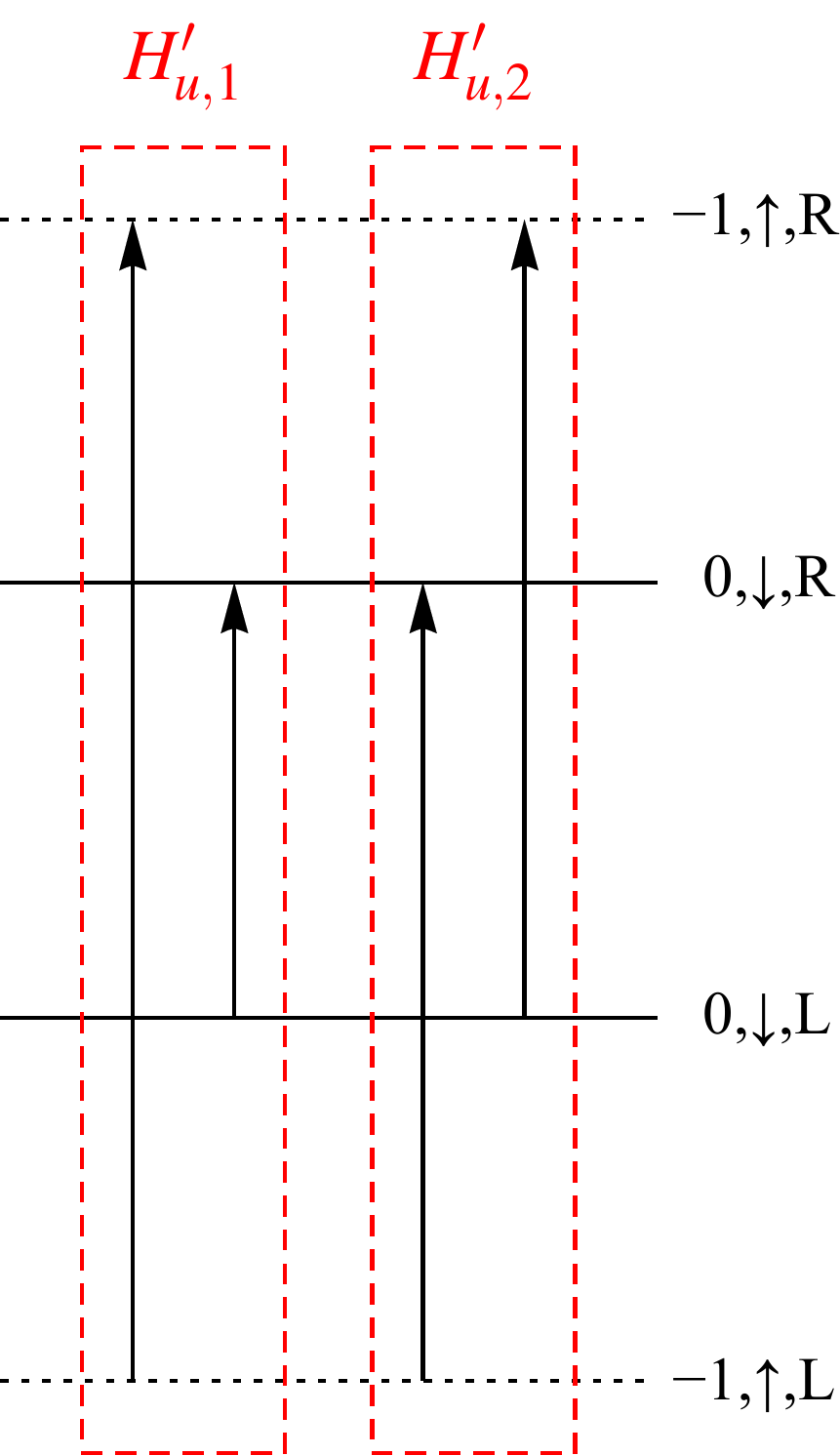} 
	\end{minipage}%
	\begin{minipage}[t]{0.32\linewidth} 
		\centering 
		\includegraphics[width=1.0in]{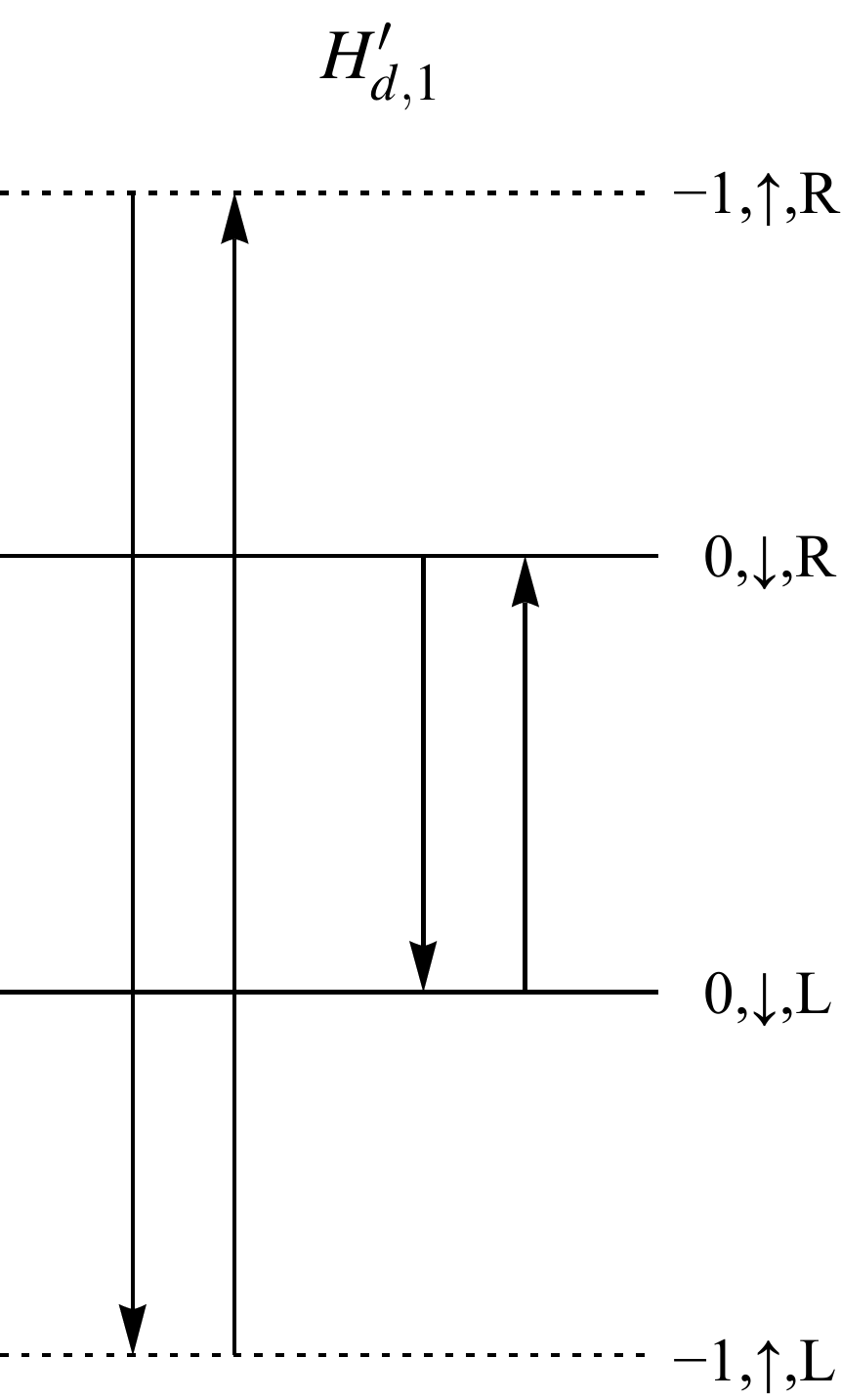} 
	\end{minipage} 
	\begin{minipage}[t]{0.32\linewidth} 
		\centering 
		\includegraphics[width=1.0in]{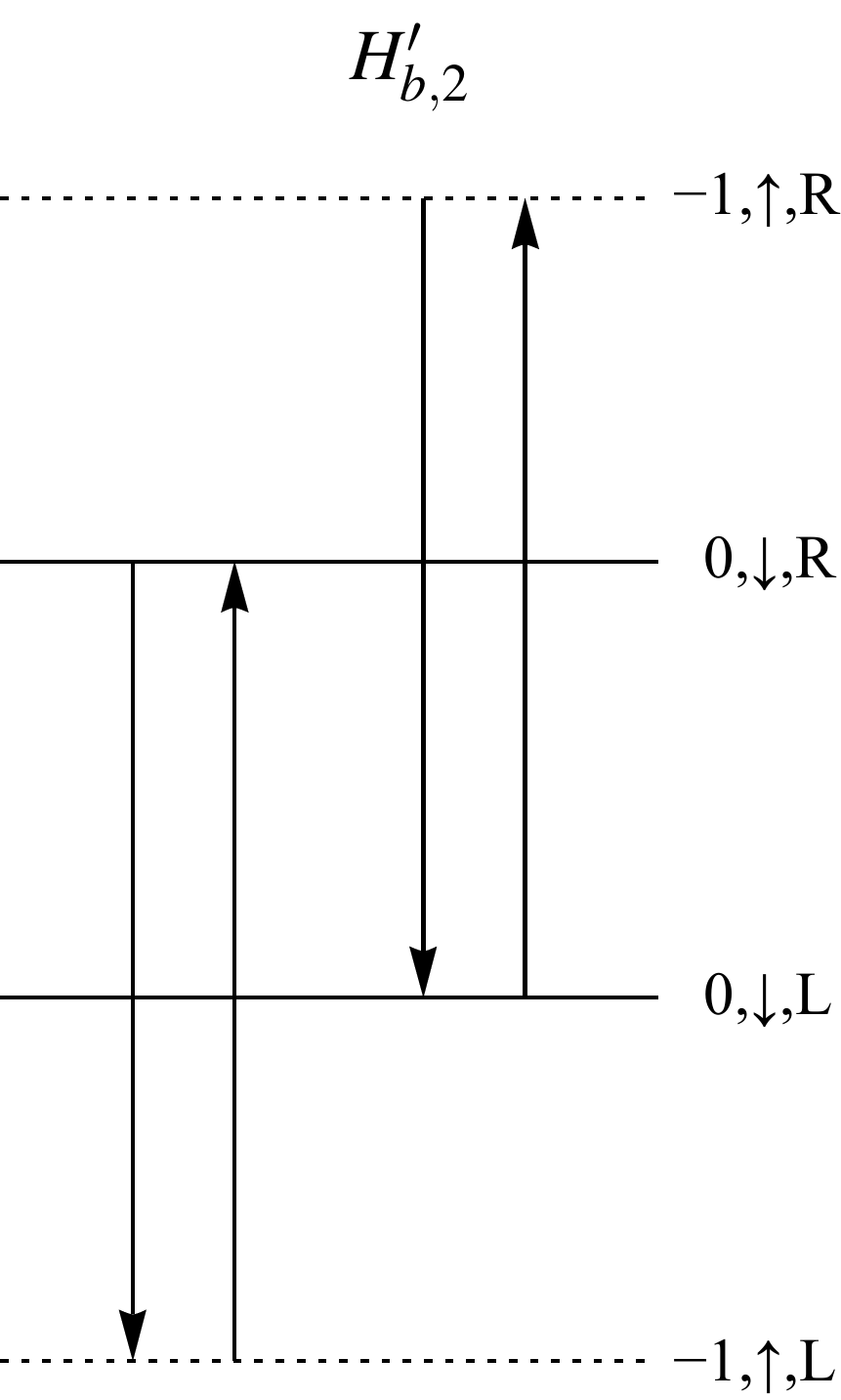} 
	\end{minipage} 
	\caption{(color online) (Left) schematic pictures of two-particle umklapp scatterings 
that are allowed in the two-pocket model under the charge neutrality condition, 
$H^{\prime}_{{\rm u},1}$ and 
$H^{\prime}_{{\rm u},2}$. They are direct ($j=n$) and exchange ($m=n$) processes 
of Eq.~(\ref{huu1}) respectively. (Middle) two-particle intra-pocket scatterings 
$H^{\prime}_{{\rm d},1}$: exchange processes ($m=n$) of Eq.~(\ref{hdd}). (Right) 
two-particle inter-pocket scatterings $H^{\prime}_{{\rm b},2}$; 
exchange processes ($m=n$) of Eq.~(\ref{hbd}). 
As in Fig.~\ref{fig:2}, the vertical axis denotes the momentum along the field direction ($k_z$), 
while the horizontal axis denotes the chain index $y_j=k_j l^2$ with 
$k_j\equiv 2\pi j/L_x$ $(j=1,2,\cdots,L_xL_y/(2\pi l^2))$.} 
	\label{fig:5}
\end{figure}

To construct effective boson theories of possible insulating phases 
stabilized by $H^{\prime}_{\rm u}$, we first assume the in-plane (graphene-plane) translational symmetry of 
the insulating phases, consider electron pairing within the same chain, and treat the inter-chain 
electron-electron interactions by the Hartree-Fock approximation. Specifically, we keep only the 
direct process (Hartree; $j=n$) and the exchange process (Fock; $m=n$) in Eqs.~(\ref{huu1},\ref{hbd},\ref{hdd}), 
and bosonize them into cosine terms;
\begin{align}
&H^{\prime}_{\rm u} + H^{\prime}_{\rm b} + H^{\prime}_{\rm d} 
= H^{\prime}_{{\rm u},1} + H^{\prime}_{{\rm u},2} + H^{\prime}_{{\rm d},1}  +  H^{\prime}_{{\rm b},2}  + \cdots,  
\nonumber \\
& H^{\prime}_{{\rm u},1} = \sum_{j,m} N^{(1)}_{j-m} \!\ \int dz \sigma_{3\overline{3},j} \sigma_{2\overline{2},m} 
\cos\big[2\phi_{3,j} + 2\phi_{2,m}\big], \label{Hu1d}\\
& H^{\prime}_{{\rm u},2} = \sum_{j,m} N^{(2)}_{j-m} \!\ \int dz \sigma_{2\overline{3},j} \sigma_{3\overline{2},m} 
\cos\big[Q^{23}_{+,j} + Q^{23}_{-,m}\big], \label{Hu2d} \\ 
& H^{\prime}_{{\rm d},1} = \sum_{j,m} O^{(1)}_{j-m} \!\ \int dz \sigma_{3\overline{3},j} \sigma_{3\overline{3},m} 
\cos\big[2\phi_{3,j} - 2\phi_{3,m}\big] \nonumber  \\
& \hspace{0.2cm} + \sum_{j,m} \overline{O}^{(1)}_{j-m} \!\ \int dz \sigma_{2\overline{2},j} \sigma_{2\overline{2},m} 
\cos\big[2\phi_{2,j} - 2\phi_{2,m}\big], \label{Hd1d}  \\
& H^{\prime}_{{\rm b},2} = \sum_{j,m} P^{(2)}_{j-m} \!\ \int dz \sigma_{2\overline{3},j} \sigma_{2\overline{3},m} 
\cos\big[Q^{23}_{-,j} - Q^{23}_{-,m}\big] \nonumber \\
&\hspace{0.2cm} + \sum_{j,m} \overline{P}^{(2)}_{j-m} \!\ \int dz \sigma_{3\overline{2},j} \sigma_{3\overline{2},m} 
 \cos\big[Q^{23}_{+,j} - Q^{23}_{+,m}\big]. \label{Hb2d}
\end{align}
Here $H^{\prime}_{{\rm u},1}$ is from the Hartree process ($j=n$) of Eq.~(\ref{huu1}), while 
$H^{\prime}_{{\rm u},2}$, $H^{\prime}_{{\rm d},1}$ and $H^{\prime}_{{\rm b},2}$ are from 
the Fock processes ($m=n$) of Eq.~(\ref{huu1}), Eq.~(\ref{hdd}) and Eq.~(\ref{hbd}) respectively (Fig.~\ref{fig:5}). 
The Hartree processes of $H^{\prime}_{\rm d}$ and $H^{\prime}_{\rm b}$ 
renormalize the Luttinger parameters and Fermi velocities in $H_0$ [Eq.~(\ref{H0})]. Especially, the Hartree term of 
$H^{\prime}_{\rm d}$ gives rise to positive $g_{2,a}$ ($a=2,3$) 
in Eqs.~(\ref{luttinger1},\ref{luttinger2}) in the presence of the repulsive electron interaction $(g>0)$. 

As in the previous section, we carried out the perturbative RG analyses on these effective boson models, 
$H_{0}+H^{\prime}_{{\rm u},1} + H^{\prime}_{{\rm u},2}+H^{\prime}_{{\rm d},1}+H^{\prime}_{{\rm b},2}$. 
At the one-loop level of the perturbative RG equations, $H^{\prime}_{{\rm u},1}$ and $H^{\prime}_{{\rm d},1}$ are coupled 
with each other, and so are $H^{\prime}_{{\rm u},2}$ and $H^{\prime}_{{\rm b},2}$. When the bare interaction 
strength $g$ is greater than critical interaction strength, respective pairs of the cosine terms grow up, to 
have larger amplitudes and the system enters strong coupling phases.  
In the following two subsections, we argue that $H^{\prime}_{{\rm u},1}$ and $H^{\prime}_{{\rm d},1}$ 
stabilize a plain superposed CDW phase, while $H^{\prime}_{{\rm u},2}$ and $H^{\prime}_{{\rm b},2}$ stabilize 
spin-nematic excitonic insulator phase.   

\subsection{a superposed CDW phase}
When $H^{\prime}_{{\rm u},1}$ and $H^{\prime}_{{\rm d},1}$ become relevant, the  
cosine terms in Eqs.~(\ref{Hu1d},\ref{Hd1d}) are maximally 
minimized by a charge density wave (CDW) phase, where a displacement 
field of the electron pocket and that of the hole pocket exhibit long-range orders individually;
\begin{align}
&2\phi_{2,j} = \Phi_{2}, \ \ 2\phi_{3,j} = \Phi_{3}, \label{phase-lock-a1} \\ 
&\Phi_{2} + \Phi_3 = \left\{\begin{array}{lc} 
2n \pi & (n_{(1)} < 0 ) \\
(2n+1)\pi  & (n_{(1)} > 0) \\
\end{array}\right., \label{phase-lock-a2}
\end{align}
with 
\begin{align}
\sigma_{2\overline{2},j} = \sigma_{3\overline{3},j} = \sigma.
\end{align} 
Such CDW is a plain superposition of a charge density wave of the electron pocket 
with $\downarrow$ spin and $\pi/k_{F,0,\downarrow}$ spatial pitch 
and that of the hole pocket with $\uparrow$ spin and $\pi/k_{F,-1,\uparrow}$ spatial 
pitch. Since this strong coupling phase is not accompanied by any long-range order of spin superconducting 
phase fields, the transition temperature of the superposed CDW phase increases 
monotonically in the magnetic field in the presence of the repulsive electron-electron interaction; 
$g_{2,a=2},g_{2,b=3}>0$.  Such behaviour of the transition temperature is {\it not} 
consistent with the graphite's experimental phase diagram; the experiment 
shows the re-entrant insulator-metal transition at $H=H_{c,2}\simeq 75$ T.

Besides, the long-range order of the {\it relative} displacement between the two 
charge density waves, $\Phi_2-\Phi_3$, is weak  
against random charged impurities, unless their spatial pitches are commensurate to the underlying lattice constant 
$c_0$~\cite{imry76,sham76,fukuyama78b,zhang17}. Namely, the impurity potentials  
induce single-particle backward scatterings within the same electron pocket  
and/or within the same hole pocket. The impurities appear spatially randomly as a 
function of the coordinate $z$. Thus, the scatterings add random U(1) phases into 
$2\phi_{2,j}$ and $2\phi_{3,j}$, unless $\Phi_2$ and $\Phi_3$ in Eq.~(\ref{phase-lock-a1}) have finite mass 
in the CDW phase. When $2k_{F,0,\downarrow}$ or 
$2k_{F,-1,\uparrow}$ is incommensurate with respect to $2\pi/c_0$, the long-range 
ordering of $\Phi_2-\Phi_3$ in Eq.~(\ref{phase-lock-a1}) 
is generally accompanied by a gapless phason excitation. Thereby, even small random charged 
impurities wipe out the long-range order of the relative 
phase between the two density waves. Meanwhile, being locked into 
the discrete values by the cosine potential in the umklapp term ($H^{\prime}_{{\rm u},1}$), 
the total displacement field, $\Phi_2+\Phi_3$, 
always has a finite mass in the superposed CDW phase. The locking is therefore  
robust against the random charged impurities, as far as their amplitudes are small.

\subsection{spin nematic excitonic insulator (SNEI-II) phase}
\begin{figure}[t]
	\centering
	\includegraphics[width=0.95\linewidth]{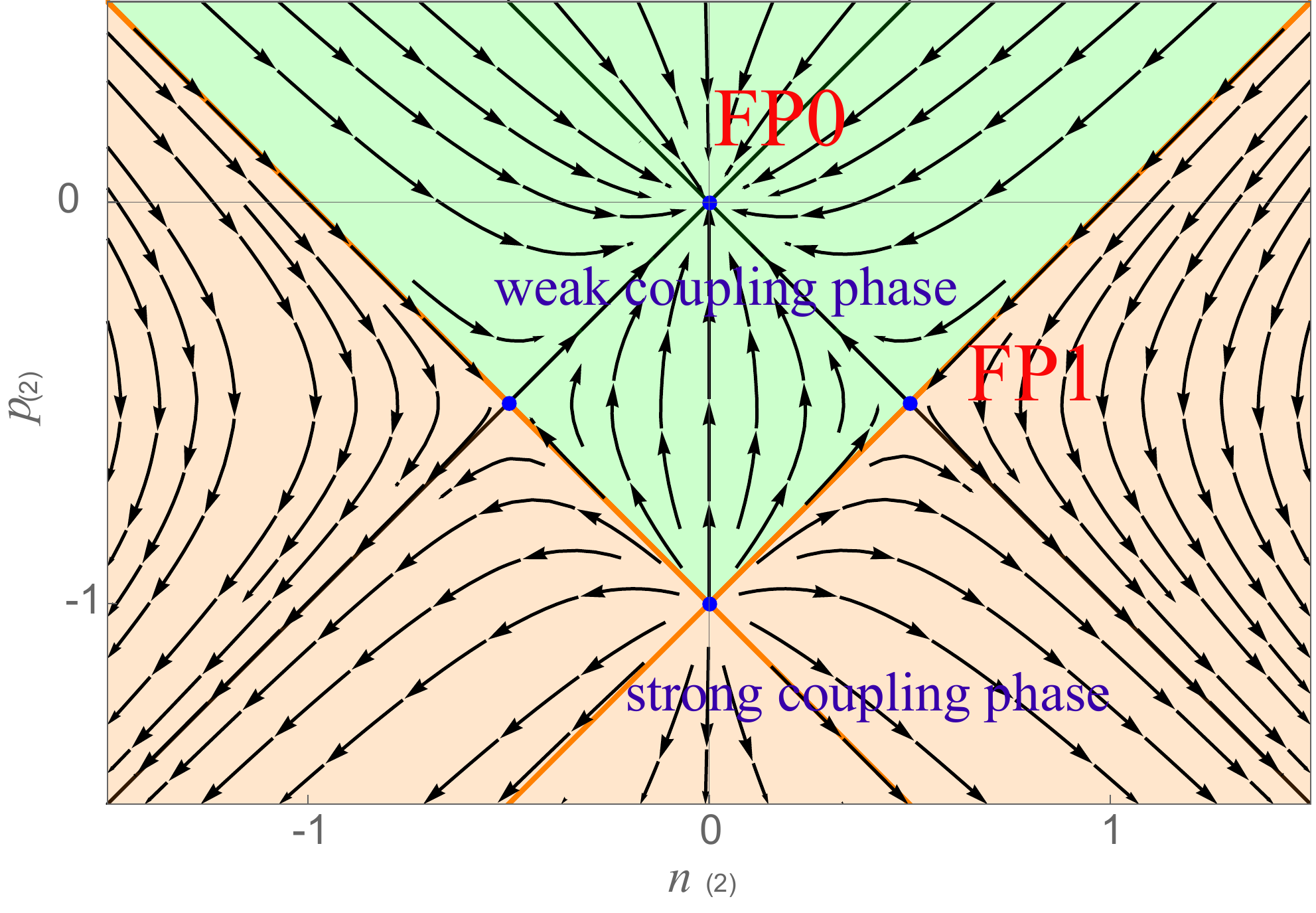}
	\caption{(color online) Renormalization group (RG) flow at $T=0$ in the two-dimensional 
parameter space subtended by $n_{(2)}$ and $p_{(2)} = \overline{p}_{(2)}$. Weak/strong 
coupling phases stand for normal metal phase/spin nematic excitonic insulator (SNEI-II) phase 
respectively. Quantum criticality of the quantum phase transition between these two are 
controlled by a fixed point named as `FP1'. The scaling dimension of the relevant 
parameter at FP1, $\nu_2$, is given in Eq.~(\ref{nu2}).} 
	\label{fig:6}
\end{figure}

When $H^{\prime}_{{\rm u},2}$ and $H^{\prime}_{{\rm b},2}$ become relevant, the 
cosine terms in Eqs.~(\ref{Hu2d},\ref{Hb2d})
are maximally minimized by the excitonic insulator phase with broken U(1) spin rotational symmetry. 
To see a nature and an RG phase diagram of this strong coupling phase, let us first reduce the inter-chain 
coupling functions in $H^{\prime}_{{\rm u},2}$ and $H^{\prime}_{{\rm b},2}$ into coupling 
constants;   
\begin{align}
n_{(2)}  &\equiv 2\pi l^2 \sum_{j} N^{(2)}_{j-m}, \nonumber \\
p_{(2)} &\equiv 2\pi l^2 \sum_{j} P^{(2)}_{j-m},  \ \  
\overline{p}_{(2)}  \equiv 2\pi l^2 \sum_{j} \overline{P}^{(2)}_{j-m}. \nonumber    
\end{align}
For the repulsive interaction case $(g>0)$, bare values of these three coupling 
constants are negative; the cosine terms in $H^{\prime}_{{\rm u},2}$ and $H^{\prime}_{{\rm b},2}$ 
are all from the exchange processes. The one-loop RG equations for these coupling constants 
take the following forms;
\begin{align}
\frac{dn_{(2)}}{d{\rm ln}b} &= A_{23} n_{(2)} - \frac{C_{23}}{\pi l^2} 
n_{(2)} \big(p_{(2)} + \overline{p}_{(2)}\big), 
\label{rg1c} \\ 
\frac{dp_{(2)}}{d{\rm ln}b} &= A_{23} p_{(2)}  - \frac{C_{23}}{\pi l^2} 
\big(n^2_{(2)} + p^2_{(2)} \big), \label{rg2c} \\ 
\frac{d\overline{p}_{(2)}}{d{\rm ln}b} &= A_{23} \overline{p}_{(2)}  
- \frac{C_{23}}{\pi l^2}  \big(n^2_{(2)} + \overline{p}^2_{(2)} \big). \label{rg3c} 
\end{align} 
Negative semi-definite $A_{23}$ and positive definite $C_{23}$ were already defined in Eq.~(\ref{Aab}) 
and appendix C3 respectively. Thanks to an inversion symmetry ($Q^{ab}_{+,j} \rightarrow -Q^{ab}_{-,j}$), 
the coupled equations as well as the bare values of the coupling constants are symmetric 
with respect to an exchange between $p_{(2)}$ and $\overline{p}_{(2)}$. This 
decouples the RG equations into 
\begin{align}
\frac{df_{\pm}}{d{\rm ln}b} = A_{23} f_{\pm} \mp \frac{C_{23}}{\pi l^2} f^2_{\pm}, \label{fpm}
\end{align}
where $f_{\pm} \equiv n_{(2)} \pm p_{(2)} = n_{(2)} \pm \overline{p}_{(2)}$. At the zero temperature, 
$A_{23}$ and $C_{23}$ have no dependence of the scale change ${\rm ln}b$. Thereby, 
the equations immediately 
give out a RG flow diagram as in Fig.~\ref{fig:6}. 
The strong and weak coupling phases at $T=0$ are defined by 
\begin{align}
\left\{\begin{array}{cl}
|n_{(2)}| - p_{(2)} > x_c  & ({\rm strong} \!\  \!\ {\rm coupling} \!\  \!\ {\rm phase}), \\ 
|n_{(2)}| - p_{(2)} < x_c & ({\rm weak} \!\ \!\  {\rm coupling} \!\ \!\ {\rm phase}),  \\ 
\end{array}\right.  \label{strong}
\end{align}
with 
\begin{align}
x_c \equiv - \frac{\pi l^2}{C_{23}} A_{23} > 0. \label{xc-snei2}
\end{align}

In the strong-coupling side, 
the cosine terms in the bosonized Hamiltonian are maximally minimized by 
\begin{align}
&\sigma_{2\overline{3},j} = \sigma_{3\overline{2},j} = \sigma, \label{snei-2-a} \\ 
&\theta_{2,j} -\theta_{3,j} = \Theta, \label{snei-2-b} \\ 
&2(\phi_{2,j} + \phi_{3,j}) = \left\{\begin{array}{lc} 2n\pi & (n_{(2)}<0) \\ 
(2n+1) \pi & (n_{(2)}>0) \\
\end{array}\right. .  \label{snei-2-c}  
\end{align}
The locking of a sum of the two displacement fields 
leads to an electrically insulating property along 
the field direction. The optical conductivity calculated within the Gaussian approximation shows a gap behavior, 
$\sigma_{zz}(\omega) = (e^2uK)/(2\pi l^2)\delta(\omega-\omega_g)$ with $uK=\sum_{a=2,3} u_a K_a$  and 
$\omega^2_g \equiv 2\pi uK \sum_{j} N^{(2)}_{j}$ (see also an inset of Fig.~\ref{fig:4}). The long-range order 
of the spin superconducting phase $\theta_2-\theta_3$ in Eq.~(\ref{snei-2-b}) breaks the global U(1) spin rotational 
symmetry. The strong coupling phase is accompanied by particle-hole pairing between the electron pocket with 
$\downarrow$ spin and the hole pocket with $\uparrow$ spin, so that we name the phase also 
as spin nematic excitonic insulator phase. Nonetheless, the phase could be symmetrically distinct from the 
spin nematic excitonic insulator discussed in the previous section, depending on a {\it spatial parity} of the 
excitonic pairing (see also Sec.~VIIIA and Sec.~XB). We thus distinguish these two by calling them as 
SNEI-I  for $H<H_0$ and SNEI-II for $H>H_0$ respectively. 

The phase boundary condition, Eqs.~(\ref{strong},\ref{xc-snei2}), explains the metal-insulator 
transition at a lower field regime.  
For simplicity, we assume that the bare values of $n_{(2)}$ and $p_{(2)}=\overline{p}_{(2)}$ 
as well as $(C_{23})_{|T=0}$ have no $H$-dependence. For a low $H$ regime, the magnetic length $l$ is 
large, so is the critical value $x_c$ in Eq.~(\ref{xc-snei2}). 
Thereby, a given bare value of $|n_{(2)}|-p_{(2)}$ can be below the critical value $x_c$ in 
lower $H$ regime (weak-coupling phase; normal metal phase). On increasing $H$, 
the magnetic length $l$ decreases and so does the critical value $x_c$. 
Thus, the bare value of $|n_{(2)}|-p_{(2)}$ 
exceeds the critical value $x_c$ at a certain critical magnetic field ($H=H^{*}_c$). For $H^{*}_{c}<H$, the 
system enters the strong coupling phase (SNEI-II phase). From a comparison with the graphite 
experiment~\cite{fauque13,akiba15,zhu17}, we assume that $H^{*}_{c}$ is 
{\it smaller} than $H_0$. In this case, the system at $T=0$ undergoes a phase 
transition from SNEI-I to SNEI-II at $H=H_0$.

The phase boundary condition Eqs.~(\ref{strong},\ref{xc-snei2}) also explains the $T=0$
insulator-metal {\it re-entrant} transition at a higher field regime.  
When the field $H$ increases further, both electron 
and hole pockets become smaller in size in the $k_z$ space. This makes their 
bare Fermi velocities $v_{F,2}$ and $v_{F,3}$ as well as 
Luttinger parameters $K_2$ and $K_3$ smaller. The smaller Luttinger parameters can 
increase the critical value $x_c$ through the dependence 
of $A_{23}$ on $K^{-1}_2$ and $K^{-1}_3$ [Eq.~(\ref{Aab})]. To be more precise, suppose 
that the electron pocket with $n=0$ LL with $\downarrow$ spin and the hole pocket with $n=-1$ LL with 
$\uparrow$ spin leave the Fermi level at $H=H_1$. When $H$ gets `close' to $H_1$ from below 
($H<H_1$), the increase of $-A_{23}$ can overcome the decrease of $l^2$ in Eq.~(\ref{xc-snei2}), such 
that $x_c$ increases again.  Namely, for $H<H_1$, $l^2$ is always bounded by $(\hbar c)/(eH_1)$ from  
below, while $K^{-1}_{2}$ and $K^{-1}_{3}$ as well as $-A_{23}$ have no upper bound in 
principle. Thus, for some magnetic field $H_{c,2}$ with $H^{*}_{c}<H_0<H_{c,2}<H_{1}$, $x_c$ 
exceeds the bare value of $|n_{(2)}|-p_{(2)}$ again and the system falls into  
the weak coupling phase (normal metal phase) again. From a set of 
reasonable parameter values used in Fig.~\ref{fig:1} (see appendix C4 for a set of 
parameters used in Fig.~\ref{fig:1}), we obtain $H_{c,2}=82$ T and $H_{1}=120$ T. 

\subsection{critical natures of the MI and re-entrant IM transitions}
The re-entrant transition point at $H=H_{c,2}$ is a zero-temperature 
continuous phase transition with dynamical exponent $z=1$. 
Toward this quantum critical point, a correlation length along the field 
direction $\xi_z$ diverges as 
\begin{align}
\xi_z \propto |H-H_{c,2}|^{-1/\nu_2}. \label{xi}
\end{align}  
A critical exponent $\nu_{2}$ is given only by the Luttinger parameters 
{\it at the critical point ($H=H_{c,2}$)},  
\begin{align}
\nu_2 = \frac{1}{2}\sum_{a=2,3} \big(K_a + K_a^{-1}\big) -2 . \label{nu2}
\end{align}
Since $z=1$, the correlation length is inversely 
proportional to the gap $\omega_g$ in the optical conductivity along the field 
direction, $\sigma_{zz}(\omega)$;
\begin{align}
\omega_g \propto (H_{c,2}-H)^{z/\nu_2} =  (H_{c,2}-H)^{1/\nu_2},  \label{wg}
\end{align}
for $H<H_{c,2}$. 
By measuring how the gap vanishes toward $H=H_{c,2}$ as a function of the field, 
one can determine the values of the Luttinger parameters at the quantum critical point. 
By seeing how much the Luttinger parameters thus determined deviate from $1$,  
one could also test a validity of our theory of the re-entrant insulator-metal transition. 

The low-$H$ metal-insulator transition between the normal metal 
and SNEI-I phases is also a quantum critical point. Toward this point, 
$H=H_{c,1}$, the gap $\omega_{g}$ in the SNEI-I phase also vanishes, 
\begin{align}
\omega_{g} \propto (H-H_{c,1})^{1/\nu_1}, \label{wg2} 
\end{align}
for $H_{c,1}<H$. The critical exponent $\nu_1$ is given by  
the Luttinger parameters at $H=H_{c,1}$;  
\begin{align}
\nu_1 = \frac{1}{2} \sum_{a} \big(K_a + K_a^{-1}\big) -2,  \label{nu1}
\end{align}
where the summation in the pocket index $a$ is taken over 
\begin{align} 
\left\{\begin{array}{cc}  a=1,4 & \ \ \  (|A_{14}| C_{23} h_{(2)} \gg |A_{23}| C_{14} \overline{h}_{(2)} ), \\
a=2,3 & \ \ \  (|A_{14}| C_{23} h_{(2)} \ll |A_{23}| C_{14} \overline{h}_{(2)}).  \\
\end{array}\right.  \label{nu1d}
\end{align}
Meanwhile, the gap in $\sigma_{zz}(\omega)$ reaches finite constant values 
at $H=H_0\pm 0$, when the phase transition from SNEI-I phase to SNEI-II phase is 
of the first order. This is the case when the {\it spatial parities} of the exctionic pairings in the 
two phases are different from each other (see also Sec.~VIIIA and Sec.~XB).

\section{in-plane resistance in the four pockets model ($H<H_0$)}    

Generally, in-plane current operators in the clean limit have finite matrix elements 
only between neighboring Landau levels. When the temperature is much lower than 
the cyclotron frequency $\hbar \omega_0$, the in-plane resistance increases on increasing
magnetic field $H$. Contrary to this expectation, the low-temperature 
in-plane resistance in graphite under high magnetic field shows an unexpected 
$H$-dependence~\cite{yaguchi98a,yaguchi98b,fauque13,akiba15}. 
It shows a broad peak around $15 \!\ {\rm T} \sim 30\!\ {\rm T}$, and then decreases slowly   
on further increasing $H$. From $H=$ 30 T to $H=H_0 \simeq 53 \!\ {\rm T}$, the 
resistance reduces by half or more. Besides, when the system enters the low-field-side out-of-plane 
insulating phase ($H_{c,1} < H < H_0$), 
the in-plane resistance shows an additional 
steep increase by $15 \%$ to $30 \%$~\cite{yaguchi98a,yaguchi98b,akiba15}. Unlike the out-of-plane 
resistivity, the additional increase amount becomes {\it smaller} for lower temperature. 

\subsection{$H$-dependence of $R_{xx}$ at $H<H_0$}
The $H$-dependence of the in-plane resistance in $30 \!\ {\rm T} < H <H_0 \simeq 53 \!\ {\rm T}$  
can be explained by charge transports along the surface chiral Fermi arc (SCFA) states. 
To see this, notice first that the electron/hole pockets in the bulk are terminated with  
SCFA states of the electron/hole type around the boundary regions of the system 
(see Fig.~\ref{fig:2} and appendix A).  A SCFA state of the electron/hole type is a bundle 
of $N_a$-number of chiral edge modes of electron/hole type respectively, where $N_a$ is a 
number of $k_z$ points within the electron/hole pocket ($a=1,2,3,4$). Here `$a$' denotes 
the pocket index; $1\equiv (0,\uparrow)$, $2\equiv (0,\downarrow)$, $3 \equiv (-1,\uparrow)$, 
and $4 \equiv (-1,\downarrow)$. The chiral edge mode enables unidirectional 
electric current flow along the boundary in a $xy$ plane. 
The chiral directions of the electric current flows of the electron-type and hole-type edge modes 
are opposite to each other. 

In the presence of short-ranged charged impurities, the current flow along the 
electron-type  edge mode with $\sigma$ spin can be scattered into the hole-type 
edge mode with the same $\sigma$ spin. In this respect, the SCFA state with 
$(-1,\sigma)$ (hole-type) and that with $(0,\sigma)$ (electron-type) cancel 
each other by the intra-surface backward scatterings 
due to the charged impurities. In the absence of any backward scatterings 
between $(0,\sigma)$ and $(-1,\overline{\sigma})$ ($(\sigma,\overline{\sigma})
=(\uparrow,\downarrow),(\downarrow,\uparrow)$; see the last paragraph in Sec.~V for 
the reasonings of the absence), both $(N_2-N_4)$-number of anticlockwise (electron-type) chiral 
edge modes with $\downarrow$ spin and $(N_3-N_1)$-number of clockwise (hole-type) chiral edge 
modes with $\uparrow$ spin {\it individually} contribute to the two-terminal conductance 
within the $xy$ plane;
\begin{align}
G_{\rm s} &= \frac{e^2}{h} \big(N_2-N_4 + N_3-N_1\big) \nonumber \\
&=  \frac{2e^2}{h} \big(N_3-N_1\big). \label{surfacec}
\end{align}
From the first line to the second line, 
we used the charge-neutrality condition; $N_1+N_2=N_3+N_4$. Importantly, the 
in-plane conductance given by Eq.~(\ref{surfacec}) usually {\it increases} on increasing $H$ for 
$H<H_0$. This is 
because a variation of $N_1$ with respect to $H$ is larger than that of $N_3$; 
$dN_1/dH<dN_3/dH<0$. For $N_3=(L_z/(2c_0))(1-H/H_1)$, 
and $N_1=(L_z/(2c_0))(1-H/H_0)$, the $H$-dependence of the resistance due to 
the surface charge transport is given by 
\begin{eqnarray}
R_{\rm s} = \frac{h}{e^2} \frac{c_0}{L_z} \frac{H_0 H_1}{H(H_1-H_0)}. \label{surfacer}
\end{eqnarray} 
The resistance is on the order of 1 $\Omega$ at $H=30$ T [$L_z=50\!\ \mu{\rm m}$, 
$c_0=0.67\!\ {\rm nm}$, $H_0=50$ T and $H_1=120$ T.] The value is on the same order 
of the experimental value ($2 \!\ \Omega \sim 3\!\ \Omega$)~\cite{fauque13}. 

\subsection{$T$-dependence of $R_{xx}$ at $H<H_0$}
The $T$-dependence of the in-plane resistance inside the low-field-side insulating phase 
($H_{c,1} < H < H_0$) can be explained 
by a coupling between the SCFA states and gapless Goldstone modes 
associated with the spin nematic order in the bulk. The spin-nematic excitonic 
insulator (SNEI-I) phase breaks two global U(1) symmetries. They are the 
U(1) spin-rotational symmetry around the field direction and a translational 
symmetry associated with a spatial polarization of the spin ($\uparrow$ or $\downarrow$) 
and pseudospin ($n=0$ LL or $n=-1$ LL) densities. 

Such SNEI-I phase has two low-energy 
gapless excitations. They are space-time fluctuations of the following two phase variables 
[Eqs.~(\ref{phi-lock},\ref{theta-lock})];    
\begin{align}
f_j(z) &\equiv \big(\theta_{3,j}(z) -\theta_{2,j}(z) \big) - 
\big(\theta_{4,j}(z) -\theta_{1,j}(z) \big) - 2\Theta_{-}, \label{spin} \\
g_j(z) &\equiv \big(\phi_{3,j}(z) +\phi_{2,j}(z) \big) - \big(\phi_{4,j}(z) + \phi_{1,j}(z) \big) - 2\Phi_{-}. \label{trans} 
\end{align} 
When they vary slowly in $z/c_0$ and $y_j/l \equiv 2\pi l j/L_x$, 
their energy dispersions become linear in the momenta; 
\begin{align}
{\cal H}_{\rm sw} &= \frac{1}{2L_z N} \sum_{\bm k} \big(B_1 k^2 + C_1 k^2_z\big) 
f^{\dagger}({\bm k}) f ({\bm k}) \nonumber \\
& \ \ \ +  \frac{1}{2L_z N} \sum_{\bm k} \big(B_2 k^2 + C_2 k^2_z\big) 
g^{\dagger}({\bm k}) g ({\bm k}), \label{sw}
\end{align}
with positive $B_t$ and $C_t$ ($t=1,2$), and ${\bm k}\equiv (k_z,k)$. 
$k_z$ and $k$ are conjugate to $z$ and $y_j \equiv 2\pi l^2 j/L_x $ respectively,    
\begin{align}
f_{j}(z) &\equiv \frac{1}{L_z N} \sum_{\bm k} e^{ik_z z + ik y_j} f({\bm k}), \nonumber \\ 
g_j(z) &\equiv \frac{1}{L_z N} \sum_{\bm k} e^{ik_z z + ik y_j} g({\bm k}). \nonumber 
\end{align}
The gapless modes couple with the SCFA states through a simple 
density-density interaction, e.g. 
\begin{align}
{\cal H}^{\prime} &= \frac{1}{L_x} \sum_{a,\tau,b}\sum_{n} \sum_{m} \int dz \!\ 
{\cal A}^{{\rm e-b}}_{(a,\tau;b)}(y_n,y_m) \nonumber \\ 
&\hspace {2.7cm} \times \rho_{a,\tau,n}(z) \!\ 
\big(\psi^{\dagger}_{b,m}(z) \psi_{b,m}(z)\big), \label{e-b}
\end{align}
with bulk density operator $\rho_{a,\tau,n}(z) \equiv \psi^{\dagger}_{a,\tau,n}(z) \psi_{a,\tau,n}(z)$ 
($|y_n|\le L_y/2$). $a,b=1,2,3,4$ denote the pocket indices, and $\tau=\pm$ is the chirality index. By definition, 
the summations over the chain indices $n$ and $m$ in Eq.~(\ref{e-b}) are restricted within the bulk 
region and edge region respectively; $L_y/2 \le |y_m|$.

When bosonized, the density operator in the bulk region  is given by a linear 
combination of the phase variables, 
$2\pi \rho_{a,\tau,n}(z) \equiv \partial_z\phi_{a,n}(z)- \tau \partial_z \theta_{a,n}(z)$. Such 
phase variables generally contain the two low-energy gapless excitations with the linear dispersions. 
Thus, the situation becomes precisely analogous to the electron-phonon interaction 
in metal~\cite{fetter03,mahan00}. The coupling gives the SCFA electrons with finite transport life 
times~\cite{mahan00}. When the temperature is on the order of a band width 
of the gapless Goldstone modes (but smaller than the transition 
temperature of the SNEI-I phase), the transport life time of the SCFA states is linear 
in temperature $T$; so is the resistivity due to the surface charge transport. This 
can explain the $T$-dependence of the in-plane resistance in the insulating 
phases in graphite ~\cite{yaguchi98a,yaguchi98b,akiba15}. 

\section{in-plane resistance in the two pockets model ($H>H_0$)}
The in-plane resistance of graphite under the high magnetic field 
stays almost constant in the field inside the high-field-side out-of-plane insulating 
phase ($H_0 < H < H_{c,2}$)~\cite{fauque13,akiba15,arnold17,zhu17,zhu18}. Above the 
re-entrant insulator-metal (IM) transition field ($H_{c,2} < H$), the resistance shows the normal 
behaviour; $R_{xx}$ increases in the field~\cite{zhu18}. 

In the following, we will argue that 
the SNEI-II phase in $H>H_0$ can be either topological~\cite{fu07,roy09,moore07} or topologically 
trivial, depending on the spatial parity of the excitonic pairing between electron pocket 
($n=0,\downarrow$) and hole pocket ($n=-1,\uparrow$). 
When the excitonic pairing field is an odd function in the momentum $k_z$, the SNEI-II phase becomes 
topological and thereby the SCFA state of electron type ($n=0,\downarrow$) 
and the SCFA state of hole type ($n=-1,\uparrow$) are reconstructed into a  
helical surface state with a gapless Dirac cone. The electric transport through 
such Dirac-cone surface state is primarily determined by carrier density doped 
in the surface region, that has little field-dependence. Thus, the reconstructed 
Dirac-cone surface state may provide a simple explanation for the field-(nearly) independent 
and metallic behaviour of the in-plane resistance observed in the high-field-side out-of-plane 
insulating phase ($H_0 < H < H_{c,2}$).   

\subsection{topological SNEI phase}

\begin{figure}
	\centering
	\includegraphics[width=0.9\linewidth]{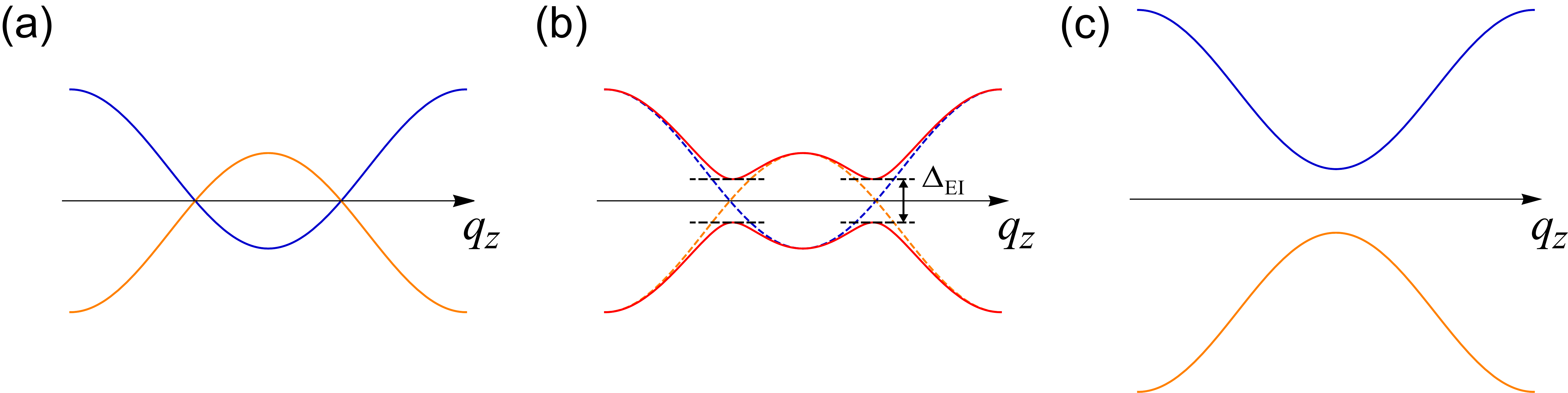}
	\caption{(color online) (a) single-particle electronic states in 
normal metal phase (two pocket model). The electron pocket (blue curve) 
is formed by the $n=0$ LL with $\downarrow$ spin, and the hole pocket (yellow curve) 
is by the $n=-1$ LL with $\uparrow$ spin. (b) single-particle electronic states with  
the excitonic pairing. (c) single-particle electronic states in the vacuum region.} 
	\label{fig:7a}
\end{figure}

The strong coupling phase discussed in Sec.~VI (SNEI-II phase) consists of 
two topologically distinct phases, depending on the sign of the umklapp term 
$n_{(2)}$. A mean-field one-dimensional electronic Hamiltonian of the strong coupling phase 
can be schematically described by the 2 by 2 Pauli matrices as 
\begin{align}
H^{\rm snei2}_{\rm mf} (q_z) &= (M-2\gamma_2 \cos(q_z c_0)) {\bm \sigma}_3
+ \Delta_{\rm EI}(q_z c_0) {\bm \sigma}_1 \nonumber \\ 
&\equiv E_{\rm EI}(q_z) \big\{N_1(q_z) {\bm \sigma}_3 
+ N_2(q_z) {\bm \sigma}_1 \big\}, \label{snei-2-fermi}
\end{align} 
with $M<2\gamma_2$, and  
\begin{align}
E_{\rm EI}(q_z) \equiv \sqrt{\big(M-2\gamma_2 \cos(q_z c_0)\big)^2+\Delta^2_{\rm EI}(q_z c_0)}. \label{snei-2-fermi2}
\end{align} 
The first and second elements of the 2 by 2 matrices correspond to the 
$n=0$ LL with $\downarrow$ spin and $n=-1$ LL with $\uparrow$ spin respectively 
(Fig.~\ref{fig:7a}(a)). For clarity, the electron pocket around $k_z=0$ is shifted by 
$\pi/c_0$ in Eq.~(\ref{snei-2-fermi}); $q_z\equiv k_z-\frac{\pi}{c_0}$. 
$\Delta_{\rm EI}(q_zc_0)$ stands for an excitonic pairing between the electron 
and hole pockets (Fig.~\ref{fig:7a}(b)). The pairing is induced by the umklapp 
$H^{\prime}_{{\rm u},2}$ and inter-pocket scattering terms $H^{\prime}_{{\rm b},2}$. 
A function form of 
$\Delta_{\rm EI}(q_z c_0)$ is determined by a value of the total displacement field, such 
as in Eq.~(\ref{snei-2-c}).

For the negative umklapp term, $n_{(2)}<0$, the excitonic pairing field $\Delta_{\rm EI}(q_z c_0)$
is an odd function in $q_z$, while, for the positive case, $n_{(2)}>0$, it is even in $q_z$. 
These two cases represent two topologically distinct phases. In the former/latter case, 
the following topological winding number defined for the bulk 1-dimensional Hamiltonian 
Eq.~(\ref{snei-2-fermi}) takes $\pm 1$/zero respectively;~\cite{ssh,wen89,sato09}
\begin{eqnarray}
Z \equiv \int^{\frac{\pi}{c_0}}_{-\frac{\pi}{c_0}} 
\frac{dq_z}{2\pi} \big(\vec{N} \times \partial_{q_z} \vec{N}\big)_3,  
\end{eqnarray}
with $\vec{N} \equiv (N_1(q_z),N_2(q_z),0)$. 

\begin{figure}
	\centering
	\includegraphics[width=0.9\linewidth]{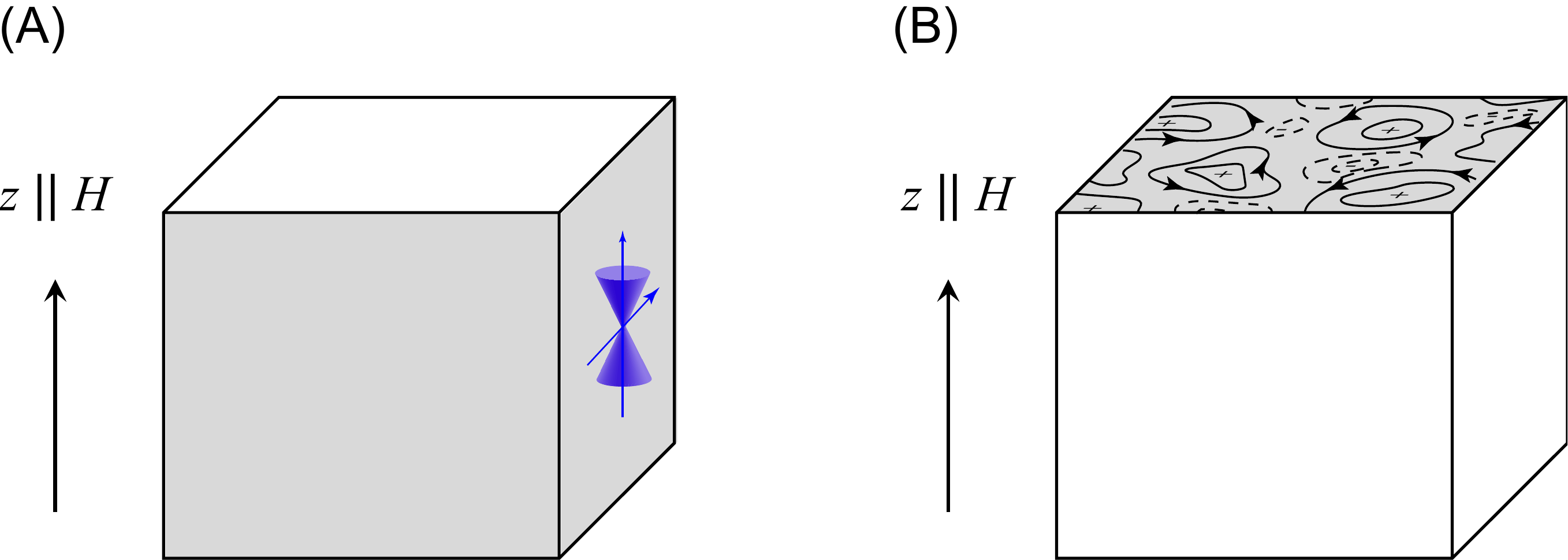}
	\caption{(color online) Schematic pictures of (A) side surfaces 
(grey area) with the two-dimensional helical surface state with a 
gapless Dirac cone. (B) top surface (grey area) with 
the two-dimensional Chalker-Coddington network model.} 
	\label{fig:7b}
\end{figure}

The non-zero bulk winding number reconstructs the SCFA state of the electron 
type and that of the hole type into a 2-d surface state with a gapless Dirac 
cone at side surfaces. The side surface is 
subtended by $z$ $(\parallel H)$ and either $x$ or 
$y$ (Fig.~\ref{fig:7b}(A)). To be concrete, impose the periodic boundary conditions along 
$z$ and $x$, put a confining potential along $y$ direction. The mass term 
$M$ in Eq.~(\ref{snei-2-fermi}) depends on the coordinate $y$. In the vacuum 
regime, $|y| > L_y/2$, the electron/hole pocket goes above/below 
the Fermi level (Fig.~\ref{fig:7a}(c)). Thereby, Eq.~(\ref{snei-2-fermi}) 
enters a normal 1-dimensional semiconductor regime, $M>2\gamma_2$; the winding 
number takes zero in the vacuum. In the bulk region, $|y| < L_y/2$, 
the gapped mean-field Hamiltonian with the negative $n_{(2)}$ is in the band-inverted regime, 
$M<2\gamma_2$; the winding number takes $\pm 1$. Such two topologically distinct 
1-dimensional gapped systems are inevitably separated by a 1-d gapless Dirac Hamiltonian, 
that should come somewhere around $|y| = L_y/2$. In other words, the side surface has a 
2-d helical surface state that forms a gapless Dirac cone as a function of $k_z$ and $y\equiv k_x l^2$ 
(Fig.~\ref{fig:7b}(A)). 

The reconstructed surface state has the {\it helical} velocities not only along the $z$-direction 
but also along the $x$-direction. To see this, notice that the velocity along the $x$-direction 
is given by a derivative of the single-particle energy with respect to the spatial coordinate $y$; 
$v_x \equiv l^2 \partial E_{\rm EI}/\partial y$. Such velocity changes its sign around $y= L_y/2$, 
where $E_{\rm EI}(q_z)$ forms the 1-d gapless Dirac dispersion; $v_{x}<0$ for $y<L_{y}/2$ 
and $v_x>0$ for $y>L_{y}/2$ (See also Figs.~\ref{fig:7a}(B,C)).

Quantitatively, the Dirac cone is highly anisotropic in its velocity within the side surface. 
Namely, the velocity along the $x$ direction is determined by a work function in the edge region;
$v_x = {\cal O}(l^2 \partial M/\partial y)$. Conventionally, the work function 
varies in energy on the order of eV within a length scale of $\AA$; 
$\partial M/\partial y = {\cal O}({\rm eV}/{\AA})$. Thus, the velocity along $x$ direction is 
much faster than that along $z$ direction, the latter of which is given by an energy scale of 
the band width ($2\gamma_2$) or the excitonic pairing ($\Delta_{\rm EI}$). 

\begin{figure}
	\centering
	\includegraphics[width=0.9\linewidth]{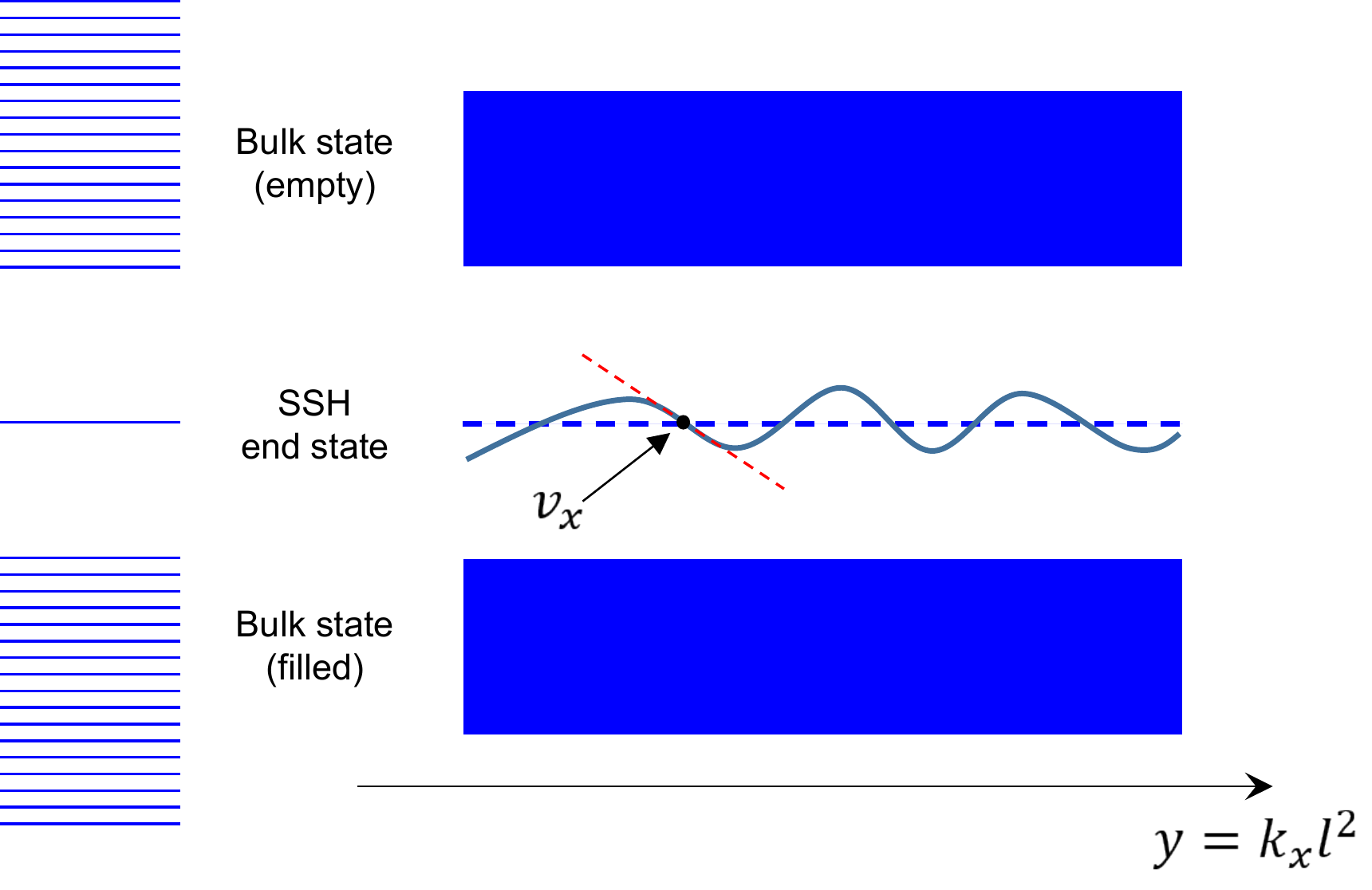}
	\caption{(color online) Schematic picture of energetically 
degenerate SSH end states within the bulk excitonic band gap (blue dotted line). 
In a generic situation, the degeneracy is lifted by an electrostatic potential (black solid curve). 
An associated spatial gradient of the end-state eigenenergy with respect to $y$ leads 
to a chiral electric current along $-x$ direction.} 
	\label{fig:7c}
\end{figure}

The 2-d helical surface state in the side surface is continuously connected to 
a 2-d critical wavefunction sitting on a top (bottom) surface. The top (bottom) 
surface is subtended by $x$ and $y$ coordinates (Fig.~\ref{fig:7b}(B)). 
Theoretically, the critical wavefunction belongs to the 2D quantum Hall 
universality class, while it is generically off the Fermi level.

To see this, impose the open boundary condition along $z$ ($\parallel H$) 
direction. The non-zero bulk winding number leads to an in-gap end state 
called as SSH (Su-Schrieffer-Heeger) state within the bulk excitonic gap 
(left figure of Fig.~\ref{fig:7c}). The end states are 
localized at the two open boundaries along $z$ direction,  
top and bottom surfaces. Due to the Landau degeneracy 
associated with the in-plane coordinate degree of freedom, each boundary has 
extensive number of such end states. In the clean limit, they are 
energetically degenerate. In the presence of charged impurities on the surface, 
the degeneracy is lifted by an electrostatic potential created by the impurities 
(right figure of Fig.~\ref{fig:7c}). The potential depends on $x$ and $y$, causing  
a finite spatial gradient of the end-state eigenenergy. 
The gradient in $x$ or $y$ 
gives rise to a chiral electric current (one-dimensional chiral mode) 
along $y$ or $-x$ direction respectively. Such chiral mode encloses 
a region with higher electrostatic potential. An uneven potential landscape 
gives rise to a group of chiral modes on the surface (Fig.~\ref{fig:7b}(B)), where 
two spatially proximate (and thus counter-propagating) modes have finite 
inter-mode hoppings. Electronic states of such surface can be 
described by the Chalker-Coddington network (CCN) model.~\cite{cc,hc} 
The previous studies on the CCN model~\cite{cc,em} conclude that a phase diagram 
as a function of the chemical potential has two localized regimes and the 
2D quantum Hall critical point intervenes between these two localized regimes. 
Thus, in-gap surface electronic states sitting on the top (bottom) surface are 
generally localized within the in-plane direction, unless the chemical potential is 
fine-tuned to the critical point.

\section{summary} 
Graphite under high magnetic field exhibits mysterious metal-insulator (MI) 
transitions as well as insulator-metal (IM) re-entrant transitions. We discuss  
these enigmatic electronic phase transitions in terms of perturbative RG 
analyses of effective boson theories. We argue that the two 
insulating phases in graphite under high field are excitonic insulators with spin 
nematic orderings. Similar conclusions were suggested by experimental 
works both for $H<H_0$~\cite{zhu17} and $H>H_0$~\cite{akiba15}. This paper 
enumerates possible umklapp terms allowed under the charge neutrality condition, 
clarifies natures of insulating states stabilized by each of them, and argues  
that excitonic insulators with long-range orderings of spin superconducting phases  
can give a possible explanation to the graphite experiments. 

Based on this, we propose a new mechanism for the re-entrant IM transition. 
When a pair of electron and hole 
pockets get smaller in size, strong quantum fluctuation 
of the spin superconducting phase destabilizes the spin-nematic excitonic insulator, causing
the re-entrant IM transition. The strength of the quantum fluctuation is 
quantified by the Luttinger parameters of the electron and hole pockets. 
We relate the Luttinger parameters with the critical exponent of the $T=0$ re-entrant IM 
transition point. We show that the exponent can be experimentally determined  
from the infrared optical spectroscopy. By determining the Luttinger parameters at the 
transition point, experimentalists can test a validity of our theory for the re-entrant IM  
transition. 
 
We attribute an `unexpected' field- and temperature-dependences of the in-plane electric 
transport in graphite under the high field as surface charge transports through surface chiral 
Fermi arc (SCFA) states and reconstructed Dirac-cone surface state. We first 
argue that a metallic temperature dependence of the in-plane transport 
observed in the low-field-side insulating phases is due to bulk-edge couplings  
between the SCFA states and gapless Goldstone modes associated with 
the spin nematic orderings. Being gapless excitations, the Goldstone modes 
in the spin-nematic excitonic insulator phases could be experimentally 
detected through ultrasound measurements~\cite{leboeuf17}.   
We also argue that the odd-parity excitonic pairing 
in the bulk reconstructs SCFA states of electron and hole into a $(2+1)$-d helical surface 
state with a gapless Dirac cone. Based on this finding, we discuss the field- (nearly) independent 
and metallic behaviour of the in-plane transport inside the high-field-side insulating 
phase~\cite{yaguchi98a,yaguchi98b,fauque13,akiba15}. 

\section{discussion}

\begin{figure}
	\centering
	\includegraphics[width=0.9\linewidth]{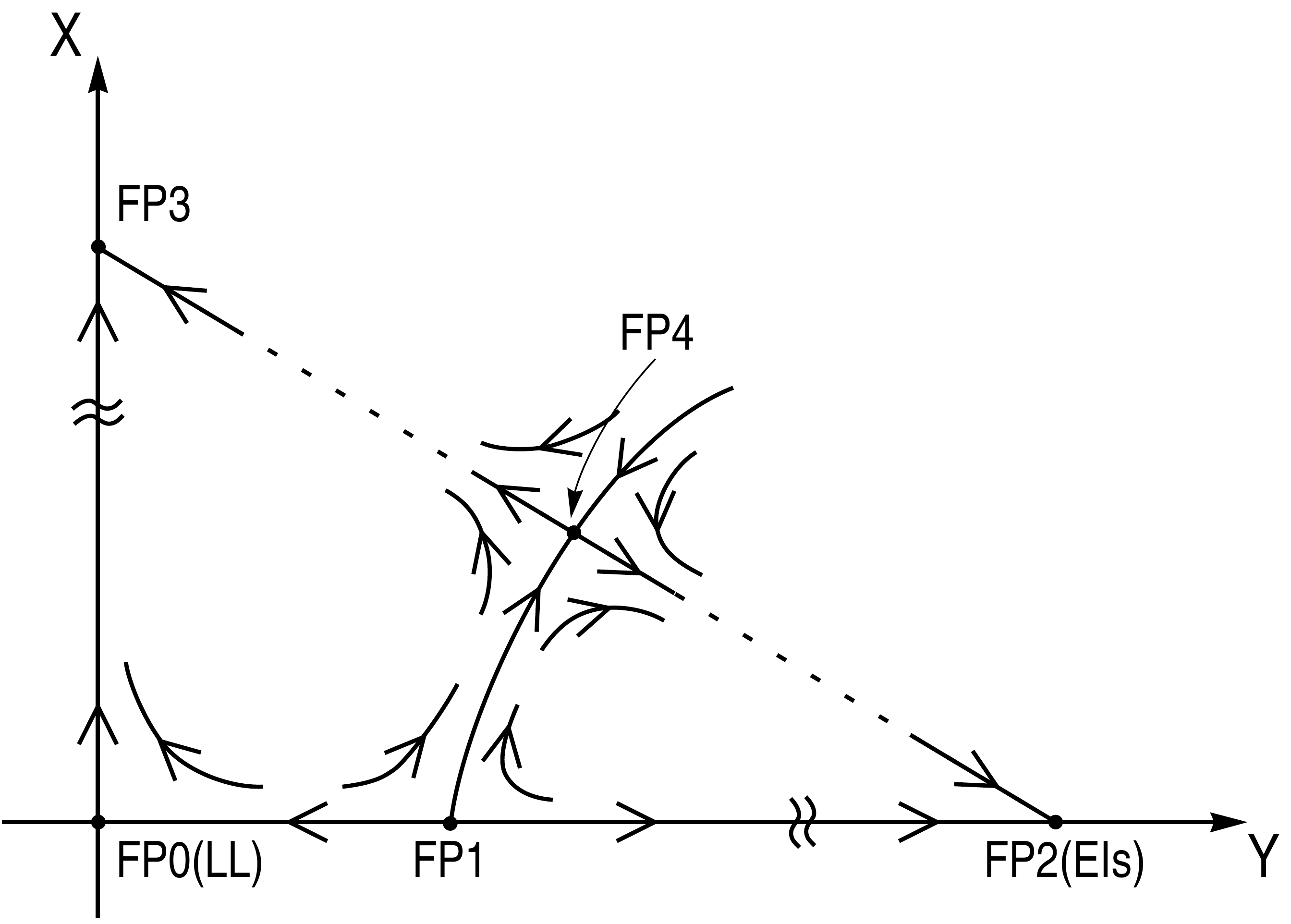}
	\caption{(color online) Schematic picture of a possible RG phase diagram 
in the presence of a relevant perturbation (denoted by `$X$' ) around the decoupled Luttinger 
liquid (LL) fixed point (denoted by `FP0'). In the presence of such relevant perturbation, 
the LL fixed point is unstable; the normal metal phase 
is characterized by a new stable fixed point (denoted by `FP3'). A horizontal axis (`$Y$') 
denotes the umklapp and inter-pocket scattering terms that drive the system into 
the excitonic insulator (EI) phases. `FP2' represents a stable fixed point charactering the EI phases.
The critical properties of the metal-insulator (MI) and re-entrant insulator-metal (IM) transitions are
characterized by a new saddle fixed point (denoted by `FP4') instead of by the FP1. 
In this schematic picture, we assume that the fixed point for the EI phases (`FP2') is locally 
stable against the small perturbation $X$.} 
	\label{fig:8}
\end{figure}

\subsection{natures of the `normal' metal phase and criticalities of metal-insulator transitions}
Our theory regards the `normal' metallic phase in the graphite 
experiment as decoupled Luttinger liquid (LL) phase, where we assume 
that inter-chain electron-electron interactions only renormalize the Luttinger 
parameters and Fermi velocities as in Eqs.~(\ref{luttinger1},\ref{luttinger2},\ref{g2a-q},\ref{g4a-q}). 
Nonetheless, it could be possible that a fixed point of the decoupled LL phase (a gaussian theory 
given by Eq.~(\ref{H0}); schematically denoted by `FP0' in Fig.~\ref{fig:8}) is {\it unstable} 
against a certain perturbation associated 
with the inter-chain interactions (denoted by `X' in Fig.~\ref{fig:8}) and, as a result, 
the `normal' metal phase is characterized by a new stable fixed point (schematically 
denoted by `FP3' in Fig.~\ref{fig:8}). The stable fixed point could be the  
Fermi-liquid fixed point~\cite{shankar91,houghton93,neto94a,neto94b,houghton94,fabrizio93} 
or the sliding Luttinger-liquid fixed point~\cite{emery00,vishwanath00}.
One of the experimental evidences that could support our theory's assumption of the decoupled 
Luttinger liquid is a $T$-linear behaviour (or at least non-Fermi-liquid behaviour) in the 
out-of-plane resistivity in the high-$T$ `normal' metal phase. To our best knowledge, however, no 
comprehensive experimental studies have been carried out so far for the temperature-dependence of 
the resistivity in the `normal' metal phase in the graphite under the high magnetic field~\cite{fauque13,zhu17}. 

When the metal phase is characterized by a new free theory instead of 
the free theory of the decoupled LL phase (a gaussian theory given by Eq.~(\ref{H0})), 
critical properties of the metal-insulator (MI) and re-entrant insulator-metal (IM) 
transitions are characterized by a new saddle-point fixed point (schematically 
denoted by `FP4' in Fig.~\ref{fig:8}), rather than by the FP1 that leads to 
the argument in Sec.~VIC. Meanwhile, having a finite charge gap, 
a fixed point of the excitonic insulator (EIs) is expected to be locally stable against the 
small perturbation. Thereby, the primary features of the two EI phases 
discussed in the paper will not change dramatically even in the presence 
of such perturbations. These features include the 
finite mobility gaps in $\sigma_{zz}(\omega)$ in the two SNEI phases, an overall 
structure of the $H$-$T$ phase diagram as well as topological Dirac-cone surface 
state in the SNEI-II phase and in-plane electric transport due to the surface state.  
    
\subsection{excitonic BCS-BEC crossover and nature of a transition between SNEI-I and SNEI-II phases}
Our theory does not include an effect of an excitonic condensation, as emphasized 
in Ref.~\cite{zhu17}. When $H$ approaches $H_0$ from above ($H > H_0$), 
electron-hole bound states formed by electron in $(n,\sigma)=(0,\uparrow)$ LL and hole in 
$(n,\sigma)=(-1,\downarrow)$ LL could undergo the Bose-Einstein condensation. Such condensation 
further assists electron-hole BCS pairings between $(n,\sigma)=(0,\downarrow)$ 
and $(n,\sigma)=(-1,\uparrow)$ LLs, through the umklapp term $H_{{\rm u},2}$. 
This leads to a phase with electrically insulating behaviour along the field direction; 
the phase is essentially same as the SNEI-I phase discussed in the paper. 
When the exciton BEC effect is included into our theory, the phase boundary 
between SNEI-I and SNEI-II phases (say $H=H_{c,3}$) will presumably 
go above $H_0$ ($H_0 < H_{c,3}$). 

For $H\ge H_{c,3}$, the long-range phase coherences defined by 
$\theta_{4,j}-\theta_{1,j} = n\pi - \Theta_{-}$ and $\phi_{4,j}+\phi_{1,j}=
(m+1)\pi - \Phi_{-}$ in Eqs.~(\ref{phi-lock},\ref{theta-lock}) fade away, 
while the other long-range phase coherences   
defined by $\theta_{3,j}-\theta_{2,j} = \Theta_{-}$ and $\phi_{3,j}+\phi_{2,j}= \Phi_{-}$  
may survive, leading to a phase similar to the spin nematic excitonic insulator phase discussed in 
Sec.~VIB. From this viewpoint, the SNEI-II phase could be regarded as 
a `partial ordered phase' derived from the SNEI-I phase. Nonetheless, it can be entirely 
possible that these two SNEI phases are symmetrically distinct from each other, depending on 
the spatial parities of the excitonic pairings in the two phases, whose importance was 
emphasized in Sec.~VIIIA. Qualitative natures of the phase transition between 
these two excitonic insulator phases need further theoretical studies.  

\appendix
\section{charge neutrality condition} 
Transverse conductivity $\sigma_{xy}$ gives a precise information of electron carrier density 
$n_e$ and hole carrier density $n_h$ in any given metal and semimetal under high magnetic field 
through the following formula;
\begin{eqnarray}
\sigma_{xy} H = ec (n_e - n_h). \label{sigmaxy-formula}
\end{eqnarray}  
$e$ $(>0)$ and $c$ are the electron charge, and the speed of light respectively. 
In the main text, we use the formula and evaluate the total number 
of $k_z$ points in the electron/hole pockets $N_e$/$N_h$ in graphite under 
the field. With the formula, the previous Hall conductivity measurement 
in the regime of $20 \!\ {\rm T} \lesssim H \lesssim 55 \!\ {\rm T}$~\cite{uji98,kopelevich09,kumar10,akiba15} 
gives $(N_e-N_h):L_z/c_0=10^{-4}:1$. Using the Kubo formula of the 
Hall conductivity, Akiba 
discussed a validity of the formula in the quasi-quantum limit in the graphite~\cite{akiba15}. 
In the following, we employ the Buttiker's theory of the Hall conductivity~\cite{halperin82,buttiker88}, to demonstrate 
a validity of the formula in a generic three-dimensional metal 
and semimetal under high field. 

Use the Landau gauge and assume 
that a given three-dimensional system is translational 
symmetric along $x$ and $z$ directions. Electrons are confined along $y$ 
direction within $|y|<L_y/2$ by a confining potential. A single-particle 
Hamiltonian comprises of 
two parts;
\begin{eqnarray}
\hat{\cal H}_{T} \equiv 
\hat{\cal H}_{0}(k_z;\hat{\kappa}_{\pm}) + 
\hat{\cal V}(k_z;\hat{\kappa}_{\pm},\hat{y}), \label{generic}
\end{eqnarray} 
with $\kappa_{\pm} \equiv (-i\partial_y) \pm i(-k_x+\frac{eH y}{c\hbar})$.  
$\hat{\cal H}_0$ is a bulk Hamiltonian that depends on the coordinate 
$y$ through $\hat{\kappa}_{+}$ and $\hat{\kappa}_{-}$. $\hat{\cal V}$ describes 
an effect of the confining potential; $\hat{\cal V}\equiv 0$ when $|y|\ll L_y/2$. 
$\hat{\cal V}$ depends on $y$ explicitly. $\hat{\cal H}_T$ in Eq.~(\ref{generic}) 
is already Fourier-transformed with respect to $x$ and $z$: they are functions of the 
conjugate momenta $k_x$ and $k_z$. In a system with multiple energy bands, 
$\hat{\cal H}_T$ takes a matrix form. For the spinless graphite case, 
$\hat{\cal H}_{T}$ is a four by four matrix; the four bases are from the 
$\pi$ orbitals in $A$, $A'$, $B$ and $B'$ carbon atoms within the unit cell. 
Using the ${\bm k}\cdot {\bm p}$ expansion, Slonczewski, Weiss and McClure  
derived ${\cal H}_0$ around the zone boundary of the first Brillouin zone of the graphite.

In the following, we only assume that ${\cal H}_0(k_z;\kappa_{\pm})$ as well 
as ${\cal V}(k_z;\kappa_{\pm},y)$ 
are given by finite order polynomials in $\kappa_{\pm}$ and $y$. Under this assumption, 
the explicit $y$-dependence of ${\cal V}$ can be rewritten into 
the $y_c$-dependence by use of  
$y\equiv (-i)(l^2/2)(\kappa_{+}-\kappa_{-})+y_c$ and $y_c \equiv k_x l^2$;
\begin{align}
\hat{\cal H}_{T}(k_z;\hat{\kappa}_{\pm},\hat{y}) 
= \hat{\cal H}^{\prime}_{T}(k_z,y_c;\hat{\kappa}_{\pm}). \label{rewrite}  
\end{align} 
Eigenstates of such $\hat{\cal H}_T$ are localized in the $y$ coordinate at $y=y_c$. 
Eigenvalues depend on $k_z$, $y_c$ and the Landau index $n$;
\begin{align}
\hat{\cal H}_T \!\ \phi_{n,k_z,y_c}(y-y_c) = E_{n}(k_z,y_c) \phi_{n,k_z,y_c}(y-y_c). \label{eigen} 
\end{align}

A single-particle velocity operator along $x$ is given by a $k_x$ derivative 
of $\hat{\cal H}_T$. With $k_x l^2 \equiv y_c$, an expectation value of 
the velocity with respect to 
the eigenstate is given by a $y_c$-derivative of the eigenvalue. Besides, the 
eigenstate is uniformly extended along $x$. Thus, an electric current 
carried by the eigenstate is given by 
\begin{eqnarray}
J_{x,n,k_z,y_c} = \frac{(-e)l^2}{\hbar L_x} \frac{\partial E_n(k_z,y_c)}{\partial y_c}. \label{current}
\end{eqnarray} 
The total current density from the $n$-the Landau level is 
a sum of $J_{x,n,k_z,y_c}$ over all the filled $k_z$ and $k_x \equiv y_c/l^2$ points;
\begin{align}
j_x &= \frac{1}{L_z L_y} \sum_{k_z} \sum_{k_x} J_{x,n,k_z,y_c} f_T(E_n(k_z,y_c)) \nonumber \\
&= \frac{(-e)}{\hbar L_y} \int^{\frac{\pi}{c_0}}_{-\frac{\pi}{c_0}} \frac{dk_z}{2\pi} 
\int^{+\infty}_{-\infty} \frac{dy_c}{2\pi} \frac{\partial E_n}{\partial y_c} f_T(E_n(k_z,y_c)). \label{B1} 
\end{align} 
$f_T(E)$ is a Fermi distribution function. At the zero temperature, this reduces to a step function,
\begin{align}
f_{T=0}(E_{n}) = \left\{\begin{array}{cl} 
\theta(\mu_{+} - E_{n}(k_z,y_c)) & \ (y_c \simeq L_y/2), \\
\theta(\mu_{-} - E_{n}(k_z,y_c)) & \ (y_c \simeq - L_y/2). \\
\end{array}\right. \label{B2} 
\end{align}
$\mu_{\pm}$ are Fermi levels around $y = \pm L_y/2$ respectively. 
In the presence of a Hall voltage $V_H$ in the $+y$ direction, $\mu_{+}-\mu_{-}=-eV_H$. 

In graphite under the high field, the two electron/hole pockets in the bulk region 
($n=0$/$n=-1$ LLs with $\uparrow$ and $\downarrow$ spins) end up with 
two electron/hole surface chiral Fermi arc (SCFA) states in the boundary region. 
Namely, 
$E_{n=0/-1,\sigma}(k_z,y_c)$ increases/decreases in energy, 
when $y_c$ goes from the bulk region to the boundary region (Fig.~\ref{fig:2});  
\begin{align}
\left\{\begin{array}{rc}
 E_{n=0,\sigma}(k_z,y_c) \!\ \!\  \nearrow & (|y_c| \!\ \nearrow), \\
 E_{n=-1,\sigma}(k_z,y_c)  \!\ \!\ \searrow & (|y_c| \!\ \nearrow). \\ 
\end{array}\right. 
\end{align}  
Accordingly, the current density 
induced by the finite Hall voltage comprises of two parts that cancel each other:
\begin{align}
j_x &= -\frac{e}{h} \frac{(\mu_{+}-\mu_{-})}{L_y} \Big(\int^{k_1}_{-k_1} \frac{dk_z}{2\pi} 
+ \int^{k_2}_{-k_2}\frac{dk_z}{2\pi} \Big) \nonumber \\ 
& \hspace{-0.2cm} -\frac{e}{h} \frac{(\mu_{-}-\mu_{+})}{L_y} \Big(\int^{\frac{2\pi}{c_0}-k_3}_{k_3} 
\frac{dk_z}{2\pi}  + \int^{\frac{2\pi}{c_0}-k_4}_{k_4}\frac{dk_z}{2\pi} \Big). \label{B3}
\end{align}  
The first part is from the two electron surface states that subtend chiral arcs from 
$k_z=-k_1$ to $k_z=k_1$ and from $k_z=-k_2$ to $k_z=k_2$ respectively. The other  
part is from the two hole surface states that subtend chiral arcs from $k_z=k_3$ 
to $2\pi/c_0-k_3$ and from $k_z=k_4$ to $k_z=2\pi/c_0-k_4$ respectively (Fig.~\ref{fig:2}).   
To have Eq.~(\ref{B3}), we assume that the hole pocket energies are same 
in the vacuum,
\begin{align}
E_{n=-1,\sigma}(k_z,y_c=-\infty) = E_{n=-1,\sigma}(k_z,y_c=+\infty). \label{assumption} 
\end{align}
Eq.~(\ref{B3}) gives the Hall conductivity as,
\begin{align}
\sigma_{xy} &= \frac{e^2}{h} \frac{1}{L_z} \Big(N_1+N_2-N_3-N_4\Big) \nonumber \\
& = \frac{ec}{H} (n_e-n_h), \label{formula2} 
\end{align}
with $(N_1+N_2)/L_z = 2\pi l^2 n_e$ and $(N_3+N_4)/L_z = 2\pi l^2 n_h$. From the previous 
Hall conductivity measurement~\cite{akiba15}, we typically have   
\begin{align}
&n_e - n_h = 5 \times 10^{15} \!\ [{\rm cm}^{-3}], \nonumber \\
&c_0 = 6.7 \times 10^{-10} \!\ [{\rm m}], \ \ l = 45 \times 10^{-10} \!\ [{\rm m}], \nonumber  
\end{align}
for $H=30$ T and 
\begin{align}
&n_e - n_h = -10 \times 10^{15} \!\ [{\rm cm}^{-3}], \nonumber \\
&c_0 = 6.7 \times 10^{-10} \!\ [{\rm m}], \ \ l = 40 \times 10^{-10} \!\ [{\rm m}], \nonumber  
\end{align}
for $H=55$ T. This gives out a ratio between $N_e-N_h$ and $L_z/c_0$ as 
\begin{eqnarray}
N_e - N_h : L_z/c_0 = \pm 3 \times 10^{-4}:1 \label{Akiba2}
\end{eqnarray}
for $30 \!\ {\rm T} < H < 55\!\ {\rm T}$. From this very small number, 
we conclude that graphite under this field regime safely satisfies the 
charge neutrality condition.

\section{renormalization of Luttinger parameters and Fermi velocities}
In the main text, we use the Hartree-Fock approximation for the four pockets model  
or two pockets model, to introduce effective boson Hamiltonians, such as 
Eqs.~(\ref{H00},\ref{H0},\ref{luttinger1},\ref{luttinger2}) with  
Eqs.~(\ref{Hu1},\ref{Hu2},\ref{Hu3},\ref{Hu4},\ref{Hb13},\ref{Hb2},\ref{Hb4})  
or with Eqs.~(\ref{Hu1d},\ref{Hu2d},\ref{Hd1d},\ref{Hb2d}). 
Thereby, the bare kinetic energy part takes a quadratic form in the phase 
variables, Eq.~(\ref{H0}), whose coefficients (Luttinger parameters and 
Fermi velocities) are further renormalized by intra-pocket forward scattering terms. 
In the following, we summarize how the intra-pocket forward scattering terms 
renormalize the Luttinger parameters and Fermi velocities.  

The electron interaction within the same pockets is given by 
\begin{align}
H_{\rm f} &= \sum_{j,m,n} \sum_{a=1,2,3,4 \!\ (2,3)}  
\int dz \int dz' e^{-\frac{(z-z')^2}{2l^2_{0,z}}} V^{(1),a}_{n-m,n-j} \nonumber \\
&\hspace{1.2cm} \psi^{\dagger}_{a,n}(z) \psi^{\dagger}_{a,j+m-n}(z') \psi_{a,m}(z') \psi_{a,j}(z),  
\label{Hfa}
\end{align} 
with $\psi_{a,n}(z) \equiv 
e^{ik_{F,a}z}\psi_{a,+,n}(z) + e^{-ik_{F,a}z}\psi_{a,-,n}(z)$. The matrix element  
$V^{(1),a}_{n,m}$ ($a=1,2,3,4$) is obtained by the substitutions of 
Eqs.~(\ref{Hint2},\ref{exp1},\ref{exp2},\ref{exp3}) into Eq.~(\ref{Hint1}). In the 
limit of short interaction length $(l_{0,z}\ll l)$, the matrix element takes a form of 
\begin{eqnarray}
V^{(1),a}_{n,m} \equiv \frac{g}{L_x} \frac{1}{l_{0,z}l} \!\ f^{(1),a}
\Big(y_n/l,y_m/l\Big).  
\label{v1anm}
\end{eqnarray}
Dimensionless functions $f^{(1),a}(x,y)$ decay quickly 
for $|x|,|y|\gg 1$. With the Hartree-Fock 
approximation, $H_{\rm f}$ is bosonized into the followings,
\begin{align}
& H_{\rm f} = \sum_{a}\sum_{j,m} \int \!\ dz \sqrt{2\pi} l_{0,z} 
\big(V^{(1),a}_{j-m,0} - V^{(1),a}_{0,j-m} \big) \nonumber \\
& \hspace{1cm} \times 
\big(\rho_{a,+,j} \rho_{a+,m} + \rho_{a,-,j} \rho_{a,-,m} \big) \nonumber \\ 
& \ \ \  + \sum_{a}\sum_{j,m} \int dz \!\ \sqrt{2\pi} l_{0,z} \big(V^{(1),a}_{j-m,0} - 
V^{(1),a}_{0,j-m} e^{-2 (k_{F,a}l_{0,z})^2} \big) \nonumber \\
& \hspace{1cm} 
\times \big(\rho_{a,+,j} \rho_{a-,m} + \rho_{a,-,j} \rho_{a,+,m} \big) \nonumber \\ 
& \ \ \  + 2 \sum_{a}\sum_{j,m} \int dz \sqrt{2\pi} l_{0,z} 
\big(V^{(1),a}_{j-m,0} e^{-2 (k_{F,a}l_{0,z})^2} - V^{(1),a}_{0,j-m} \big) \nonumber \\
& \  \times \eta_{a,+,j}\eta_{a,-,j} \eta_{a,-,m} \eta_{a,+,m} \!\ 
\cos\big[2(\phi_{a,j}(z) - \phi_{a,m}(z))\big] \nonumber \\
& \ \ \   + \cdots, \label{Hf-boson} 
\end{align}
where $\rho_{a,\pm,j}(z)$ stands for an electron density in the right ($+$) or 
left ($-$) branch in the $a$-th pocket ($a=1,2,3,4$) of the $j$-th chain 
($j=1,2\cdots,\frac{S}{2\pi l^2}$);
\begin{align}
\rho_{a,\pm,j}(z) \equiv \psi^{\dagger}_{a,\pm,j} \psi_{a,\pm,j} 
= -\frac{1}{2\pi} \big(\partial_z \phi_{a,j} \mp \partial_{z} \theta_{a,j}\big). \nonumber 
\end{align} 
The third term in Eq.~(\ref{Hf-boson}) represents a rigidity between two displacement fields 
in different chains in the same pocket. When the corresponding inter-chain interaction is 
negative definite, this could result in charge density wave orders with broken  
translational symmetry along the field direction. An interplay 
between this inter-chain rigidity term and one of the 
umklapp term is discussed for the two-pocket model case (see Sec.~V). 

The first two terms in Eq.~(\ref{Hf-boson}) lead to the renormalizations of the Luttinger 
parameters and Fermi velocities. To quantify them, we employ a 
gradient expansion with respect to the chain index, 
\begin{align}
\rho_{a,\tau,m} &= \rho_{a,\tau,j} + (y_m-y_j) \partial_{y_j} \rho_{a,\tau,j} \nonumber \\
& \ \ \ \ + \frac{1}{2} (y_m-y_j)^2 \partial^2_{y_j} \rho_{a,\tau,j} 
+ \cdots, \label{gradient}
\end{align} 
to keep only the leading order. This leads to 
\begin{align}
H_{\rm f} &= \sum_{a} \sum_{j} \int \!\ dz \nonumber \\
&\hspace{-0.3cm} \bigg\{ \frac{g_{2,a}+g_{4,a}}{(2\pi)^2} 
\big(\partial_z \phi_{a,j}\big)^2 + \frac{-g_{2,a}+g_{4,a}}{(2\pi)^2} 
\big(\partial_z \theta_{a,j}\big)^2 \bigg\} + \cdots, \label{Hf-boson2}
\end{align} 
with 
\begin{align}
g_{2,a} &= 2\sqrt{2\pi} l_{0,z} \sum_{m} \big(V^{(1),a}_{m,0} 
- V^{(1),a}_{0,m} e^{-2 (k_{F,a}l_{0,z})^2}\big) \nonumber \\
&= \sqrt{\frac{2}{\pi}} \frac{g}{l^2} \int \!\ dx \big( f^{(1),a}(x,0)  
- f^{(1),a}(0,x) e^{-2 (k_{F,a}l_{0,z})^2} \big), \label{g2a-q} \\
g_{4,a} &= 2\sqrt{2\pi} l_{0,z} \sum_{m} \big(V^{(1),a}_{m,0} 
- V^{(1),a}_{0,m}\big) \nonumber \\
&= \sqrt{\frac{2}{\pi}} \frac{g}{l^2} \int \!\ dx \big( f^{(1),a}(x,0)  
- f^{(1),a}(0,x) \big). \label{g4a-q} 
\end{align}
When combined with the bare kinetic energy part;
\begin{eqnarray}
H_{\rm kin} = \sum_{a,j} \frac{v_{F,a}}{2\pi} \int  dz \Big\{  
\big(\partial_z \phi_{a,j}\big)^2 +  
\big(\partial_z \theta_{a,j}\big)^2 \Big\}, \label{H-kin-boson1}
\end{eqnarray}
Eq.~(\ref{Hf-boson2}) gives out Eq.~(\ref{H0}) with Eqs.~(\ref{luttinger1},\ref{luttinger2}). 

\section{derivation of renormalization group (RG) equations}
In the main text, we employ one-loop RG equations, Eqs.~(\ref{rg1},\ref{rg2},\ref{rg3}), and 
clarify possible insulating phases as well as natures of 
$T=0$ metal-insulator and insulator-metal transition points in graphite 
under the high field. We solve the RG equations numerically 
to obtain a finite temperature phase diagram as in Fig.~\ref{fig:1}. 
The RG equations are derived 
perturbatively by use of the standard momentum-shell renormalization 
method~\cite{giamarchi03}. In the following, 
we briefly summarize how to derive the one-loop RG equations for $H_{{\rm u},2}$ and $H_{{\rm b},2}$, 
Eqs.~(\ref{rg1},\ref{rg2},\ref{rg3}). 

We begin with a partition function of the effective field theory; 
\begin{eqnarray}
Z = \sum_{\sigma_{\cdots}}\int {\cal D}\phi {\cal D}\theta e^{-S_0[\phi,\theta]-S_1[\phi,\theta]}.   
\end{eqnarray}
An action $S$ comprises of a gaussian part $S_0$ and non-gaussian part $S_{1}$;
\begin{align}
S_{0} &= 
\int_{0}^{\beta}d\tau\int dz\sum_{a,j} \frac{1}{2\pi} \Big\{-2i\partial_z\theta_{a,j} 
(\bm{r})\partial_{\tau}\phi_{a,j}(\bm{r}) \nonumber \\ 
&\hspace{1.2cm} 
+u_{a}K_{a}[\partial_z\theta_{a,j}(\bm{r})]^2 + 
\frac{u_{a}}{K_{a}}[\partial_z\phi_{a,j}(\bm{r})]^2\Big\}, \label{S0} \\ 
S_{1} &= 
\int_{0}^{\beta} d\tau \Big\{ H_{{\rm u},2} + H_{{\rm b},2} + \cdots \Big\}.  
\label{S1} 
\end{align} 
Here $a$ is the pocket index ($a=1,2,3,4$). The summation over Ising variables 
$\sigma_{\cdots}$ represent traces over two-dimensional Hilbert spaces 
subtended by two Klein factors associated with the bosonization. 
With ${\bm r}=(z,\tau)$, ${\bm q}=(k_z,i\omega_n)$ 
and Matsubara frequency $\omega_n=2n\pi/\beta$, the Fourier transforms of 
$\phi_{j,a}(z,\tau)$ and $\theta_{j,a}(z,\tau)$ are given by 
\begin{align}
\phi_{j,a}({\bm r}) = \frac{1}{\beta L_z} \sum_{i\omega_n} \sum_{|k_z|<\Lambda} 
e^{ik_z z - i\omega_n \tau} \phi_{j,a}({\bm q}). \label{FT}
\end{align} 
$\Lambda$ is a cutoff in the momentum space. We decompose  
the field operators into a slow mode and a fast mode in the momentum space,
\begin{align}
\phi_{j,a}({\bm r}) &= \phi^{<}_{j,a}({\bm r}) + \phi^{>}_{j,a}({\bm r}), \nonumber \\
\phi^{<}_{j,a}({\bm r}) & = \frac{1}{\beta L_z} \sum_{i\omega_n} \sum_{|k_z|<\Lambda'} 
e^{i{\bm q}\cdot {\bm r}} \phi_{j,a}({\bm q}), \nonumber \\
\phi^{>}_{j,a}({\bm r}) & = \frac{1}{\beta L_z} \sum_{i\omega_n} \sum_{\Lambda'<|k_z|<\Lambda} 
e^{i{\bm q}\cdot {\bm r}} \phi_{j,a}({\bm q}), \nonumber 
\end{align}
with $\Lambda'=\Lambda b^{-1}$. $b$ ($>1$) denotes a scale change.    

First integrate out the fast mode $\phi^{>}$ and $\theta^{>}$ in the partition 
function and rescale spatial and temporal length scales as 
\begin{align}
z_{\rm new} = z_{\rm old} b^{-1}, \ \tau_{\rm new} = \tau_{\rm old} b^{-1}, \ \beta_{\rm new} = \beta_{\rm old} b^{-1}.   
\label{rescale}
\end{align}
This gives a partition function for the slow mode. The partition function  
takes essentially the same form as Eqs.~(\ref{S0},\ref{S1}), while 
the interchain interactions in Eq.~(\ref{S1}) are renormalized. 
The renormalization is calculated with respect to an infinitesimally 
small scale change $\ln b$ ($\ll 1$). This gives 
the RG equations for the interactions as in Eqs.~(\ref{rg1},\ref{rg2},\ref{rg3}). 

We derive the partition function for the slow mode perturbatively in the non-gaussian part $S_{1}$. 
We do so up to the second order in $S_{1}$;
\begin{align}
Z &= Z^{>}_{0} \int {\cal D}\phi^{<} {\cal D}\theta^{<} e^{-S^{<}_{0}} e^{- \langle S_{U} \rangle_{>}} 
+ {\cal O}(S^3_{i}), \label{Su1} 
\end{align} 
where 
\begin{align}
\langle S_{U} \rangle_{>} &= \langle S_{1} \rangle_{>} - \frac{1}{2} \big(\langle S^2_{1} \rangle_{>} 
- \langle S_{1} \rangle^2_{>}\big), \label{Su2} 
\end{align}
and 
\begin{align}
\langle \cdots \rangle_{>} &= \frac{1}{Z^{>}_{0}} \int {\cal D}\phi^{>} {\cal D}\theta^{>} 
\cdots e^{-S^{>}_{0}}, \nonumber \\
S^{<}_{0} &= \frac{1}{2\beta L_z} \sum_{a,j} \sum_{i\omega_n} \sum_{|k_z|<\Lambda'} \cdots, \nonumber \\
S^{>}_{0} & = \frac{1}{2\beta L_z} \sum_{a,j} \sum_{i\omega_n} \sum_{\Lambda'<|k_z|<\Lambda} \cdots,  \nonumber  
\end{align}
with $Z^{>}_{0}  = \langle 1 \rangle_{>}$. ``$\cdots$" in the right-hand sides 
of $S^{</>}_{0}$ is a Fourier transform of the integrand in Eq.~(\ref{S0}). The first term in Eq.~(\ref{Su2}) 
gives a tree-level renormalization to the interchain interactions, while the second term gives a 
one-loop level renormalization. 

\subsection{tree-level renormalization}   
$\langle S_1\rangle_{>}$ in Eq.~(\ref{Su2}) gives the tree-level renormalization to the interchain 
interactions;   
\begin{align}
\Big\langle \int_{0}^{\beta} H_{{\rm u},2} d\tau \Big\rangle_{>} 
&= \frac{1}{2} \int d^2{\bm r} \sum_{j,m} \sum_{\epsilon=\pm} 
\sum_{\eta=\pm} \nonumber \\
&\hspace{-0.5cm} M^{(2)}_{j-m} \sigma^{\eta}_{j} \tau^{\eta}_{m} e^{i\epsilon M^{\eta,<}_{jm}({\bm r})} 
e^{-\frac{1}{2} \big\langle M^{\eta,>}_{jm}({\bm r})^2 \big\rangle_{>}} \label{M-tree} \\
\Big\langle \int_{0}^{\beta} H_{{\rm b},2} d\tau \Big\rangle_{>} 
& = \frac{1}{2} \int d^2{\bm r} \sum_{j \ne m} \sum_{\epsilon=\pm} 
\sum_{\eta=\pm} \nonumber \\
&\hspace{-0.9cm} \Big\{ H^{(2)}_{j-m} \sigma^{\eta}_{j} \sigma^{\eta}_{m} e^{i\epsilon H^{\eta,<}_{jm}({\bm r})} 
e^{-\frac{1}{2} \big\langle H^{\eta,>}_{jm}({\bm r})^2 \big\rangle_{>}} \nonumber \\ 
&\hspace{-1.2cm} + \overline{H}^{(2)}_{j-m} \tau^{\eta}_{j} \tau^{\eta}_{m} e^{i\epsilon \overline{H}^{\eta,<}_{jm}({\bm r})} 
e^{-\frac{1}{2} \big\langle \overline{H}^{\eta,>}_{jm}({\bm r})^2 \big\rangle_{>}} \Big\} \label{HH-tree}
\end{align}
where 
\begin{align}
M^{\eta}_{jm}({\bm r}) &\equiv Q^{23}_{\eta,j}({\bm r}) + Q^{14}_{\eta,m}({\bm r}), \nonumber \\
H^{\eta}_{jm}({\bm r}) &\equiv Q^{23}_{\eta,j}({\bm r}) - Q^{23}_{\eta,m}({\bm r}), \nonumber \\
\overline{H}^{\eta}_{jm}({\bm r}) &\equiv Q^{14}_{\eta,j}({\bm r}) - Q^{14}_{\eta,m}({\bm r}), \nonumber 
\end{align}
and 
\begin{align}
&\sigma^{+}_{j} \equiv \sigma_{3\overline{2},j}, \ \ \sigma^{-}_{j} \equiv \sigma_{\overline{3}2,j}, \nonumber \\
&\tau^{+}_{j} \equiv \sigma_{4\overline{1},j}, \ \ \tau^{-}_{j} \equiv \sigma_{\overline{4}1,j}. \nonumber 
\end{align}
As the leading order in the infinitesimally small $\ln b$, we obtain, 
\begin{align}
\langle M_{jm}^{\eta,>}(\bm{r})^2\rangle_> 
&=\sum_{a=1,2,3,4}\frac{1}{2}\big(K_{a}+\frac{1}{K_{a}}\big)\coth\frac{\beta u_{a} \Lambda}{2} \ln b,\nonumber\\
\langle H_{jm}^{\eta,>}(\bm{r})^2\rangle_>
&=2\sum_{a=2,3}\frac{1}{2}\big(K_{a}+\frac{1}{K_{a}}\big)\coth\frac{\beta u_a \Lambda}{2}\ln b,\nonumber\\
\langle \overline{H}_{jm}^{\eta,>}(\bm{r})^2\rangle_>
&=2\sum_{a=1,4}\frac{1}{2}\big(K_{a}+\frac{1}{K_{a}}\big)\coth\frac{\beta u_a \Lambda}{2} \ln b,\nonumber
\end{align}
This leads to the tree-level RG equation as 
\begin{align}
\frac{dM^{(2)}_{j-m}}{d\ln b}&=\Big[2-\frac{1}{4}\sum_{a=1,2,3,4}\big(K_a+\frac{1}{K_a}\big) 
\coth\frac{u_a \Lambda}{2T}\Big]M^{(2)}_{j-m}, \nonumber\\
\frac{dH^{(2)}_{j-m}}{d\ln b}&=\Big[2-\frac{1}{2}\sum_{a=2,3}\big(K_a+\frac{1}{K_a}\big)\coth 
\frac{u_a \Lambda}{2T}\Big]H^{(2)}_{j-m},\nonumber\\
\frac{d\overline{H}^{(2)}_{j-m}}{d\ln b}&=\Big[2-\frac{1}{2}\sum_{a=1,4}\big(K_a+\frac{1}{K_a}\big) 
\coth\frac{u_a \Lambda}{2T}\Big]\overline{H}^{(2)}_{j-m}. \nonumber
\end{align}

\subsection{one-loop level renormalization} 
$\langle S^2_{1}\rangle_{>,c} \equiv \langle S^2_{1} \rangle_{>} - \langle S_{1} \rangle^2_{>} $ 
in Eq.~(\ref{Su2}) gives the one-loop level renormalization 
to the interchain interactions. The one-loop renormalization comprises of products between 
different interactions;
\begin{align}
S_{1} & = S_M + S_H + S_{\overline{H}} + \cdots, \nonumber \\ 
S^2_1 &= S^2_{M} + S^2_{H} + S^2_{\overline{H}} \nonumber \\
& \ \ \ \ \  + 2S_M S_H + 2S_M S_{\overline{H}} 
 + 2S_H S_{\overline{H}} + \cdots, \label{square} 
\end{align}
where $S_{M}$, $S_H$ and $S_{\overline{H}}$ are defined as follows,
\begin{align}
S_I &\equiv \frac{1}{2} \int d^2{\bm r} \sum_{j\ne m}\sum_{\epsilon=\pm} 
\sum_{\eta=\pm}  \nonumber \\
& \hspace{0.8cm}   I^{(2)}_{j-m}(\cdots)^{\eta}_{j}(\cdots)^{\eta}_{m} e^{i\epsilon I^{\eta,<}_{jm}({\bm r})} 
e^{i\epsilon I^{\eta,>}_{jm}({\bm r})},  
\end{align}
with $I=M, H,\overline{H}$. The products of two interaction terms take 
forms of
\begin{align}
\langle S_I S_J \rangle_{>,c} &= \frac{1}{4} \int d^2{\bm r} \int d^2{\bm r}' 
\sum_{i\ne j} \sum_{m \ne n} \sum_{\epsilon,\epsilon',\eta,\eta'} I^{(2)}_{i-j} J^{(2)}_{m-n} \nonumber \\
&\hspace{-1.4cm} (\cdots)^{\eta}_{i}
(\cdots)^{\eta}_{j} (\cdots)^{\eta'}_{m} (\cdots)^{\eta'}_{n} \langle e^{i\epsilon I^{\eta}_{ij}({\bm r})}
e^{i\epsilon' J^{\eta'}_{mn}({\bm r}')} \rangle_{>,c} \label{SISJ}
\end{align}  
where $\langle AB \rangle_{>,c} \equiv \langle AB \rangle_{>}-\langle A\rangle_{>}\langle B\rangle_{>}$. 
When $i\ne m,n$ and $j\ne m,n$ in Eq.~(\ref{SISJ}), the right hand side vanishes identically. 
The terms with $i=m$ and $j=n$ or those with $i=n$ and $j=m$ are negligibly smaller than 
the others in the larger $L_x$ limit. We thus consider only those terms in 
Eq.~(\ref{SISJ}) with $i=m,n$ and $j\ne m,n$ and/or those terms with $i\ne m,n$ and $j=m,n$. 

The one-loop renormalization in Eq.~(\ref{SISJ}) generates $S_M$, $S_H$ and $S_{\overline{H}}$ 
as well as other types of cosine terms. Nonetheless, tree-level scaling dimensions of all the 
other cosine terms thus generated are negatively much larger than those of $S_M$, $S_H$ 
and $S_{\overline{H}}$. 
Namely, they are much more irrelevant than $S_M$, $S_H$ and $S_{\overline{H}}$ at 
the tree-level renormalization 
group flow. Thus, we only keep those terms in Eq.~(\ref{SISJ}) that generate $S_M$, $S_H$ and 
$S_{\overline{H}}$. $S^2_{M}$ with $\epsilon=-\epsilon'$, $\eta=\eta'$ and $i=m$ (or $j=n$) 
generates $S_{\overline{H}}$ (or $S_H$) respectively. $S^2_{H}$ ($S^2_{\overline{H}}$) 
with $\epsilon=-\epsilon'$, $\eta=\eta'$, and $i=m$ or $j=n$ or with 
$\epsilon=\epsilon'$, $\eta=\eta'$, and $i=n$ or $j=m$ generates $S_H$ ($S_{\overline{H}}$) 
respectively. $S_M S_H$ ($S_M S_{\overline{H}}$) with 
$\epsilon=\epsilon'$, $\eta=\eta'$, and $i=n$ ($j=n$) or with 
$\epsilon=-\epsilon'$, $\eta=\eta'$, and $i=m$ ($j=m$) generates $S_M$. 
$S_H S_{\overline{H}}$ does not generate any of $S_M$, $S_H$ and $S_{\overline{H}}$. In the 
following, we only demonstrate how $S^2_{M}$ generates $S_{\overline{H}}$. 

With $\epsilon=-\epsilon'$, $\eta=\eta'$ and $i=m$, Eq.~(\ref{SISJ}) with $I=J=M$ reduces to 
\begin{align}
\langle S^2_M \rangle_{>,c}
& = \frac{1}{4} \int d^2{\bm r} \int d^2{\bm r}' 
\sum^{j\ne n}_{j,n} \sum_{i=m} \sum_{\epsilon,\eta} \tau^{\eta}_{j} \tau^{\eta}_{n} M^{(2)}_{i-j} 
M^{(2)}_{i-n} \nonumber \\
&\hspace{-1.0cm}   e^{i\epsilon (M^{\eta,<}_{ij}({\bm r}) 
- M^{\eta,<}_{in}({\bm r}'))} \big\langle e^{i\epsilon M^{\eta,>}_{ij}({\bm r})}  
e^{- i\epsilon M^{\eta,>}_{in}({\bm r}')} \big\rangle_{>,c} \nonumber \\
& = \frac{1}{2} \int d^2{\bm r} \int d^2{\bm r}' 
\sum^{j\ne n}_{j,n} \sum_{i=m} \sum_{\epsilon,\eta} \tau^{\eta}_{j} \tau^{\eta}_{n} M^{(2)}_{i-j} 
M^{(2)}_{i-n} \nonumber \\
&\hspace{-1.2cm}  \cos\big[ M^{\eta,<}_{ij}({\bm r}) 
- M^{\eta,<}_{in}({\bm r}')) \big] \big\langle 
M^{\eta,>}_{ij}({\bm r}) M^{\eta,>}_{in}({\bm r}') \big\rangle_{>} \label{SMSM} 
\end{align} 
where 
\begin{align}
&\cos\big[ M^{\eta,<}_{ij}({\bm r}) 
- M^{\eta,<}_{in}({\bm r}')) \big] = \nonumber \\
&\  \cos \big[Q^{14,<}_{\eta,j}({\bm r}) - Q^{14,<}_{\eta,n}({\bm r}')\big] 
\cos \big[Q^{23,<}_{\eta,i}({\bm r}) - Q^{23,<}_{\eta,i}({\bm r}')\big] \nonumber \\ 
& \ - \sin \big[Q^{14,<}_{\eta,j}({\bm r}) - Q^{14,<}_{\eta,n}({\bm r}')\big] 
\sin \big[Q^{23,<}_{\eta,i}({\bm r}) - Q^{23,<}_{\eta,i}({\bm r}')\big]. \label{cosine}
\end{align}
The largest part of the contribution comes from ${\bm r}={\bm r}'$. 
In this case, the second term in Eq.~(\ref{cosine}) vanishes (see the next subsection 
for a justification of this approximation). For the first term with $j\ne n$, we replace 
$\cos [Q^{23,<}_{\eta,i}({\bm r}) - Q^{23,<}_{\eta,i}({\bm r}')]$ 
by its normal ordering with use of a formula 
$\cos \Phi = :\cos\Phi: \exp[-\langle \Phi^2\rangle/2]$~\cite{giamarchi03,nozieres87}. 
Within the normal order, we employ a Taylor expansion with respect to small 
${\bm r}'-{\bm r}$. At the leading order expansion, Eq.~(\ref{cosine}) becomes 
\begin{align}
&\cos\big[ M^{\eta,<}_{ij}({\bm r}) - M^{\eta,<}_{in}({\bm r}') \big] \simeq \nonumber \\ 
&\ \ \cos \big[Q^{14,<}_{\eta,j}({\bm r}) - Q^{14,<}_{\eta,n}({\bm r})\big]  
e^{-\frac{1}{2} \langle (Q^{23,<}_{\eta,i}({\bm r}) - Q^{23,<}_{\eta,i}({\bm r}'))^2\rangle_{<}}.  
\label{taylor-exp}
\end{align}
Thereby, we have 
\begin{align}
\langle S^2_M \rangle_{>,c}
& =\int d^2{\bm r} 
\sum^{j\ne n}_{j,n} 
\sum_{\epsilon,\eta} \tau^{\eta}_{j} \tau^{\eta}_{n}   \cos\big[\overline{H}^{\eta,<}_{jn}({\bm r})\big]  \nonumber \\ 
& \hspace{1.2cm}  \times C_{23} \sum_{i}  M^{(2)}_{i-j} M^{(2)}_{i-n} \ln b, \label{1-loop-1}
\end{align}
where 
\begin{align}
&C_{cd} \ln b \equiv \nonumber \\
&\frac{1}{2}\int d{\bm r}' 
e^{-\frac{1}{2} \langle (Q^{cd,<}_{\eta,i}({\bm r}) - Q^{cd,<}_{\eta,i}({\bm r}'))^2\rangle_{<}} 
\langle Q^{cd,>}_{\eta,i}({\bm r}) Q^{cd,>}_{\eta,i}({\bm r}') \rangle_{>}, \label{Ccd}
\end{align}
with $c,d=1,2,3,4$. Note that the integrand in Eq.~(\ref{Ccd}) is short-ranged in ${\bm r}-{\bm r}'$ 
and $C_{cd}$ is a positive definite real-valued quantity (see the next 
subsection). Eq.~(\ref{1-loop-1}) in combination 
with Eqs~(\ref{Su2},\ref{square}) dictates 
that $\overline{H}^{(2)}_{j-n}$ acquires the following one-loop renormalization,  
\begin{align}
\frac{d \overline{H}^{(2)}_{j-n}}{d\ln b} = \cdots -\frac{C_{23}}{2} \sum_{i} M^{(2)}_{i-j} M^{(2)}_{i-n} + \cdots.  
\end{align} 
Since $M^{(2)}_{i-j} = M^{(2)}_{j-i}$, this is nothing 
but the first term of the one-loop renormalization in Eq.~(\ref{rg3}).  Similarly, one can show 
all the other terms of the one-loop renormalizations in Eqs.~(\ref{rg1},\ref{rg2},\ref{rg3}). A  
factor ``$4$" in the second term of the one-loop renormalization in Eq.~(\ref{rg3}) is due to 
the four distinct contributions to $S_{\overline{H}}$ from $S^2_{\overline{H}}$;  
(i) $\epsilon=-\epsilon'$, $\eta=\eta'$, $i=m$, (ii) $\epsilon=-\epsilon'$, $\eta=\eta'$, $j=n$, 
(iii) $\epsilon=\epsilon'$, $\eta=\eta'$, $i=n$, (iv) $\epsilon=\epsilon'$, $\eta=\eta'$, $j=m$ 
in Eq.~(\ref{SISJ}). Likewise, $2S_MS_{H}$ ($2S_MS_{\overline{H}}$) has two distinct contributions 
to $S_M$, giving rise to the first (second) term of the one-loop renormalization in Eq.~(\ref{rg1}); 
(i) $\epsilon=\epsilon'$, $\eta=\eta'$, $i=n$ ($j=n$), (ii) $\epsilon=-\epsilon'$, $\eta=\eta'$, $i=m$ 
($j=m$) in Eq.~(\ref{SISJ}). This completes the derivation of Eqs.~(\ref{rg1},\ref{rg2},\ref{rg3}).

\subsection{evaluation of $C_{cd}$}
$C_{ab}$ is defined in Eq.~(\ref{Ccd}). Let us first calculate the integrand in Eq.~(\ref{Ccd});
\begin{align}
&\big\langle Q^{ab,>}_{\eta,i}({\bm r}) Q^{ab,>}_{\eta,i}({\bm r}') \big\rangle_{>} 
= \nonumber \\
& \frac{1}{(\beta L_z)^2} \sum_{\Lambda'<|k_z|<\Lambda} 
\sum_{i\omega_n} e^{i{\bm q}({\bm r}-{\bm r}')} 
\big\langle {Q^{ab,>}_{\eta,i}({\bm q})}^{*} Q^{ab,>}_{\eta,i}({\bm q}) \big\rangle_{>}, 
\nonumber \\ 
&\big\langle \big(Q^{ab,<}_{\eta,i}({\bm r}) - Q^{ab,<}_{\eta,i}({\bm r}') \big)^2\big\rangle_{<} = 
\nonumber \\
& \hspace{-0.1cm} 
\frac{1}{(\beta L_z)^2} \sum_{|k_z|<\Lambda'} \sum_{i\omega_n} 2(1- e^{i{\bm q}({\bm r}-{\bm r}')}) 
 \big\langle {Q^{ab,<}_{\eta,i}({\bm q})}^{*} Q^{ab,<}_{\eta,i}({\bm q}) \big\rangle_{<}, 
\nonumber 
\end{align}
where 
\begin{align}
&\big\langle {Q^{ab,>/<}_{\eta,i}({\bm q})}^{*} Q^{ab,>/<}_{\eta,i}({\bm q}) \big\rangle_{>/<} 
= \nonumber \\
& \ \sum_{c=a,b}  \Big\{\langle \phi^{*}_{c,i}({\bm q}) \phi_{c,i}({\bm q}) \rangle_{>/<} + 
\langle \theta^{*}_{c,i}({\bm q}) \theta_{c,i}({\bm q})\rangle_{>/<} \nonumber \\
& \hspace{-0.2cm} + \eta 
(-1)^{c} \big( \langle \phi^{*}_{c,i}({\bm q}) \theta_{c,i}({\bm q}) \rangle_{>/<} 
+  \langle \theta^{*}_{c,i}({\bm q}) \phi_{c,i}({\bm q}) \rangle_{>/<} \big) 
\Big\}. 
\end{align}
with $(-1)^a=1$ and $(-1)^b=-1$. We used Fourier transform in Eq.~(\ref{FT}). The 
gaussian integrals over the fast/slow modes lead to  
\begin{align}
&\langle \phi^{*}_{c,i}({\bm q}) \phi_{c,i}({\bm q}) \rangle_{>/<} = \frac{\beta L_z \pi u_{c} K_c}{u^2_c k^2_z + \omega^2_n}, \nonumber \\
&\langle \theta^{*}_{c,i}({\bm q}) \theta_{c,i}({\bm q}) \rangle_{>/<} = \frac{\beta L_z \pi u_{c} K^{-1}_c}{u^2_c k^2_z + \omega^2_n}, 
 \nonumber \\
& \langle \phi^{*}_{c,i}({\bm q}) \theta_{c,i}({\bm q}) \rangle_{>/<} = 
- \frac{\beta L_z i\pi \omega_n}{k_z (u^2_c k^2_z + \omega^2_n)}.  \nonumber 
\end{align}  
Accordingly, we have 
\begin{align}
&\langle Q^{ab,>}_{\eta,i}({\bm r}) Q^{ab,>}_{\eta,i}({\bm r}') \rangle_{>}  
= \frac{1}{2} \sum_{c=a,b} \big(K_c + K^{-1}_c\big) M_{c}({\bm r}-{\bm r}') 
\nonumber \\
&\hspace{2.7cm} + \sum_{c=a,b} \eta (-1)^c F^{\prime}_{2,c}({\bm r}-{\bm r}'),  \nonumber \\
&\big\langle (Q^{ab,<}_{\eta,i}({\bm r}) - Q^{ab,<}_{\eta,i}({\bm r}') )^2\big\rangle_{<} = 
\frac{1}{2} \sum_{c=a,b} \big(K_c + K^{-1}_c\big) F_{1,c}({\bm r}-{\bm r}') 
\nonumber \\
&\hspace{2.7cm} + \sum_{c=a,b} \eta (-1)^c F_{2,c}({\bm r}-{\bm r}'),   
\end{align}
with  
\begin{align}
M_{c}({\bm r}) &\equiv \int_{\Lambda'<|k_z|<\Lambda} dk_z \frac{1}{\beta}  \sum_{i\omega_n}
\frac{u_c e^{i{\bm q}{\bm r}} }{\omega^2_n + u^2_c k^2_z} \nonumber \\
& =  \cos (\Lambda z) e^{-u_c\Lambda |\tau|}  \ln b, \nonumber \\
F^{\prime}_{2,c}({\bm r}) &\equiv -\int_{\Lambda'<|k_z|<\Lambda} dk_z 
 \frac{1}{\beta}  \sum_{i\omega_n}  \frac{i\omega_n}{k_z} 
\frac{e^{i{\bm q}{\bm r}}}{\omega^2_n + u^2_c k^2_z} \nonumber \\
& = -i \!\ {\rm sgn} (\tau) \sin (\Lambda z) e^{-u_c \Lambda |\tau|} \ln b, \nonumber \\
F_{1,c}({\bm r}) &\equiv \int_{|k_z|<\Lambda'} dk_z 
 \frac{1}{\beta}  \sum_{i\omega_n}   
\frac{2(1-\cos({\bm q}{\bm r})) u_c }{\omega^2_n + u^2_c k^2_z} \nonumber \\
& = \log \big[(x^2+y^2_c)/\alpha^2 \big], \nonumber \\
F_{2,c}({\bm r}) &\equiv \int_{|k_z|<\Lambda'} dk_z 
 \frac{1}{\beta}  \sum_{i\omega_n}  \frac{i\omega_n}{k_z} 
\frac{2\cdot e^{i{\bm q}{\bm r}}}{\omega^2_n + u^2_c k^2_z} \nonumber \\
& = 2i {\rm Arg} \big[y_c + ix\big] \equiv 2i \theta_{c}({\bm r}), \nonumber 
\end{align}
and $y_c \equiv u_c \tau + \alpha {\rm sgn}(\tau)$. In the right hand side, 
$M_{c}({\bm r})$, $F^{\prime}_{2,c}({\bm r})$, $F_{1,c}({\bm r})$ and $F_{2,c}({\bm r})$ 
are evaluated 
at the zero temperature. Substituting these into Eq.~(\ref{Ccd}), we obtain 
$C_{ab}$ at $T=0$ as, 
\begin{align}
&C_{ab,T=0}  =  \sum_{c=a,b} \int^{\infty}_{-\infty} d\tau \int^{\infty}_{-\infty} dz 
\bigg(\frac{\alpha^2}{z^2+y^2_a}\bigg)^{\lambda_a} 
\bigg(\frac{\alpha^2}{z^2+y^2_b}\bigg)^{\lambda_b} \nonumber \\ 
& \hspace{0.8cm}
e^{-u_{c}\Lambda |\tau|} \Big\{ \lambda_c  
\cos (\Lambda z) \cos (\Delta_{ab}({\bm r})) \nonumber \\
&\hspace{2cm} 
+ \frac{\eta}{2} (-1)^c \sin(\Lambda z) {\rm sgn}(\tau) \sin (\Delta_{ab}({\bm r})) 
\Big\} \nonumber \\
& \hspace{1.2cm} \simeq \sum_{c=a,b} \lambda_c 
\int^{\infty}_{-\infty} d\tau \!\  e^{-u_{c}\Lambda |\tau|}  \!\ 
\int^{\infty}_{-\infty} dz \nonumber \\ 
&\hspace{1.9cm} 
\bigg(\frac{\alpha^2}{z^2+y^2_a}\bigg)^{\lambda_a} 
\bigg(\frac{\alpha^2}{z^2+y^2_b}\bigg)^{\lambda_b} 
\cos (\Lambda z), 
 \label{Cab-2}
\end{align}
with $y^2_c \equiv (u_c |\tau| + \alpha)^2$, $\lambda_a\equiv \frac{1}{4}(K_a+K^{-1}_a)$
and $\Delta_{ab}({\bm r}) \equiv \theta_{a}({\bm r})-\theta_{b}({\bm r})$. The integrand 
in the first line is short-ranged in ${\bm r}$, justifying a posteriori the approximations 
made in Eqs.~(\ref{cosine},\ref{taylor-exp}). Based on the same spirit, we 
approximate $\Delta_{ab}({\bm r})$ by zero, to obtain the second line.  

$C_{ab}$ is positive definite. One can show this by  
carrying out the $z$-integral formally, 
\begin{align}
C_{ab,T=0} = \sum_{c=a,b} \lambda_c \int^{\infty}_{-\infty} 
d\tau \!\ G(\tau)  e^{-u_{c}\Lambda |\tau|}, \label{Cab-3}
\end{align} 
and  
\begin{align}
G(\tau) &\equiv \int^{\infty}_{-\infty} d\xi F_{a}(\xi;\tau) F_{b}(\Lambda-\xi;\tau) \!\ d\xi, \label{G} \\
F_{a}(\xi;\tau) & \equiv \int^{\infty}_{-\infty} dz \!\ e^{i\xi z} 
\Big(\frac{\alpha^2}{z^2+y^2_a}\Big)^{\lambda_a},  \nonumber \\
&= 2\sqrt{\pi} \alpha^{2\lambda_a} 
\Big(\frac{|\xi|}{2|y_{a}|}\Big)^{\lambda_a-\frac{1}{2}} \frac{K_{\lambda_a-\frac{1}{2}}(|y_a| |\xi|)}{\Gamma(\lambda_a)},  
\label{Fa}
\end{align}
with the Bessel function $K_{\nu}(x)$ and the Gamma function 
$\Gamma(x)$. Since $\lambda_a>1/2$, $F_{a}(\xi,\tau)$ is positive definite 
and so is $G(\tau)$. With Eq.~(\ref{Cab-3}), this assures the positive definiteness 
of $C_{ab,T=0}$.  

$C_{ab,T=0}$ in Eq.~(\ref{Cab-2}) depends on the Luttinger parameters $K_a$ and $K_b$. 
Nonetheless, the dependence is much weaker than that of $A_{ab}$ in Eq.~(\ref{Aab}). 
One can see this, by evaluating an upper bound of $C_{ab,T=0}$, 
\begin{align}
C_{ab,T=0} &< \sum_{c=a,b} \lambda_{c}  \int dz  
\bigg(\frac{\alpha^2}{z^2+\alpha^2}\bigg)^{\lambda_a+\lambda_b} 
\int d\tau e^{-u_{c}\Lambda |\tau|} \nonumber \\
&=\sum_{c=a,b} \frac{\alpha \lambda_c}{\Lambda_{\cal E}} 
\frac{\Gamma(\frac{1}{2}) \Gamma(\lambda_a+\lambda_b-\frac{1}{2})}{\Gamma(\lambda_a+\lambda_b)} 
\equiv C_{\rm u}. \nonumber 
\end{align}
$\Lambda_{\cal E}$ denotes a {\it finite} high-energy cutoff in the energy scale, 
$\Lambda_{\cal E}= \Lambda \times {\rm max}_{c=a,b}(u_c)$. When the Luttinger parameters 
get much smaller/larger than 1, $\lambda_{a}+\lambda_b \rightarrow +\infty$, 
the upper bound of $C_{ab,T=0}$ as well as $|A_{ab,T=0}|$ diverge;
\begin{align}
&C_{\rm u} \rightarrow \frac{\alpha}{\Lambda_{\cal E}}  \Gamma\Big(\frac{1}{2}\Big) 
\big(\lambda_{a}+\lambda_b\big)^{\frac{1}{2}}, \nonumber \\
&|A_{ab,T=0}| \rightarrow 2 \big(\lambda_{a}+\lambda_b\big).  
\end{align}  
Meanwhile, $C_{ab,T=0}/|A_{ab,T=0}|$ goes to the zero in the limit of 
$\lambda_{a}+\lambda_b \rightarrow +\infty$. For simplicity, 
we assume that $C_{ab,T=0}$ does not depend on the magnetic field $H$ in the main text.  
A typical value of $C_{ab,T=0}$ is evaluated in a simple case with $K_a=K_b=1$ 
and $u_a=u_b=u$;
\begin{align}
&C_{ab,T=0,K_{a,b}=1,u_{a,b}=u}  \nonumber \\
&\ = \int d\tau e^{-u\Lambda|\tau|} 
\int dz \frac{\alpha^2}{z^2 + (u|\tau|+\alpha)^2} e^{i\Lambda z} \nonumber \\
& \ = e^{-\Lambda \alpha} \frac{2\alpha^2}{u} \int^{\infty}_{0} dx \frac{e^{-2\Lambda x}}{x+\alpha} 
 =  e^{\Lambda \alpha} \frac{2\alpha^2}{u} E_1(2\Lambda\alpha). \label{intexp}
\end{align} 
$E_1(x)$ is the exponential integral. $\alpha$ is a lattice constant along the $z$-direction while 
$\Lambda$ is a high energy cutoff in the momentum space; $\Lambda \alpha={\cal O}(1)$. 

\subsection{parameters used in Fig.~\ref{fig:1}} 
To obtain theoretical phase diagram at finite temperature as in Fig.~\ref{fig:1}, 
we solved numerically the RG equations Eqs.~(\ref{rg1a},\ref{rg2a},\ref{rg3a}) for 
$H<H_0$ and Eqs.(\ref{rg1c},\ref{rg2c},\ref{rg3c}) for $H_{0}<H<H_1$. Thereby, a 
set of parameters in the 
RG equations are chosen in the following way. 

$C_{ab}$ has an engineering dimension of [length]/[energy]. From 
Eq.~(\ref{intexp}), we set 
\begin{eqnarray}
C_{ab} = \frac{2\alpha}{\Lambda_{\cal E}}, 
\end{eqnarray} 
for any $a,b=1,2,3,4$. $\alpha$ is the lattice constant of the graphite along 
the $c$-axis, $\alpha=c_0=6.7 {\rm \AA}$. 
$\Lambda_{\cal E}$ is a high energy cutoff in the energy scale. We set this to be a 
band width of the four pockets, $\Lambda_{\cal E}=40$ [meV]. 

According to Eqs.~(\ref{m2-i},\ref{h2-i},\ref{h2d-i}), 
$m_{(2)}$, $h_{(2)}$, $\overline{h}_{(2)}$, $n_{(2)}$, $p_{(2)}$, and $\overline{p}_{(2)}$ have the 
same engineering dimension as $\tilde{g} \equiv g/\alpha^2$, where $g$ represents 
an interaction strength as in Eq.~(\ref{Hint2}). For initial values of 
$m_{(2)}$, ... ,$\overline{p}_{(2)}$ in the RG flow, we set 
\begin{align}
\left\{\begin{array}{c}
\big(m_{(2)},h_{(2)},\overline{h}_{(2)}\big) = \tilde{g} \!\ (3,-1.25,-1.25), \\ 
\big(n_{(2)},p_{(2)},\overline{p}_{(2)} \big) = \tilde{g} \!\ (1.1, -1.25, -1.25). \\
\end{array}\right. 
\end{align}
A value of $\tilde{g}$ is set in the following way. We consider that  
the interaction is from the Coulomb interaction and therefore its 
typical interaction energy scale is given by  
\begin{align}
E_{\rm int} = \frac{e^2}{\epsilon l}.
\end{align}
The magnetic length $l$ depends on the magnetic field and the relative 
permittivity $\epsilon$ is set to $13$ for graphite. 
We regard that the Coulomb interaction ranges over the magnetic length in 
the $xy$ plane, and ranges over the Tohmas-Fermi screening length 
along the $z$ direction $\lambda_{\rm TF}$. We thus 
compare $E_{\rm int}$ with $g/(l^2 \lambda_{\rm TF})$ [see Eq.~(\ref{Hint2})]. 
This leads to  
\begin{align}
\tilde{g}  = \frac{g}{\alpha^2} = \frac{e}{\epsilon l} \frac{l^2}{\alpha^2} \lambda_{\rm TF}. \label{gtil}
\end{align} 
The screening length along the $c$-axis is set to 
 $\lambda_{\rm TF}=c_0/\sqrt{6}$. 

$A_{ab}$ in the RG equations is given by Eq.~(\ref{Aab}). 
$u_{c}\Lambda$ in Eq.~(\ref{Aab}) ($c=1,2,3,4$) is set to the high-energy 
cutoff in the energy scale, $\Lambda_{\cal E}=40$ [meV] .  
For the Luttinger parameters $K_a$ in Eq.~(\ref{Aab}), 
we use Eq.~(\ref{luttinger2}). The intra-pocket forward scattering strengths 
in Eq.~(\ref{luttinger2}) are set as,  
\begin{align}
g_{4,a=1} = g_{4,a=4} &= \tilde{g}, \nonumber \\
g_{4,a=2} = g_{4,a=3}&= \tilde{g}, \nonumber \\ 
g_{2,a=1} = g_{2,a=4}&= \tilde{g}/1.6, \nonumber \\ 
g_{2,a=2} = g_{2,a=3}&= \tilde{g}/1.1,  \nonumber 
\end{align} 
where $\tilde{g}$ is given in Eq.~(\ref{gtil}). The bare Fermi velocity  
in Eq.~(\ref{luttinger2}) $v_{F,a}$ is a $k_z$ derivative of the energy 
dispersion of the four pockets given in Eq.~(\ref{Enk});
\begin{align}
v_{F,a} &= \frac{\partial E_{n,\sigma}(k_z)}{\partial k_z}_{|k_z = k_{F,n,\sigma}} \nonumber \\
&\equiv - 2\gamma_2 c_0 \sin (2\pi \xi_{n,\sigma}) \label{vns}
\end{align} 
with $a=(n,\sigma)$; $1=(0,\uparrow)$, $2=(0,\downarrow)$, $3=(-1,\uparrow)$, 
and $4=(-1,\downarrow)$. We set $2\gamma_2=40$ [meV], and 
\begin{align}
\left\{\begin{array}{c}
\xi_{0,\uparrow} = \frac{1}{4} - \frac{H}{200 [{\rm T}]}, \\ 
\xi_{0,\downarrow} = \frac{1}{4} - \frac{H}{480 [{\rm T}]}, \\ 
\xi_{-1,\uparrow} = \frac{1}{4} + \frac{H}{480 [{\rm T}]}, \\ 
\xi_{-1,\downarrow} = \frac{1}{4} + \frac{H}{200 [{\rm T}]}. \\ 
\end{array}\right. \label{xi-} 
\end{align}
Eq.~(\ref{xi-}) realizes $H_0=50$ [T] and $H_1=120$ [T].

\section{calculation of optical conductivity $\sigma_{zz}(\omega)$} 
In the main text, we describe how the longitudinal 
optical conductivity along the field direction behaves in the SNEI phases 
as well as the metal-insulator transition points at $H=H_{c,1}$ and $H=H_{c,2}$. According to the 
linear response theory, the conductivity is given by a retarded correlation function between 
an electron polarization operator $\hat{P}_z$ and current operator $\hat{J}_z$. In the bosonization 
language, the former is a sum of the displacement fields over the pocket index ($a$) 
and the chain index ($j$), 
\begin{align}
\hat{P}_z = -\frac{e}{\pi} \sum_{j} \sum_{a} \int dz \phi_{a,j}(z). \label{Pz}
\end{align} 
The latter is a sum of the current density fields, 
\begin{align}
\hat{J}_z = \frac{e}{\pi} \sum_{j} \sum_{a} u_a K_a \int dz \partial_z \theta_{a,j}(z). \label{Jz}
\end{align} 

The correlation function is calculated with respect to a mean field action for the SNEI phases. 
For the mean field action, we employ a Gaussian approximation for $H_{{\rm u},2}$ and 
$H^{\prime}_{{\rm u},2}$, to replace their cosine terms by proper quadratic terms,
\begin{align}
H_{{\rm u},2} \simeq &\sum_{j,m} M^{(2)}_{j-m} \int dz \!\ 
\Big\{\big(\phi_{2,j} + \phi_{3,j} + \phi_{1,m} + \phi_{4,m}\big)^2 
\nonumber \\ 
& \hspace{1.5cm} + \big(\theta_{2,j} - \theta_{3,j} + \theta_{1,m} - \theta_{4,m}\big)^2  \Big\} \nonumber \\
H^{\prime}_{{\rm u},2} \simeq & \frac{1}{2} \sum_{j,m} N^{(2)}_{j-m} \int dz \!\   
\Big\{\big(\phi_{2,j} + \phi_{3,j} + \phi_{2,m} + \phi_{3,m}\big)^2 
\nonumber \\ 
& \hspace{1.5cm} 
+ \big(\theta_{2,j} - \theta_{3,j} - \theta_{2,m} + \theta_{3,m}\big)^2  \Big\}. \nonumber 
\end{align}
This in combination with $H_0$ in Eq.~(\ref{H0}), gives a gaussian (`mean-field') action that takes a form of 
\begin{eqnarray}
\mathcal{S}_{\rm MF}=\frac{1}{2\beta L_zN}\sum_{\bm{K}}\begin{pmatrix}
\vec{\phi}^{\dagger}_{\bm K} & \vec{\theta}^{\dagger}_{\bm K}
\end{pmatrix}[{\bm M}_{0,{\bm K}}]\begin{pmatrix}
\vec{\phi}_{\bm K} \\ \vec{\theta}_{\bm K}
\end{pmatrix}, 
\end{eqnarray}
with ${\bm K}\equiv (k_z,k,i\omega_n)$. The Fourier transform is taken with respect to the 
spatial coordinate $z$, imaginary time $\tau$ and 
the chain index $j$ ($y_{j}\equiv 2\pi l^2 j/L_x$);
\begin{eqnarray}
\phi_{a,j}(z,\tau) \equiv \frac{1}{\beta L_z N} \sum_{\bm K} e^{ik_z z + ik y_j - i\omega_n \tau} \phi_{a,{\bm K}}.  
\label{FF3} 
\end{eqnarray}
In the following, we briefly summarize how to calculate the retarded correlation function with respect to 
${\cal S}_{\rm MF}$ in the SNEI-I phase with/without disorder. 

For the model with two electron pockets and two hole pockets, the gaussian action is described 
by a 8 by 8 matrix, 
\begin{align}
[{\bm M}_{0,{\bm K}}] \equiv \left[\begin{array}{cc} 
{\bm A}_{\bm K} & {\bm B}_{\bm K} \\
{\bm C}_{\bm K} & {\bm D}_{\bm K} \\
\end{array}\right]. \label{8b8}
\end{align}
A 4 by 4 matrix ${\bm A}_{\bm K}$ is for the displacement fields of the four 
pockets $\phi_{a}$ ($a=1,2,3,4$), and 4 by 4 matrix 
${\bm D}_{\bm K}$ is for the superconducting phase 
fields of the four pockets $\theta_{a}$ ($a=1,2,3,4$). They are given by
\begin{widetext}
\begin{align}
{\bm A}_{\bm K} &\equiv \left[\begin{array}{cccc}
\frac{u_1}{\pi K_1}k_z^2+2M(0) & 2M(0) & 2M^*(k) & 2M^*(k) \\
2M(0) & \frac{u_4}{\pi K_4}k_z^2+2M(0) & 2M^*(k) & 2M^*(k) \\
2M(k) & 2M(k) & \frac{u_2}{\pi K_2}k_z^2+2M(0) & 2M(0) \\
2M(k) & 2M(k) & 2M(0) & \frac{u_3}{\pi K_3}k_z^2+2M(0) \\ 
\end{array}\right], \label{A-mat} \\ 
{\bm D}_{\bm K} &\equiv \left[\begin{array}{cccc}
\frac{u_1K_1}{\pi}k_z^2+2M(0) & -2M(0) & 2M^*(k) & -2M^*(k) \\
-2M(0) & \frac{u_4K_4}{\pi}k_z^2+2M(0) & -2M^*(k) & 2M^*(k) \\
2M(k) & -2M(k) & \frac{u_2K_2}{\pi}k_z^2+2M(0) & -2M(0) \\
-2M(k) & 2M(k) & -2M(0) & \frac{u_3K_3}{\pi}k_z^2+2M(0) \\ 
\end{array}\right], \label{D-mat}
\end{align}
\end{widetext}
where $M(k)\equiv \sum_jM^{(2)}_j e^{iky_j}$.
The other 4 by 4 matrices ${\bm B}_{\bm K}$ and ${\bm C}_{\bm K}$ 
connect the four $\phi$ fields and the four $\theta$ fields,
\begin{align}
{\bm B}_{\bm K} = {\bm C}_{\bm K} =  
\frac{ik_z \omega_n}{\pi} {\bm 1}_{4\times 4}. \label{BC-mat} 
\end{align}
${\bm 1}_{4 \times 4}$ stands for the 4 by 4 unit matrix. 

For later convenience, we introduce a new basis with respect to the 
pocket index;
\begin{align}
&\vec{\Phi} \equiv \left[\begin{array}{c}
\Phi_{+} \\
\Phi_{I} \\
\Phi_{II} \\
\Phi_{III} \\
\end{array}\right] \equiv   \frac{1}{2} \left[\begin{array}{cccc}
1 & 1 & 1 & 1 \\
1 & 1 & -1 & -1 \\
1 & -1 & 1 & -1 \\
1 & -1 & -1 & 1 \\
\end{array}\right] \left[\begin{array}{c}
\phi_{1} \\
\phi_{4} \\
\phi_{2} \\
\phi_{3} \\
\end{array}\right] \equiv {\bm T} \vec{\phi}, \nonumber \\
&\vec{\Theta} \equiv {\bm T} \vec{\theta}. \label{T-mat} 
\end{align}
In the right hand side, we omitted the subscript ${\bm K}$ for 
the $\phi$, $\theta$, $\Phi$ and $\Theta$ fields. 
With the new basis, the gaussian action is given by 
\begin{align}
\mathcal{S}_{\rm MF}=\frac{1}{2\beta L_zN}\sum_{\bm{K}}\begin{pmatrix}
\vec{\Phi}^{\dagger}_{\bm K} & \vec{\Theta}^{\dagger}_{\bm K}
\end{pmatrix}[{\bm M}_{c,{\bm K}}]\begin{pmatrix}
\vec{\Phi}_{\bm K} \\ \vec{\Theta}_{\bm K}
\end{pmatrix}, \nonumber 
\end{align} 
and 
\begin{align}
[{\bm M}_{c,{\bm K}}] \equiv \left[\begin{array}{cc} 
{\bm T}\!\ {\bm A}_{\bm K} \!\ {\bm T} & {\bm B}_{\bm K} \\
{\bm C}_{\bm K} & {\bm T} \!\ {\bm D}_{\bm K} \!\ {\bm T} \\
\end{array}\right]. \label{8b8-c}
\end{align}
We consider that the total displacement field $\Phi_{+}$ couples with 
a disorder potential through;
\begin{align}
\hat{H}_{\rm imp} = \sum_{j} \int dz \!\ \epsilon_{j}(z) \Phi^2_{+,j}(z). \label{imp-4}
\end{align}
Physically, such disorder potential $\epsilon_{j}(z)$ is nothing but a local fluctuation of
the dielectric constant. 
We take a quenched average over the local fluctuation as 
\begin{eqnarray}
\overline{\cdots} \equiv 
\frac{\int d{\epsilon}_{j}(z) \cdots e^{-\frac{1}{g_y} 
\sum_{j} \int dz \!\ \epsilon^2_{j}(z)}}{\int d{\epsilon}_{j}(z)  \!\ e^{-\frac{1}{g_y} 
\sum_{j} \int dz \!\ \epsilon^2_{j}(z)}}. \label{disorder-ave}
\end{eqnarray}
$g_y$ stands for a disorder strength associated with spatially (but {\it not} 
temporally) fluctuating dielectric constant. 

We first calculate an imaginary-time time-ordered correlation function between $\hat{P}_z$ and 
$\hat{J}_z$, and then take an analytic continuation, $i\omega_n \rightarrow \omega+i\eta$. 
This gives the retarded correlation function. The real part of the retarded correlation function 
is nothing but the optical conductivity $\sigma_{zz}(\omega)$;
\begin{align}
\sigma_{zz}(\omega) &= {\rm Re} \!\ 
\Big\{\overline{\sigma_{zz}(i\omega_n)}_{|i\omega_n =\omega+i\eta}\Big\}, \nonumber \\
\overline{\sigma_{zz}(i\omega_n)} &= {\vec{e}_{+}}^{\!\ T} \!\ 
{\bm U}^{-1} \!\ {\bm T} \!\ \overline{{\bm Q}^{c}_{zz}(i\omega_n)} \!\  
{\bm T} \!\ \vec{e}_{+}, \label{G-formula}
\end{align} 
with $\vec{e}_{+} \equiv (1,1,1,1)^T$. ${\bm U}^{-1}$ and ${\bm Q}^c_{zz}(i\omega_n)$ 
($\overline{{\bm Q}^c_{zz}(i\omega_n)}$ is the quenched average of ${\bm Q}^c_{zz}(i\omega_n)$) 
as well as ${\bm T}$ are 4 by 4 matrices,
\begin{align}
{\bm U}^{-1} \equiv \left[\begin{array}{cccc} 
u_1 K_1 & & & \\
& u_4 K_4 & & \\
& & u_{2} K_2 & \\
& & & u_3 K_3 \\
\end{array}\right]. 
\end{align} 
${\bm Q}^{c}_{zz}(i\omega_n)$ is 
a Fourier transform of the imaginary-time time-ordered correlation function between four 
$\Phi$ fields and four $\Theta$ fields,
\begin{align}
{\bm Q}^{c}_{zz}(i\omega_n) &= \int^{\beta}_{0} d\tau \!\ {\bm Q}^{c}_{zz}(\tau) \!\ e^{i\omega_n \tau}, \nonumber \\
[{\bm Q}^{c}_{zz}(\tau)]_{\alpha\beta} &\equiv \frac{e^2}{\pi^2 V} \sum_{j,m} \int dz 
\int dz' [{\bm R}^{c}_{jm}(\tau,z|0,z')]_{\alpha\beta}, \nonumber \\
 [{\bm R}^{c}_{jm}(\tau,z|0,z')]_{\alpha\beta} &\equiv \frac{\int d\vec{\Phi}_ d\vec{\Theta} 
e^{-{\cal S}_{\rm MF}} 
\partial_z \Theta_{\alpha,j}(z,\tau) \Phi_{\beta,m}(z,0)}{\int d\vec{\Phi} d\vec{\Theta} 
e^{-{\cal S}_{\rm MF}}},  \label{R-def} 
\end{align} 
with $\alpha,\beta=+,I,II,III$ and the chain index $j,m=1,\cdots,S/(2\pi l^2)$.    

With use of a Born approximation~\cite{zhang17},  
we can take the quenched average of ${\bm Q}^c_{zz}(i\omega_n)$,
\begin{align}
&\overline{Q^{c}_{zz}(-i\omega_n)}^T = \frac{2e^2}{\pi^2 V} \sum_{m} \int dz^{\prime\prime} 
\sum_{\bm k} e^{-ik_z z^{\prime\prime}-ik y_m} (-ik_z) \nonumber \\ 
&\hspace{0.4cm} 
\Big[{\bm 1}_{4\times 4} 
- \big[{\bm M}_{c,{\bm K}}^{-1}\big]_{\Phi\Phi} \!\ \big[{\bm P}(i\omega_n)\big] \Big]^{-1}  
\big[{\bm M}_{c,{\bm K}}^{-1}\big]_{\Phi\Theta}, \label{born}
\end{align}
where $[{\bm M}_{c,{\bm K}}^{-1}]_{\Phi\Phi}$, $[{\bm M}_{c,{\bm K}}^{-1}]_{\Phi\Theta}$, 
and $[{\bm P}(i\omega_n)]$ are 4 by 4 matrices. 
$[{\bm M}_{c,{\bm K}}^{-1}]_{\Phi\Phi}$ and $[{\bm M}_{c,{\bm K}}^{-1}]_{\Phi\Theta}$ are 
4 by 4 blocks of an inverse of the 8 by 8 matrix $[{\bm M}_{c,{\bm K}}]$ that connects  
$\Phi$ and $\Phi$ and that connects $\phi$ and $\Theta$ respectively;
\begin{align}
&[{\bm M}_{c,{\bm K}}^{-1}]_{\Phi\Phi} \equiv {\bm T} \!\ \big({\bm A} - {\bm B} 
{\bm D}^{-1} {\bm C}\big)^{-1} \!\ {\bm T}, \label{pp} \\
& [{\bm M}_{c,{\bm K}}^{-1}]_{\Phi\Theta} \equiv {\bm T} \!\ \big({\bm A} - {\bm B} 
{\bm D}^{-1} {\bm C}\big)^{-1} \!\ {\bm B} {\bm D}^{-1}\!\ {\bm T}. \label{pt}
\end{align} 
4 by 4 matrices ${\bm A}$, ${\bm B}$, ${\bm C}$, ${\bm D}$ and ${\bm T}$ in 
the right hand sides are given by Eqs.~(\ref{A-mat},\ref{D-mat},\ref{BC-mat},\ref{T-mat}). 
$[{\bm P}(i\omega_n)]$ is a 4 by 4 diagonal matrix that represents an effect of the disorder,
\begin{align}
[{\bm P}(i\omega_n)] \equiv 
\left[\begin{array}{cccc} 
g_y m(i\omega_n) &  &  &  \\
 & 0 &  &  \\
 &  & 0 &  \\
 &  &  & 0 \\
\end{array}\right].  \label{P-def} 
\end{align}
$m(i\omega_n)$ is a sum of the $(\Phi_{+},\Phi_{+})$-component of the inverse of the 8 by 8 matrix 
$[{\bm M}_{c,{\bm K}}]$ over ${\bm k}\equiv (k_z,k)$;
\begin{align}
m(i\omega_n) \equiv \frac{2}{L_z N} \sum_{\bm k} [{\bm M}_{c,{\bm K}}^{-1}]_{\Phi_{+}\Phi_{+}}. 
\label{m-def} 
\end{align}
Note that $m(i\omega_n)$ is an even function of $\omega_n$ (see below). 

One may rewrite Eq.~(\ref{born}) into 
\begin{align}
&\overline{Q^{c}_{zz}(-i\omega_n)}^T \nonumber \\
&\ \ \ = \frac{2e^2}{\pi^2 V} \sum_{m} \int dz^{\prime\prime} 
\sum_{\bm k} e^{-ik_z z^{\prime\prime}-ik y_m} (-\omega_n) \nonumber \\
&\hspace{0.6cm} 
\!\ {\bm T}\!\ \Big[ \frac{\pi^2}{k^2_z} \big({\bm D}{\bm A} - {\bm D}{\bm T}{\bm P}{\bm T}\big) 
+ \omega^2_n {\bm 1}_{4\times 4} \Big]^{-1} {\bm T} \nonumber \\ 
& \ \ \ = -\frac{e^2\omega_n}{\pi^2 l^2} \!\ 
{\bm T}\!\ \Big[ \frac{\pi^2}{k^2_z} \big({\bm D}{\bm A} - {\bm D}{\bm T}{\bm P}{\bm T}\big) 
+ \omega^2_n {\bm 1}_{4\times 4} \Big]^{-1}_{|{\bm k}={\bm 0}} {\bm T} \label{born-2}
\end{align}
From the first to the second line, we took the sum over the chain index $l$ and 
the integral over $z''$;
\begin{eqnarray}
\frac{1}{V} \sum_{m} \int dz'' e^{-ik_z z'' -ik y_m} = \frac{1}{2\pi l^2} \delta^2_{{\bm k},{\bm 0}}. 
\label{sum-zdd-l} 
\end{eqnarray}
Substituting Eq.~(\ref{born-2}) into Eq.~(\ref{G-formula}), we obtain the imaginary-time optical 
conductivity as 
\begin{align}
&\overline{\sigma_{zz}(i\omega_n)} = \nonumber \\
& \ \ \ \ \frac{e^2 \omega_n}{\pi^2 l^2} {\vec{e}_{+}}^{\!\ T} 
\Big[\frac{\pi^2}{k^2_z} \big({\bm D}{\bm A} - {\bm D}{\bm T}{\bm P}{\bm T}\big) 
+ \omega^2_n {\bm 1}_{4\times 4} \Big]^{-1}_{|{\bm k}={\bm 0}} 
{\bm U}^{-1} \!\ \vec{e}_{+}. \label{born-3} 
\end{align}
The ${\bm k}={\bm 0}$ limit in the integrand is well-defined. 
To see this, use Taylor expansions of ${\bm A}$ and ${\bm D}$ in 
small $k$;
\begin{align}
{\bm A}_{\bm K} = 2M(0) {\bm A}_0 + k^2_z {\bm A}_1 + {\cal O}(k), \nonumber \\
{\bm D}_{\bm K} = 2M(0) {\bm D}_0 + k^2_z {\bm D}_1 + {\cal O}(k), \nonumber 
\end{align}
with 
\begin{align}
{\bm A}_0 \equiv \left[\begin{array}{cccc}
1 & 1 & 1 & 1 \\
1 & 1 & 1 & 1 \\
1 & 1 & 1 & 1 \\
1 & 1 & 1 & 1 \\
\end{array}\right], \  
{\bm D}_0 \equiv \left[\begin{array}{cccc}
1 & -1 & 1 & -1 \\
-1 & 1 & -1 & 1 \\
1 & -1 & 1 & -1 \\
-1 & 1 & -1 & 1 \\
\end{array}\right],   \nonumber 
\end{align}
and 
\begin{align}
&{\bm A}_1 \equiv \frac{1}{\pi}\left[\begin{array}{cccc}
\frac{u_1}{K_1} &  &  &  \\
 & \frac{u_4}{K_4} &  &  \\
 &  & \frac{u_2}{K_2} &  \\
 &  &  & \frac{u_3}{K_3} \\
\end{array}\right], \nonumber \\  
&{\bm D}_1 \equiv \frac{1}{\pi} \left[\begin{array}{cccc}
u_1 K_1 &  &  &  \\
 & u_4 K_4 &  & \\
 &  & u_2 K_2 &  \\
 &  &  & u_3 K_3 \\
\end{array}\right]. \nonumber   
\end{align}
Since ${\bm D}_0{\bm T}{\bm P}={\bm 0}$ and ${\bm D}_0{\bm A}_0 = {\bm 0}$, 
the integrand in the ${\bm k}=0$ limit takes a finite value; 
\begin{align}
&\lim_{k_z \rightarrow 0} 
\lim_{k \rightarrow 0} \frac{1}{k^2_z}\big( {\bm D}_{\bm K}{\bm A}_{\bm K} 
- {\bm D}_{\bm K}{\bm T}{\bm P}{\bm T}\big)  \nonumber \\
&\ \ \ \ = 2M(0) \big({\bm D}_0 {\bm A}_1 + {\bm D}_1 {\bm A}_0\big) 
- {\bm D}_1 {\bm T}{\bm P}{\bm T}  \nonumber \\ 
&  \ \ \ \ = 2M(0) {\bm D}_0 {\bm A}_1 + \frac{1}{\pi} \Big(2M(0) - \frac{g_y m(i\omega_n)}{4} \Big) 
{\bm U}^{-1} {\bm A}_{0}. \nonumber 
\end{align}
From the second to the last line, we used ${\bm T}{\bm P}{\bm T}=\frac{g_y m(i\omega_n)}{4}{\bm A}_0$ 
and $\pi {\bm D}_1 = {\bm U}^{-1}$. 

The imaginary-time optical conductivity is further calculated 
from Eq.~(\ref{born-3}) as,
\begin{align}
&\overline{\sigma_{zz}(i\omega_n)} = \frac{e^2 \omega_n}{\pi^2 l^2} \!\ {\vec{e}_{+}}^{\!\ T} 
\bigg[2\pi^2 M(0) {\bm D}_0{\bm A}_1 
\nonumber \\
&\hspace{0.0cm} + \pi \Big(2M(0)-\frac{g_y m(i\omega_n)}{4}\Big) 
{\bm U}^{-1} {\bm A}_0  
+ \omega^2_n {\bm 1}_{4\times 4}\bigg]^{-1} 
{\bm U}^{-1} \!\ \vec{e}_{+} \nonumber \\
& \ = \frac{e^2 \omega_n}{\pi^2 l^2} \!\ {\vec{e}_{+}}^{\!\ T} 
\bigg[\pi \Big( 2M(0) - \frac{g_y m(i\omega_n)}{4} \Big) {\bm U}^{-1} \vec{e}_{+} {\vec{e}_{+}}^{\!\ T}  
\nonumber \\
&\hspace{3cm} + \omega^2_n {\bm 1}_{4\times 4} \bigg]^{-1} 
{\bm U}^{-1} \!\ \vec{e}_{+}, \nonumber \\
& \ = \frac{e^2 uK}{\pi^2 l^2} \frac{\omega_n}{\omega^2_n + \pi uK \big(2M(0) - \frac{g_y m(i\omega_n)}{4}\big)},    
 \label{born-4} 
\end{align}
with $uK \equiv \sum_{a=1,2,3,4} u_a K_a$. From the first to the second line, we used 
${\bm A}_0 {\bm D}_0={\bm 0}$, ${\vec{e}_{+}}^{\!\ T}{\bm D}_0 = 0$ and 
${\bm A}_0 = \vec{e}_{+} {\vec{e}_{+}}^{\!\ T}$. From the second to the last line, we used 
${\vec{e}_{+}}^{\!\ T} {\bm U}^{-1} \vec{e}_{+} = uK$. 
In the clean limit ($g_y=0$), 
this gives $\sigma_{zz}(\omega)=(e^2 uK)/(2\pi l^2) \delta(\omega-\omega_g)$ 
with $\omega_g \equiv 2\pi uK \sum_{j} M^{(2)}_{j}$ after the analytic continuation.  

The effect of the disorder average is included in $m(i\omega_n)$. To see this effect in 
$\sigma_{zz}(\omega)$, let us take $u_1=u_4$, $K_1=K_4$, $u_2=u_3$ and $K_2=K_3$ for 
simplicity. With use of $M(k)=0$ for $k\gg 1/l$~\cite{zhang17}, we 
 obtain the following expression for $m(i\omega_n)$,
\begin{align}
m(i\omega_n) = \frac{\pi}{2} \Big(\frac{K_1}{\sqrt{\omega^2_n+\omega^2_1}} 
+ \frac{K_2}{\sqrt{\omega^2_n+\omega^2_2}}\Big), \label{m-cal}
\end{align}
with $\omega^2_1\equiv 4\pi M(0) u_1 K_1<4\pi M(0) u_2 K_2 \equiv \omega^2_2$. 
After the analytic continuation, we finally obtain the optical conductivity as follows,
\begin{align}
\sigma_{zz}(\omega) = \left\{\begin{array}{cc} 
\frac{e^2 uK}{\pi l^2} \frac{|\omega_{*}|}{|g'(\omega_*)|} \delta(\omega - \omega_{*}) & 0<\omega < \omega_1,  \\
\frac{e^2 uK}{\pi^2 l^2} \frac{\omega b_1(\omega)}{a^2_1(\omega)+b^2_1(\omega)} & \omega_1 < \omega < \omega_2, \\
\frac{e^2 uK}{\pi l^2} \frac{\omega b_2(\omega)}{a^2_2(\omega)+b^2_2(\omega)} & \omega_1 < \omega_2 < \omega, \\
\end{array}\right. \label{cond-1}
\end{align}
where 
\begin{align}
&g(\omega) = -\omega^2 + \omega^2_g \nonumber \\
& \ \ \ - \frac{g_y}{4}\frac{\pi^2 uK}{2} \bigg( \frac{K_1}{\sqrt{\omega^2_1-\omega^2}} 
+ \frac{K_2}{\sqrt{\omega^2_2-\omega^2}} \bigg), \nonumber 
\end{align}
and 
\begin{align}
a_{1}(\omega) 
& \equiv -\omega^2 + \omega^2_g - 
\frac{g_y}{4}\frac{\pi^2 uK}{2} \frac{K_2}{\sqrt{\omega^2_2-\omega^2}}, \nonumber \\
b_1(\omega) & \equiv \frac{g_y}{4} \frac{\pi^2 uK}{2} \frac{K_1}{\sqrt{\omega^2 -\omega^2_1}}, \nonumber 
\end{align}
and 
\begin{align}
a_2(\omega) &\equiv -\omega^2 + \omega^2_g, \nonumber \\
b_2(\omega) & \equiv \frac{g_y}{4} \frac{\pi^2 uK}{2} \bigg( 
\frac{K_1}{\sqrt{\omega^2 -\omega^2_1}}  
+ \frac{K_2}{\sqrt{\omega^2 -\omega^2_2}} \bigg).  
\nonumber 
\end{align}
Note that $\omega=\omega_{*} (<\omega_g)$ in Eq.~(\ref{cond-1}) is one and only one  
solution of $g(\omega)=0$ within $0<\omega<\omega_{1}$. The renormalized 
gap $\omega_{*}$ becomes progressively smaller, when the disorder strength increases. 
There exists a critical value of the disorder,  
\begin{eqnarray}
g_{y,c} \equiv \frac{1}{\pi^2 uK} \frac{8 \omega^2_{g} \omega_1 \omega_2}{K_1 \omega_2 + K_2 \omega_1}. \label{criti}
\end{eqnarray} 
When $g_y$ approaches the critical value, the renormalized gap $\omega_{*}$ reduces to 
zero continuously. At $g_y=g_{y,c}$, the system undergoes a quantum phase transition from 
the SNEI-I phase ($g_y<g_{y,c}$) to a disorder-driven phase ($g_y>g_{y,c}$). To obtain 
Fig.~\ref{fig:4}, we use the same parameter sets as in the appendix C4. We set 
$u_1=u_4$ and $u_2=u_3$ by Eq.~(\ref{luttinger1}). We set $g_y$ to be smaller than $g_{y,c}$. 

\section{magnetism and spin nematicity in SNEI phases}
 SNEI phases introduced in the main text are characterized by particle-hole pairings between 
$n=0$ LL with $\uparrow$ ($\downarrow$) spin and $n=-1$ LL with $\downarrow$ ($\uparrow$) 
spins. The phases break the U(1) spin rotational symmetry around the field direction. Nonetheless, 
neither $A$-carbon site $\pi$-orbital electron spin nor $B$-carbon site electron spins 
exhibit magnetic 
order in the SNEI phases;
\begin{align}
\langle S_{A,+}({\bm r}) \rangle &= \langle \psi^{\dagger}_{\uparrow}({\bm r},A) \psi_{\downarrow}({\bm r},A) \rangle 
= 0, \nonumber \\
\langle S_{B,+}({\bm r}) \rangle &= \langle \psi^{\dagger}_{\uparrow}({\bm r},B) \psi_{\downarrow}({\bm r},B) \rangle 
 \nonumber \\ 
&= \frac{1}{L_x} \sum_{j} \Big(Y_{1,j}(y) Y_{0,j}(y)\Big) \times \nonumber \\
& \hspace{-0.9cm} 
\times \sum_{\tau=\pm} \Big( 
\gamma^{*}_{B,\uparrow} \eta_{B,\downarrow} e^{-i\tau (k_{F,1}+k_{F,4})z} \langle \psi^{\dagger}_{1,\tau,j} \psi_{4,-\tau,j} \rangle 
\nonumber \\ 
& \hspace{-0.6cm} + \eta^{*}_{B,\uparrow} \gamma_{B,\downarrow} e^{-i\tau (k_{F,2}+k_{F,3})z} 
\langle \psi^{\dagger}_{3,\tau,j}\psi_{2,-\tau,j} \rangle \Big) \nonumber \\
& = 0, \label{magn-SNEI}
\end{align} 
because  
\begin{align}
&\lim_{L_x \rightarrow \infty} 
\frac{1}{L_x} \sum_{j} Y_{1,j}(y) Y_{0,j}(y) \nonumber \\ 
&= \frac{1}{2\pi l^2} \int dy Y_{1,j}(y) Y_{0,j}(y) = 0. \nonumber 
\end{align} 

Magnetism of the SNEI-I phase is most explicitly manifested by a long-range order of a symmetric part of 
a 2nd rank spin tensor composed of spin-$\frac{1}{2}$ moment of $A$-carbon-site $\pi$-orbital 
electron and that of $B$-carbon-site. Such 2nd rank spin tensor has two components, 
\begin{align}
Q^{AB}_{+-}({\bm r}) &\equiv \langle S_{A,+}({\bm r}) S_{B,-}({\bm r}) \rangle, \nonumber \\
Q^{AB}_{++}({\bm r}) &\equiv \langle S_{A,+}({\bm r}) S_{B,+}({\bm r}) \rangle. \nonumber 
\end{align}  
In the SNEI-I phase, $Q^{AB}_{+-}({\bm r})$ vanishes identically, while 
$Q^{AB}_{++}({\bm r})$ exhibits both a ferro-type and a density-wave-type order;
\begin{align}
Q^{AB}_{++}({\bm r}) &= \langle \psi^{\dagger}_{\uparrow}({\bm r},A) \psi_{\downarrow}({\bm r},A) 
\psi^{\dagger}_{\uparrow}({\bm r},B) \psi_{\downarrow}({\bm r},B) \rangle \nonumber \\ 
& = \frac{1}{L_x} \Big(\sum_{j} Y^2_{0,j}(y) \Big) \frac{1}{L_x} \Big(\sum_{m} Y^2_{0,m}(y) \Big) \nonumber \\
& \hspace{-1.5cm}  
\times \Big\{ \gamma^{*}_{A,\uparrow} \gamma_{A,\downarrow} \eta_{B,\downarrow} \eta^{*}_{B,\uparrow} 
e^{-2i\Theta_{-}} \nonumber \\
& \hspace{-0.8cm} + \gamma^{*}_{A,\uparrow} \gamma_{A,\downarrow} \eta_{B,\downarrow} \eta^{*}_{B,\uparrow} 
e^{-2i\Theta_{-}} \nonumber \\ 
& \hspace{-0.6cm} + \gamma^{*}_{A,\uparrow} \gamma_{A,\downarrow} \eta_{B,\downarrow} \eta^{*}_{B,\uparrow} 
e^{i\Delta K z}e^{-i2\Phi_{-}-2i\Theta_{-}} \nonumber \\
& \hspace{-0.4cm}  + \gamma^{*}_{A,\uparrow} \gamma_{A,\downarrow} \eta_{B,\downarrow} \eta^{*}_{B,\uparrow} 
e^{-i\Delta K z}e^{i2\Phi_{-}-2i\Theta_{-}} \Big\}, \nonumber 
\end{align}
with $\Delta K \equiv k_{F,2}+k_{F,3}-k_{F,1}-k_{F,4}$. 
Here we used Eqs.~(\ref{phi-lock},\ref{theta-lock},\ref{Ising-lock}) and 
\begin{align}
\langle \psi^{\dagger}_{1,+,j}(z) \psi_{4,-,m}(z) \rangle &= \delta_{jm} i\sigma_{\overline{4}1,m} 
e^{i(\phi_{1}+\phi_4)+i(\theta_4-\theta_1)},  \nonumber \\
\langle \psi^{\dagger}_{1,-,j}(z) \psi_{4,+,m}(z) \rangle &= \delta_{jm} i\sigma_{4\overline{1},m} 
e^{-i(\phi_{1}+\phi_4)+i(\theta_4-\theta_1)},  \nonumber \\
\langle \psi_{2,+,j}(z) \psi^{\dagger}_{3,-,m}(z) \rangle &= \delta_{jm} i\sigma_{\overline{3}2,m} 
e^{-i(\phi_{2}+\phi_3)-i(\theta_3-\theta_2)}, \nonumber \\
\langle \psi_{2,-,j}(z) \psi^{\dagger}_{3,+,m}(z) \rangle &= \delta_{jm} i\sigma_{3\overline{2},m} 
e^{i(\phi_{2}+\phi_3)-i(\theta_3-\theta_2)}.  \nonumber
\end{align}
The spatial inversion symmetry generally allows 
\begin{align} 
\gamma^{*}_{A,\uparrow} \gamma_{A,\downarrow} \eta_{B,\downarrow} \eta^{*}_{B,\uparrow} = u  
\end{align}
This gives 
\begin{align}
Q^{ab}_{++}({\bm r}) = \frac{e^{-2i\Theta_{-}} }{(\pi^2 l^2)^2} 
\Big(u +  u\cos \big(\Delta K z - 2\Phi_{-} \big)\Big). \label{Qab++}
\end{align}

Note also that the SNEI phases could be accompanied by a long-range ordering 
of small magnetic moments within the $xy$ plane. Nonetheless, the moment does exist 
only in those spatial regions in the unit cell where two $\pi$-orbitals of $A$-carbon 
site and $B$-carbon site overlap. This statement 
is suggested by Eq.~(\ref{magn-SNEI}) and finite expectation values of the following 
two quantities in the SNEI phases;
\begin{align} 
\langle \psi^{\dagger}_{\uparrow}({\bm r},A) \psi_{\downarrow}({\bm r},B) \rangle 
&= \frac{\sqrt{2} i v}{\pi l^2} e^{-i\Theta_{-}} \cos\big((k_{F,1}+k_{F,4})z + \Phi_{-}\big), \nonumber \\
\langle \psi^{\dagger}_{\downarrow}({\bm r},A) \psi_{\uparrow}({\bm r},B) \rangle 
&= \frac{\sqrt{2} i w}{\pi l^2} e^{i\Theta_{-}} \cos\big((k_{F,2}+k_{F,3})z - \Phi_{-}\big), \nonumber 
\end{align}
with  
\begin{align}
v &\equiv \gamma^{*}_{A,\uparrow} \eta_{B,\downarrow} \langle \sigma_{1\overline{4},j}\rangle \ne 0, \nonumber \\
w &\equiv \gamma^{*}_{A,\downarrow} \eta_{B,\uparrow} \langle \sigma_{2\overline{3},j}\rangle \ne 0. \nonumber 
\end{align}

\begin{acknowledgements}
RS appreciate helpful discussion with  Zengwei Zhu, Benoit Fauque,  Kamran Behnia, John Singleton, 
Miguel A. Cazalilla, Kazuto Akiba, Masashi Tokunaga, Toshihito Osada, Gang Chen, and 
Yoshihiro Iwasa. This work was supported by NBRP of China Grants No.~2014CB920901, 
No.~2015CB921104, and No.~2017A040215. 
\end{acknowledgements}


\end{document}